\documentclass[a4paper,11pt]{article}
\pdfoutput=1 
\usepackage{jheppub} 
\usepackage{graphicx}
\usepackage{xcolor,colortbl}
\usepackage{subfig}
\usepackage{dcolumn}
\usepackage{mathtools}
\usepackage{bm}
\usepackage{amssymb}
\usepackage{amsmath}
\usepackage{epsfig}
\usepackage[toc,page]{appendix}
\usepackage{slashed}
\usepackage{hhline}
\usepackage{color}
\usepackage{multirow}
\usepackage[normalem]{ulem}
\usepackage{tikz}
\usepackage{tikz-feynman}
\usepackage{pifont}
\usetikzlibrary{snakes}
\usepackage{blindtext}
\usepackage{hyperref}
\usepackage{forest}
\usepackage{url}
\usepackage[shortlabels]{enumitem}
\usepackage{empheq}
\usepackage[most]{tcolorbox}
\newtcbox{\mymath}[1][]{nobeforeafter, math upper, tcbox raise base, enhanced, colframe=blue!30!black, colback=blue!30, boxrule=1pt,  #1}

\usepackage[T1]{fontenc} 
\def\sb{\color{blue}}
\title{\boldmath Detection possibility of a Pseudo-FIMP in presence of a thermal WIMP}
\author[a]{Subhaditya Bhattacharya, }
\author[b]{Jayita Lahiri, }
\author[a]{Dipankar Pradhan}

\newcommand{\bmt}{\begin{pmatrix}}
\newcommand{\emt}{\end{pmatrix}}
\newcommand{\ba}{\begin{array}{c}}
\newcommand{\ea}{\end{array}}
\newcommand{\be}{\begin{equation}}
\newcommand{\ee}{\end{equation}}
\newcommand{\bea}{\begin{eqnarray}}
\newcommand{\eea}{\end{eqnarray}}

\newcommand{\bi}{\begin{itemize}}
\newcommand{\ei}{\end{itemize}}

\newcommand{\baz}{\begin{array}{cc}}
\newcommand{\cmark}{\ding{51}}
\newcommand{\xmark}{\ding{55}}

\newcommand{\lsim}{\raisebox{-0.13cm}{~\shortstack{$<$ \\[-0.07cm]
      $\sim$}}~}

\newcommand\Scaption[1]{%
\refstepcounter{mysfig}%
\vskip.5\abovecaptionskip
  \sbox\@tempboxa{\small\themysfig~#1}%
  \ifdim \wd\@tempboxa >\hsize
    \small\themysfig~#1\par
  \else
    \global \@minipagefalse
    \hb@xt@\hsize{\hfil\box\@tempboxa\hfil}%
  \fi
  \vskip\belowcaptionskip}
\makeatother
\affiliation[a]{Department of Physics, Indian Institute of Technology Guwahati,\\	North Guwahati, Assam-781039, India,}
\affiliation[b]{II. Institut f{\"u}r Theoretische Physik, Universit{\"a}t Hamburg, 22761 Hamburg, Germany.}
\emailAdd{subhab@iitg.ac.in}
\emailAdd{jayita.lahiri@desy.de}
\emailAdd{d.pradhan@iitg.ac.in}
\abstract{A dark matter (DM) having feeble interaction with the visible sector can thermalise via substantial interaction with a 
Weakly Interacting Massive Particle (WIMP). Such DM candidates are categorised as pseudo-FIMP (pFIMP). pFIMP can provide 
both direct and indirect search prospects via WIMP loop. This work focuses into such possibilities. We provide all such one loop 
graphs involving scalar, fermion and vector boson particles via which pFIMP can interact with the Standard Model assuming 
both of them are stabilised via $\mathbb{Z}_2\otimes \mathbb{Z}_2^{\prime}$ symmetries. 
We elaborate upon a model where a fermion DM acts as WIMP and a scalar singlet acts as pFIMP having negligible Higgs portal interaction and substantial 
conversion via Yukawa interaction. We study in details the loop induced direct and indirect search prospects of the pFIMP in the relic density allowed region of the model.}

\keywords{Models for Dark Matter, Particle Nature of Dark Matter, Specific BSM Phenomenology.}

\DeclareUnicodeCharacter{2212}{-}
\begin{document} 
\maketitle
\flushbottom

%%%%%%%%%%%%%%%%%%%
\section{Introduction}
\label{sec:intro}
%%%%%%%%%%%%%%%%%%%
Particle dark matter (DM) has long been studied as it caters to most of the astrophysical and cosmological observations in a consistent manner. Apart from the electromagnetic charge neutrality and stability over the scale of universe's life time, there are not many unique characteristics that can be assigned to a DM particle. There exists a plethora of possibilities based on how it saturates the observed relic density as given by the PLANCK data ($\rm\Omega_{DM} h^2=0.1200\pm0.0012$ \cite{Planck:2018vyg}). In one class of models, the DM remains in equilibrium with thermal bath due to sizable interaction with the visible sector. Then it freezes out as the universe expands and cools down. If the depletion of DM occurs via $2_{\rm DM} \to 2_{\rm SM}$ interactions, then the annihilation cross-section required for the DM  to attain correct relic density is of the order of electroweak interaction ($\sim 10^{-10}~{\rm GeV}^{-2}$). Hence, such class of particles are dubbed as Weakly Interacting Massive Particle (WIMP) \cite{Gondolo:1990dk,Jungman:1995df}. In cases where the number changing processes are governed mainly within the dark sector via $3_{\rm DM}\to 2_{\rm DM}$ or $4_{\rm DM}\to 2_{\rm DM}$ processes, the annihilation cross-section requires to be much higher to adjust the additional phase factor suppression and such class of DM particles are called Strongly Interacting Massive Particle (SIMP) \cite{Hochberg:2014dra}. The other possibility is to assume the DM having a feeble interaction with the visible sector, so that the DM remains out of equilibrium and gets produced via decay or annihilation of thermal bath particles and saturates when the temperature drops below the DM mass. Such particles are called Feebly Interacting Massive Particles (FIMP) \cite{Hall:2009bx}. Several other possibilities like SIDM \cite{McDonald:2001vt,Kaplinghat:2013yxa}, cannibal DM \cite{Pappadopulo:2016pkp}, have also been ideated. Our discussion will mostly be centred around WIMP and FIMP. 

Detectability of these different DM particles often provide the key distinctive features amongst them. WIMPs having sizeable interaction with the SM have been explored in direct DM search experiments via nuclear/electron recoil in XENON \cite{XENON:2018voc,XENON:2020kmp,XENONCollaboration:2022kmb}, PandaX \cite{PandaX-II:2021nsg,PandaX-4T:2021bab}, LUX-ZEPLIN \cite{LZ:2022ufs} etc although not found yet. Similar signal for FIMP or SIMP is often difficult just because of the small interaction cross-section. Apart from the direct search, collider search of DM has been extensively studied where the DM carries away missing momentum or energy in presence of some visible leptons, photons or jets, both in context of LHC \cite{Goodman:2010yf,Goodman:2010ku,Rajaraman:2011wf,Fox:2011pm,Buchmueller:2013dya,Petrov:2013nia,Altmannshofer:2014cla,Capdevilla:2017doz,Bell:2015sza} and ILC \cite{Yu:2013aca,Essig:2013vha,Kadota:2014mea,Yu:2014ula,Freitas:2014jla,Dutta:2017ljq,Habermehl:2020njb}. Again, WIMPs have the best bet to provide such signals \cite{Baltz:2006fm,Jungman:1995df}, while for FIMP one has to look for disappearing charge track or displaced vertex signal, see for example, \cite{Belanger:2018sti}. However, null observation in both direct and collider searches put bounds on the available parameter space. Apart from these terrestrial DM search experiments, indirect search for DM stemming from its annihilation into photon \cite{Boehm:2003bt,Boehm:2002yz,Tylka:1989wt}, anti-proton \cite{STEPHENS198955,PhysRev.171.1344,osti_6276777,Evoli:2011id} or positron \cite{Delahaye:2007fr,Bergstrom:2013jra,Tylka:1989xj} via their excess fluxes in the centre of galaxies have been studied. Again, FIMP type models are difficult to probe in such cases as well.    

Dark sector constituting of more than one DM components is the focus of the current paper. Two-component DM models have been proposed in various cases \cite{Cao:2007fy,Zurek:2008qg,Profumo:2009tb,Bhattacharya:2013hva} and extensively studied in many~\cite{DiazSaez:2021pfw,Belanger:2011ww,Maity:2019hre}. They provide many new features, the pivotal point of which encompass around the issue of DM-DM interaction or conversion. WIMP-WIMP models show features like modified freeze-out and modified direct search prospects \cite{Bhattacharya:2013hva}. Two component FIMP models have recently been studied where the interaction can cater to structure formation issues \cite{Ghosh:2021wrk}. WIMP-FIMP interaction on the other hand, when feeble, helps FIMP production from the WIMP, but doesn't have too much phenomenological advantage. However, when WIMP-FIMP interaction is of the weak interaction strength, it brings the FIMP to thermal bath, as proposed recently \cite{Bhattacharya:2022dco}, called pseudo-FIMP or pFIMP. pFIMPs always rely on the conversion for the freeze-out and this makes them distinct from their WIMP partner. pFIMP possibilities have been explored in literature~\cite{Belanger:2011ww,Bhattacharya:2013hva,DiazSaez:2021pfw}, but without detailing upon its potential characteristics or phenomenological consequences.

We study a model with two DM components, a scalar and a fermion. The minimal version of such a framework has already been discussed in \cite{Bhattacharya:2013hva}, where the fermion DM acts as pFIMP, while the scalar acts as WIMP. However, the allowed parameter space turns very constrained. We rather choose the scalar DM to be pFIMP, while the fermion sector is enlarged with a vector like doublet and a singlet, where the lightest one after mixing via electroweak symmetry breaking (EWSB) serves as the WIMP component having gauge interaction with the visible sector. One of the most interesting aspects of pFIMP is the possibility of bringing them under direct search scanner via WIMP loop, which is the highlight of this paper. The possibilities in indirect search experiments have also been discussed. 

The paper is organised as follows. In Section~\ref{sec: WIMP-pFIMP}, we present a general discussion on the WIMP-pFIMP scenario and possible interactions between WIMP and pFIMP, that can give rise to interesting prospects in detection experiments. In Section~\ref{sec:models}, we propose a model constituting two-component DM, that can give rise to WIMP-pFIMP. The dark matter phenomenology of this model including the its relic density and prospect of direct and indirect detection are discussed in detail in Section~\ref{sec:phenomenology}. In Section~\ref{conclusion}, we summarize our discussion and conclude.

%%%%%%%%%%%%%%%%%%%%%%%%%%%%%%%
\section{WIMP-pFIMP ensemble}
\label{sec: WIMP-pFIMP}
%%%%%%%%%%%%%%%%%%%%%%%%%%%%%%%

We have demonstrated in~\cite{Bhattacharya:2022dco}, that a FIMP like DM can thermalise via sizable interaction with an WIMP, when it is called a pFIMP. 
Importantly, most of the pFIMP characteristics can be described in a model independent way. In the following, we will have a short account of it before going into 
different possible frameworks where pFIMP can be realised. 

%%%%%%%%%%%%%%%%%%%%%%%%%%%%%%%%
\subsection{A generic discussion}
\label{sec:generic}
%%%%%%%%%%%%%%%%%%%%%%%%%%%%%%%%

The freeze-out pattern of pFIMP is governed via a generic coupled Boltzmann Equations (cBEQ) as given by, 

\begin{eqnarray}\begin{split}
 \frac{dY_1}{dx}=&-\frac{2\pi^2\rm{M_{pl}}}{45\times 1.67}\frac{g_{\star}^{ s}}{ \sqrt{g_{\star}^{ \rho}}}\frac{\mu_{12}}{x^2}\Biggl[\langle\sigma v\rangle_{11\to\rm{SM}}\Bigl(Y_1^2-Y_{1}^{\rm{eq}^2}\Bigr)\\&+\langle\sigma v\rangle_{11\to22}\Bigl(Y_1^2-\frac{Y_1^{\rm eq^2}}{Y_2^{\rm eq^2}}Y_2^2\Bigr)\Biggr],\\
\frac{dY_2}{dx}=&\frac{2\rm{M_{pl}}}{1.67\times \sqrt{g_{\star}^{\rho}}}\frac{x}{\mu_{12}^2}\langle\Gamma_{\rm SM\to22}\rangle (Y_{\rm SM}^{\rm eq }-\frac{Y_2^2}{Y_2^{\rm eq^2}}Y_{\rm SM}^{\rm eq})\\&+\frac{4\pi^2\rm{M_{pl}}}{45\times 1.67}\frac{g_{\star}^{ s}}{ \sqrt{g_{\star }^{\rho}}}\frac{\mu_{12}}{x^2}\Biggl[\langle\sigma v\rangle_{\rm{SM}\to 22}\Bigl(Y_{\rm SM}^{\rm eq^2}-\frac{Y_2^2}{Y_2^{\rm eq^2}}Y_{\rm SM}^{\rm eq^2}\Bigr)\\&+\langle\sigma v\rangle_{11\to22}\Bigl(Y_1^2-\frac{Y_{1}^{\rm eq^2}}{Y_{2}^{\rm eq^2}}Y_2^2\Bigr)\Biggr]\,.\end{split}
\label{eq:cbeq}
\end{eqnarray}
In the above and also in the rest of the draft, subscripts $1,~2$  denote WIMP and FIMP (pFIMP) components respectively. The interaction channels which crucially govern the freeze-out/freeze-in 
of the DM components are:
\begin{itemize}
 \item $\rm DM_1, DM_1\to SM, SM$: annihilation/depletion of the WIMP to the SM states, denoted by $\langle\sigma v\rangle_{11\to\rm{SM}}$,
 \item $\rm DM_1, DM_1\to DM_2, DM_2$: conversion of the WIMP to the FIMP (pFIMP) or vice versa, denoted by $\langle\sigma v\rangle_{11\to22}$,
 \item $\rm SM, SM\to DM_2, DM_2$: production of the FIMP (pFIMP) from the thermal (SM) bath denoted by $\langle\sigma v\rangle_{\rm{SM}\to 22}$,
% \item $\rm DM_2, DM_2\to DM_1, DM_1$: annihilation of the FIMP to the WIMP (depletion of $\rm DM_2$).
 \item $\rm SM \to  DM_2, DM_2$: decay of the bath particles to the FIMP (pFIMP), denoted by $\langle\Gamma_{\rm SM\to22}\rangle$.
\end{itemize}
We note here that the cBEQ for a two component WIMP case is no different than WIMP-FIMP case as shown in Eq.~\ref{eq:cbeq}. 
The difference lies in the strength of the DM-SM interactions ($\langle\sigma v\rangle_{\tt {\tt WIMP}} \sim 10^{-8}~{\rm GeV}^{-2}$, whereas 
$\langle\sigma v\rangle_{\tt {\tt FIMP}} \sim 10^{-20}~{\rm GeV}^{-2}$). The other difference lies 
in the initial conditions on yield, for WIMP: $Y_1|_{x \sim 0}=Y_1^{\rm eq}\sim x^{3/2}e^{-x}$, while for FIMP: $Y_2|_{x \sim 0}=0$. 
The pFIMP solution is obtained when 
\bea
{\rm pFIMP}:~ \langle\sigma v\rangle_{\rm{SM}\to 22}, \langle\Gamma_{\rm SM\to22}\rangle \ll\langle\sigma v\rangle_{11 \to 22}\sim \langle\sigma v\rangle_{11\to\rm{SM}}\sim 10^{-12}~ {\rm GeV}^{-2}\,.
\eea
The pFIMP becomes WIMP when $\langle\sigma v\rangle_{\rm{SM}\to 22}, \langle\Gamma_{\rm SM\to22}\rangle \sim \langle\sigma v\rangle_{11 \to 22}\sim \langle\sigma v\rangle_{11\to\rm{SM}}$.
We further note that the cBEQ is written in terms of yields $Y_{1,2}=\frac{n_{1,2}}{s}$, where $s$ refers to the entropy density (per co-moving volume) as, 
\bea
s = \frac{2 \pi^2}{45} g_{\star }^{s} (T) T^3 \,; \quad
g_{\star}^{ s}(T) = \sum_k \mathcal{C}_k g_k \left (\frac{T_k}{T} \right )^3\theta (T-m_k).
\label{eq:gstars}
\eea
Here $k$ runs over all particles, $T_k$ is the temperature of particle $k$, $g_k$ its number of internal degrees of freedom and $ \mathcal{C}_k=1\,(7/8)$ when $k$ is a boson(fermion). 
We also define the Hubble parameter as
\bea
\mathcal{H}\left(T\right) = 1.67 \sqrt{g_{\star}^{\rho}}\frac{T^2}{\rm{M_{pl}}}; ~ g_{\star}^{\rho}(T)=\sum_{i=bosons} g_i \Biggl(\frac{T_i}{T}\Biggr)^4+ \frac{7}{8} \sum_{i=fermions}g_i \Biggl(\frac{T_i}{T}\Biggr)^4\,.
\eea
We will assume the relativistic degrees of freedom (DOF) $g_{\star}^{\rho,s}\approx 106.7$ to be approximately constant as the temperature during which the FIMP freezes in or the 
WIMP freezes out is rather high. Note that, since two DM with different masses $m_{1,2}$ are involved, we define a common variable $x=\mu_{12}/T$ where 
$\mu_{12}=m_1 m_2/\left(m_1+m_2\right)$ is the reduced mass of the system of two DMs. This is possible in particular to the pFIMP solution, when both pFIMP and WIMP share 
the same temperature. With the redefined $x$ we can write the equilibrium yield \cite{Gondolo:1990dk} as,
\bea
Y_{i}^{\rm eq}\left(x\right) = \frac{45}{4\pi^4}\frac{g_i}{g_{\star}^{ s}}\Biggl(\frac{m_i}{\mu_{12}}x\Biggr)^{2}K_2\Biggl(\frac{m_i}{\mu_{12}}x\Biggr)\,. 
\eea
The expressions of thermal average of annihilation cross-section is given by, 
\bea
\langle \sigma v \rangle= \frac{1}{8m^4T}\frac{1}{K_2(m/T)^2}\int_{4m^2}^\infty \sigma(s)(s-4m^2)\sqrt{s}K_1\left(\frac{\sqrt{s}}{T}\right) ds \,,
\eea
where $v=1/(E_1E_2)(\sqrt{(p_1.p_2)^2-m_1^2m_2^2})$ denotes Mollar velocity, $s$ denotes center-of-mass (c.o.m) energy and $m$ denotes DM mass. 
We further note that the conversion from one DM species to the other are related by: 
\bea
\langle \sigma v \rangle_{11\to22}=\langle \sigma v \rangle_{22\to11}\left( \frac{Y_2^{eq}}{Y_1^{eq}}\right)^2\,.
\eea
The above relation basically indicates that, when WIMP is heavier than pFIMP, i.e. $m_1>m_2$ (hierarchy 1), the conversion from WIMP to pFIMP
is kinematically allowed, while the reverse process is Boltzmann suppressed by a factor $\sim e^{-2x\delta m}$, 
where $\delta m$ is the mass difference between the two DM's. The converse is true when pFIMP is heavier than WIMP, i.e. $m_2>m_1$ (hierarchy 2).
This plays an important role to distinguish the allowed parameter space of the pFIMP-WIMP case for different mass hierarchies.
The key features of pFIMP freeze out are discussed in details in \cite{Bhattacharya:2022dco}, a summary of which is as follows:
\begin{itemize}
\item pFIMP freezes out together or before WIMP, so the relic density of pFIMP is always larger than the WIMP partner\footnote{Recall that $\Omega_{\tt WIMP} h^2 \sim 1/{\langle \sigma v\rangle}$.}. 
When the conversion cross-section is of similar order to that of WIMP annihilation to SM, both pFIMP and WIMP share similar relic densities. 
\item When the conversion rate is higher than the WIMP annihilation, the freeze out and resultant relic density of pFIMP remains constant in 
hierarchy 1, while the WIMP relic becomes much smaller. In hierarchy 2, this is exactly the other way round. With larger conversion, the WIMP 
relic remains constant and pFIMP relic drops sharply.
\item The mass splitting between WIMP and pFIMP needs to be small for the relic density and direct search allowed parameter space, with 
$\delta m\sim 10$ GeV, for $\sim$100 GeV WIMP or pFIMP.
\end{itemize}

%%%%%%%%%%%%%%%%%%%
\subsection{Possible pFIMP-WIMP interactions}
\label{sec:model}
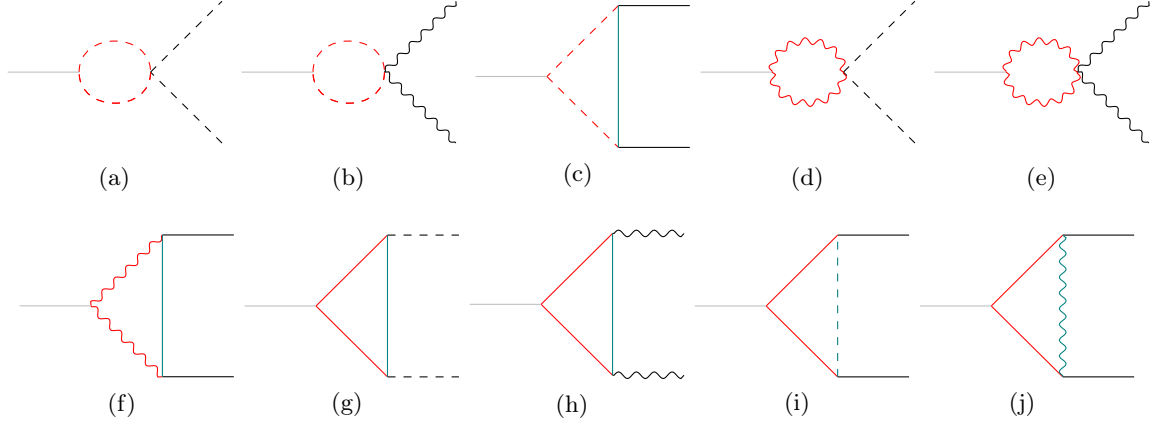
\begin{figure}[!htpb]
\centering
\subfloat[]{
	\begin{tikzpicture}[baseline={(current bounding box.center)},style={scale=0.47, transform shape}]
	\begin{feynman}
	\vertex (a);
	\vertex[above right=2cm and 2cm of a] (a2){\(\rm \)};
	\vertex[below right=2cm and 2cm of a] (a1){\(\rm \)}; 
	\vertex[left=2cm of a] (c); 
	\vertex[left=2cm of c] (d); 	
	\diagram* {
		(a1) -- [scalar] (a),(a) -- [scalar] (a2),(a) -- [scalar,half left,scalar,style=red] (c) [crossed dot] -- [scalar,half left,style=red] (a),(d) -- [plain,style=gray!50,edge label={\(\rm\textcolor{black}{ }\)}] (c)};
	\end{feynman}
	\end{tikzpicture}\label{dm-dd-a}}\subfloat[]{
	\begin{tikzpicture}[baseline={(current bounding box.center)},style={scale=0.47, transform shape}]
	\begin{feynman}
	\vertex (a);
	\vertex[above right=2cm and 2cm of a] (a2){\(\rm \)};
	\vertex[below right=2cm and 2cm of a] (a1){\(\rm \)}; 
	\vertex[left=2cm of a] (c); 
	\vertex[left=2cm of c] (d); 	
	\diagram* {
		(a1) -- [boson] (a),(a) -- [boson] (a2),(a) -- [scalar,half left,scalar,style=red] (c) [crossed dot] -- [scalar,half left,style=red] (a),(d) -- [plain,style=gray!50,edge label={\(\rm\textcolor{black}{ }\)}] (c)};
	\end{feynman}
	\end{tikzpicture}\label{dm-dd-b}}\subfloat[]{
	\begin{tikzpicture}[baseline={(current bounding box.center)},style={scale=0.47, transform shape}]
	\begin{feynman}
	\vertex (a);
	\vertex[above right=2cm and 2cm of a] (d1);
	\vertex[below right=2cm and 2cm of a] (d2); 
	\vertex[right=2cm of d1] (a1);
	\vertex[right=2cm of d2] (a2); 
	\vertex[left=2cm of a] (b); 
	\diagram*{(d1) -- [plain,edge label={\(\rm \)},style=black] (a1),(d2) -- [plain,edge label={\(\rm \)},style=black] (a2),(d1) --[scalar,style=red] (a),(a)--[scalar,style=red] (d2),(b) -- [plain,style=gray!50,edge label={\(\rm\textcolor{black}{ }\)}] (a) ,(d2)--[plain,style=teal] (d1)};
	\end{feynman}
	\end{tikzpicture}\label{dm-dd-c}}\subfloat[]{	
	\begin{tikzpicture}[baseline={(current bounding box.center)},style={scale=0.47, transform shape}]
	\begin{feynman}
	\vertex (a);
	\vertex[above right=2cm and 2cm of a] (a2){\(\rm \)};
	\vertex[below right=2cm and 2cm of a] (a1){\(\rm \)}; 
	\vertex[left=2cm of a] (c); 
	\vertex[left=2cm of c] (d); 	
	\diagram* {
		(a1) -- [scalar] (a),(a) -- [scalar] (a2),(a) -- [boson,half left,style=red] (c) [crossed dot] -- [boson,half left,style=red] (a),(d) -- [plain,style=gray!50,edge label={\(\rm\textcolor{black}{ }\)}] (c)};
	\end{feynman}
	\end{tikzpicture}\label{dm-dd-d}}\subfloat[]{
	\begin{tikzpicture}[baseline={(current bounding box.center)},style={scale=0.47, transform shape}]
	\begin{feynman}
	\vertex (a);
	\vertex[above right=2cm and 2cm of a] (a2){\(\rm \)};
	\vertex[below right=2cm and 2cm of a] (a1){\(\rm \)}; 
	\vertex[left=2cm of a] (c); 
	\vertex[left=2cm of c] (d); 	
	\diagram* {
		(a1) -- [boson] (a),(a) -- [boson] (a2),(a) -- [boson,half left,style=red] (c) [crossed dot] -- [boson,half left,style=red] (a),(d) -- [plain,style=gray!50,edge label={\(\rm\textcolor{black}{ }\)}] (c)};
	\end{feynman}
	\end{tikzpicture}\label{dm-dd-e}}
	
	\subfloat[]{
	\begin{tikzpicture}[baseline={(current bounding box.center)},style={scale=0.47, transform shape}]
	\begin{feynman}
	\vertex (a);
	\vertex[above right=2cm and 2cm of a] (d1);
	\vertex[below right=2cm and 2cm of a] (d2); 
	\vertex[right=2cm of d1] (a1);
	\vertex[right=2cm of d2] (a2); 
	\vertex[left=2cm of a] (b); 
	\diagram*{(d1) -- [plain,edge label={\(\rm \)},style=black] (a1),(d2) -- [plain,edge label={\(\rm \)},style=black] (a2),(d1) --[boson,style=red] (a),(a)--[boson,style=red] (d2),(b) -- [plain,style=gray!50,edge label={\(\rm\textcolor{black}{ }\)}] (a) ,(d2)--[plain,style=teal] (d1)};
	\end{feynman}
	\end{tikzpicture}\label{dm-dd-f}}\subfloat[]{
	\begin{tikzpicture}[baseline={(current bounding box.center)},style={scale=0.47, transform shape}]
	\begin{feynman}
	\vertex (a);
	\vertex[above right=2cm and 2cm of a] (d1);
	\vertex[below right=2cm and 2cm of a] (d2); 
	\vertex[right=2cm of d1] (a1);
	\vertex[right=2cm of d2] (a2); 
	\vertex[left=2cm of a] (b); 
	\diagram*{(d1) -- [scalar,edge label={\(\rm \)},style=black] (a1),(d2) -- [scalar,edge label={\(\rm \)},style=black] (a2),(d1) --[plain,style=red] (a),(a)--[plain,style=red] (d2),(b) -- [plain,style=gray!50,edge label={\(\rm\textcolor{black}{ }\)}] (a) ,(d2)--[plain,style=teal] (d1)};
	\end{feynman}
	\end{tikzpicture}\label{dm-dd-g}}\subfloat[]{
	\begin{tikzpicture}[baseline={(current bounding box.center)},style={scale=0.47, transform shape}]
	\begin{feynman}
	\vertex (a);
	\vertex[above right=2cm and 2cm of a] (d1);
	\vertex[below right=2cm and 2cm of a] (d2); 
	\vertex[right=2cm of d1] (a1);
	\vertex[right=2cm of d2] (a2); 
	\vertex[left=2cm of a] (b); 
	\diagram*{(d1) -- [boson,edge label={\(\rm \)},style=black] (a1),(d2) -- [boson,edge label={\(\rm \)},style=black] (a2),(d1) --[plain,style=red] (a),(a)--[plain,style=red] (d2),(b) -- [plain,style=gray!50,edge label={\(\rm\textcolor{black}{ }\)}] (a) ,(d2)--[plain,style=teal] (d1)};
	\end{feynman}
	\end{tikzpicture}\label{dm-dd-h}}\subfloat[]{	
	\begin{tikzpicture}[baseline={(current bounding box.center)},style={scale=0.47, transform shape}]
	\begin{feynman}
	\vertex (a);
	\vertex[above right=2cm and 2cm of a] (d1);
	\vertex[below right=2cm and 2cm of a] (d2); 
	\vertex[right=2cm of d1] (a1);
	\vertex[right=2cm of d2] (a2); 
	\vertex[left=2cm of a] (b); 
	\diagram*{(d1) -- [plain,edge label={\(\rm \)},style=black] (a1),(d2) -- [plain,edge label={\(\rm \)},style=black] (a2),(d1) --[plain,style=red] (a),(a)--[plain,style=red] (d2),(b) -- [plain,style=gray!50,edge label={\(\rm\textcolor{black}{ }\)}] (a) ,(d2)--[scalar,style=teal] (d1)};
	\end{feynman}
	\end{tikzpicture}\label{dm-dd-i}}\subfloat[]{
	\begin{tikzpicture}[baseline={(current bounding box.center)},style={scale=0.47, transform shape}]
	\begin{feynman}
	\vertex (a);
	\vertex[above right=2cm and 2cm of a] (d1);
	\vertex[below right=2cm and 2cm of a] (d2); 
	\vertex[right=2cm of d1] (a1);
	\vertex[right=2cm of d2] (a2); 
	\vertex[left=2cm of a] (b); 
	\diagram*{(d1) -- [plain,edge label={\(\rm \)},style=black] (a1),(d2) -- [plain,edge label={\(\rm \)},style=black] (a2),(d1) --[plain,style=red] (a),(a)--[plain,style=red] (d2),(b) -- [plain,style=gray!50,edge label={\(\rm\textcolor{black}{ }\)}] (a) ,(d2)--[boson,style=teal] (d1)};
	\end{feynman}
	\end{tikzpicture}\label{dm-dd-j}}
	\caption{Sample Feynman diagrams showing pFIMP interaction with SM via WIMP loops where WIMP, pFIMP, SM $(h/Z/\gamma)$ and a heavy thermal bath 
	particle are represented by red, black, grey and teal color lines respectively. All possible combinations of scalar (dashed), fermion (solid) and vector boson 
	(wavy lines) particles are shown assuming the WIMP and pFIMP to transform under different stabilizing symmetries.}\label{dm-dd}
\end{figure}

%%%%%%%%%%%%%%%%%%%

pFIMPs do not have a sizeable interaction with the SM particles, but thanks to the interaction with WIMPs, they can have a one loop interaction to SM. 
We first discuss the possible scenarios in a model-independent manner under which a pFIMP can interact with SM states and thereby can produce 
DM signal at the future direct and indirect search experiments. Possible pFIMP-SM interactions via WIMP loop are shown in Figure~\ref{dm-dd}. 
Here WIMPs are denoted by red lines, pFIMPs by black lines and the WIMP portals ($Z/h$ or heavy NP particles) with SM by grey lines. 
The dashed, solid and wavy lines indicate scalar, fermion and vector bosons. While drawing these vertices, we have kept the spin conservation 
in mind. Also, pFIMPs and WIMPs are expected to be stabilized under separate symmetries. Therefore, the particles denoted by teal color lines 
are expected to transform suitably under both the symmetries.  

\begin{table}[!htpb]
\centering\renewcommand{\arraystretch}{1.5}\small
\begin{tabular}{cccccc}\hline\hline
\multirow{2}{1.2cm}{\centering Scenarios}&\multicolumn{3}{c}{\centering Two component DM Model}&\multirow{2}{1cm}{\centering Relic} &\multirow{2}{2cm}{\centering Detection \\Possibility}\\ \cline{2-4}
 &WIMP&  pFIMP &\centering WIMP-pFIMP Interaction&  &\\\hline
 (a)&Scalar$(\phi)$&  Scalar$(\Phi)$& $\phi^2\Phi^2$&~~~\cmark\cite{DiazSaez:2021pfw,Bhattacharya:2022dco}&\xmark\\
 (b)&Scalar$(\phi)$&  Vector$(V)$& $\phi^2V^{\mu}V_{\mu}$&\xmark&\xmark \\
 (c)&Scalar$(\phi)$&  Fermion$(\chi)$&$\overline{\chi}\chi^{\prime}\phi$ &~~\cmark\cite{Bhattacharya:2013hva}&\cmark\cite{Bhattacharya:2013hva}\\
 (d)&Vector$(X)$&  Scalar$(\Phi)$& $X^{\mu}X_{\mu}\Phi^2$&\xmark&\xmark\\
 (e)&Vector$(X)$&  Vector$(V)$& $X^{\mu}X_{\mu}V^{\nu}V_{\nu}$&\xmark&\xmark\\
 (f)&Vector$(X)$&  Fermion$(\chi)$&$\overline{\chi}\gamma^{\mu}\chi^{\prime}X_{\mu}$ &\xmark&\xmark\\
 (g)&Fermion$(\psi)$&  Scalar$(\Phi)$&$\overline{\psi}\psi^{\prime}\Phi$ &This work&This work\\
 (h)&Fermion$(\psi)$&  Vector$(V)$&$\overline{\psi}\gamma^{\mu}\psi^{\prime}V_{\mu}$ &\xmark&\xmark\\
 (i,j)&Fermion$(\psi)$&  Fermion$(\chi)$&$\overline{\psi}\chi\phi^{\prime},~\overline{\psi}\gamma^{\mu}\chi X^{\prime}_{\mu}$ &~~~\cmark\cite{Belanger:2011ww}&\xmark\\
\hline\hline
\end{tabular}
\caption{The possible two-component WIMP-pFIMP set-ups (respective Feynman diagrams in Fig.\ref{dm-dd}). Prime particles 
$(\chi^{\prime},\psi^{\prime},\phi^{\prime},X^{\prime}_{\mu})$ are $\rm\mathbb{Z}_2-odd ~and ~\mathbb{Z}_2^{\prime}-odd$ bath particles connecting both WIMP-pFIMP.}
  \label{tab:WIMP-pFIMP}
\end{table} 

\noindent
In \autoref{tab:WIMP-pFIMP}, we have shown possible renormalizable interactions between WIMP and pFIMP in two component DM scenarios stabilized by 
$\mathbb{Z}_2\otimes \mathbb{Z}_2^{\prime}$ for all combinations of scalar, fermion and vector boson particles. The scenarios have one to one correspondence 
to the vertices shown in Fig.~\ref{dm-dd}. WIMPs are odd under $\mathbb{Z}_2$ and even under $\mathbb{Z}_2^{\prime}$, while pFIMPs are odd under 
$\mathbb{Z}_2^{\prime}$, and even under $\mathbb{Z}_2$. Note that particles denoted by prime like $(\chi^{\prime},\psi^{\prime},\phi^{\prime},X^{\prime}_{\mu})$, 
charged under both $\mathbb{Z}_2\otimes \mathbb{Z}_2^{\prime}$, connect WIMP and pFIMP states, as shown by teal color lines in Fig.\ref{dm-dd}. 
This list excludes dark sector particles having non-trivial SM charges and SM particles having dark charges as well as more complicated spin configurations 
and higher order operators having mass dimension larger than four. The right and cross signs in the relic and detection possibilities in 
\autoref{tab:WIMP-pFIMP} indicate whether such possibilities have been studied before or not. A short account of these models and their interactions are as follows:

%Now we demonstrated all of this possible situation in Fig. \ref{dm-dd} where only WIMP-pFIMP interaction has represented as pFIMP phenomenology dominantly is governed by this interaction term. Here we will discuss only the simplest possibility of two component dark matter scenario extending the SM gauge symmetry by $\mathbb{Z}_2\otimes \mathbb{Z}_2^{\prime}$.

\begin{enumerate}[(i)]
\item Figures \ref{dm-dd-a}, \ref{dm-dd-b}, \ref{dm-dd-d} and \ref{dm-dd-e} correspond to two-component scalar-scalar, scalar-vector, vector-scalar, vector-vector DM scenarios. In these 
class of models, WIMPs and pFIMPs are connected without a third particle.
\item Figure \ref{dm-dd-c} and \ref{dm-dd-f} correspond to two-component DM set up, where the WIMP are scalar or vector-boson while 
pFIMP is a Dirac-fermion, connected by another thermal bath fermion odd under both $\rm \mathbb{Z}_2,~and~ \mathbb{Z}_2^{\prime}$.
\item Figure \ref{dm-dd-g} and \ref{dm-dd-h} denote a scenario where WIMP is a Dirac fermion and pFIMPs are
scalar or vector-boson particles, again connected by a thermal bath fermionic particle odd under both $\rm \mathbb{Z}_2,~and~ \mathbb{Z}_2^{\prime}$. 
\item Finally, Figures \ref{dm-dd-i} and \ref{dm-dd-j} correspond to a situation where both WIMP and pFIMPs are fermions connected by 
a scalar (Fig. \ref{dm-dd-i}) or a vector-boson (Fig. \ref{dm-dd-j}) particle, odd under both $\rm \mathbb{Z}_2,~and~\mathbb{Z}_2^{\prime}$. 
\end{enumerate}

%\begin{itemize}
%\item A brief discussion on pFIMP mechanism. 
%\item scalar singlet DM as pFIMP.
%\item Choice of WIMP: Why scalar WIMP is not possible?
%\item Possible solution: A scenario with fermion WIMP.
%\end{itemize}

%%%%%%%%%%%%%%%%%%%%
\section{Model example of a pFIMP-WIMP scenario}
\label{sec:models}
%%%%%%%%%%%%%%%%%%%

 Amongst the possibilities described in Table~\ref{tab:WIMP-pFIMP}, the WIMP-pFIMP phenomenology has been explored in the case of two-component scalar DM 
 model~\cite{Bhattacharya:2022dco, DiazSaez:2021pfw}. It has been pointed out that when pFIMP couples to the SM states via a scalar WIMP loop, 
 it is not possible to achieve a relic density allowed parameter space where the scalar pFIMP can be detected in future direct detection experiment. 
 We will elaborate more on this later. In \cite{Bhattacharya:2013hva}, the direct detection prospect of a fermion pFIMP was studied in a scalar-fermion set up as 
 in Fig.~\ref{dm-dd-c}, without elaborating upon the pFIMP characteristics. The resulting parameter space of this model is highly constrained by the 
 recent most direct search results. In~\cite{Belanger:2011ww}, pFIMP phenomenon was discussed partially, but detectability of pFIMP via WIMP loop has been neglected. 
 We will focus on a WIMP-pFIMP set up where the direct detection possibility of pFIMP is achieved in next generation experiment 
 and make connections with indirect detection as well. Unlike the model-independent approach taken in~\cite{Bhattacharya:2022dco}, 
 here we elaborate on the channels through which the freeze-out can occur for both the DM components taking 
 temperature-dependence of all the annihilation cross-sections and decay widths into consideration. 
 
 %%%%%%%%%%%%%%%%%%%
%\subsection{Two Component Fermion-Scalar DM}
%\label{sec:fsmodel}
%%%%%%%%%%%%%%%%%%%
Our model consists of (i) a real scalar-singlet DM $\phi$, which acts like pFIMP and (ii) the lightest admixture of a vector-like fermion doublet $\psi=\left(\psi^0~~\psi^-\right)^{T}$ 
\cite{delAguila:1989rq} and a vector-like singlet fermion $\psi_1$, which behaves like WIMP. We additionally introduce another vector-like singlet fermion $\psi_2$, 
which acts as a messenger between the two DM sectors. Stability of both DM components can be ensured by a ${\rm \mathbb{Z}_2 \otimes \rm \mathbb{Z}^\prime_2 }$ symmetry. 
The quantum numbers of all the relevant fields are given in Table \ref{tab:tab1}.

\begin{table}[!htpb]
\begin{center}
\begin{tabular}{cccccc}
\rowcolor{gray!20}{\bf Dark Fields}&$\hspace{0.8cm} \rm{SU(3)_c \times SU(2)_L \times U(1)_Y}\times \rm{\rm \mathbb{Z}_2 \times {\rm \mathbb{Z}^\prime}_2}$\\\rowcolor{gray!16}  
\hspace{0.2cm}$\psi=\begin{pmatrix} \psi^0 \\  \psi^- \end{pmatrix}$ &$ \hspace{1.3cm} 1 \hspace{1.3cm} 2 \hspace{1.15cm} -1 \hspace{0.8cm} - \hspace{0.4cm} +$\\\rowcolor{gray!12} 
 \hspace{1cm}${\psi}_1$ &$ \hspace{1.3cm} 1 \hspace{1.3cm} 1 \hspace{1.6cm} 0 \hspace{0.8cm}-\hspace{0.4cm}+$\\ \rowcolor{gray!8}
 \hspace{1cm}${\psi}_2$ &$ \hspace{1.3cm} 1 \hspace{1.3cm} 1 \hspace{1.6cm} 0 \hspace{0.8cm} +\hspace{0.4cm}-$\\\rowcolor{gray!4} 
 \hspace{1cm}$\phi$ &$ \hspace{1.3cm} 1 \hspace{1.3cm} 1 \hspace{1.6cm} 0 \hspace{0.8cm} -\hspace{0.4cm} -$
\end{tabular}
\end{center}
\caption{Dark sector fields and their corresponding quantum numbers.}
  \label{tab:tab1}
\end{table}

The corresponding Lagrangian can be written as:,
%%%%%%%%%%%%%%%%%%%%%%%%%%%%%
\bea
\mathcal{L}\supset\mathcal{L}_{\rm{Scalar}}+\mathcal{L}_{\rm{VF}}~,
\eea
where,
\bea\begin{split}
\mathcal{L}_{\rm{Scalar}}=\frac{1}{2}|\partial_{\mu}\phi|^2-\frac{1}{2}\mathfrak{m}_{\phi}^2\phi^2-\frac{1}{4!}\lambda_{\phi}\phi^4-\frac{1}{2}\lambda_{\phi H}\phi^2H^{\dagger}H~ ,
\end{split}\eea
\bea\begin{split}
\mathcal{L}_{\rm{VF}}=&\overline{\psi}\left[i\gamma^{\mu}\left(\partial_{\mu}+i g\frac{\sigma^a}{2}W^{a}_{\mu}+i g^{\prime}\frac{Y}{2}B_{\mu}\right)-m_{\psi}\right]\psi+\sum_{\alpha=1,2}\overline{\psi}_{\alpha}\left(i\gamma^{\mu}\partial_{\mu}-m_{\psi_{\alpha}}\right)\psi_{\alpha}\\&-(Y_1\overline{\psi}\widetilde{H}\psi_1+Y_2\overline{\psi}_2\psi_1\phi+\rm{h.c.}) ~;
\end{split}\label{eq: lagvf}\eea
%%%%%%%%%%%%%%%%
After Electroweak Symmetry Breaking(EWSB), SM Higgs $H$ acquires vacuum expectation value(VEV) $(v=246~\rm{GeV})$ and in unitarity gauge we can write,  
$H=\left(0~~\frac{1}{\sqrt{2}}(v+h)\right)^T$. After symmetry breaking the physical mass term of $\phi$ can be written as 
$m_{\phi}^2=\mathfrak{m}_{\phi}^2+\frac{1}{2}\lambda_{\phi H}v^2$. $\phi$ is a stable DM candidate and serves as pFIMP with negligible $\lambda_{\phi H}$.
From Eq.\ref{eq: lagvf}, it is straight-forward to calculate the mass terms for the vector-like fermions. 
The mass eigenstates $(\chi_1,\chi_2)$ can be obtained via diagonalization of the fermion mass matrix through a unitary transformation from the flavour basis $(\psi_1,\psi^0)$.
\bea\begin{split}
-\mathcal{L}_{\rm{mass}}=m_{\chi_{1}}\overline{\chi}_1\chi_1+m_{\chi_{2}}\overline{\chi}_2\chi_2+m_{\psi}\psi^+\psi^-\,;
\end{split}\eea
where,
\bea\begin{split}
\chi_1&=\cos\theta\psi_1+\sin\theta\psi^0\,,\\
\chi_2&=-\sin\theta\psi_1+\cos\theta\psi^0\,,\\
m_{\chi_1}&=\sin^2\theta m_{\psi}+\cos^2\theta m_{\psi_1}+\frac{Y_1 v}{\sqrt{2}}\sin 2\theta\,,\\
m_{\chi_2}&=\cos^2\theta m_{\psi}+\sin^2\theta m_{\psi_1}-\frac{Y_1 v}{\sqrt{2}}\sin 2\theta\,.
\end{split}\eea
The mixing angle $\theta$ can be written as,
\bea
\tan2\theta=\frac{\sqrt{2}Y_1v}{m_{\psi_1}-m_{\psi}}\,.
\label{eq: mixing_angle}
\eea
Using Eq. \ref{eq: mixing_angle}, we can easily write,
\bea 
\label{eq:massdiff}
Y_1=\frac{\sin 2\theta}{\sqrt{2}v}(m_{\chi_1}-m_{\chi_2})\,,\\
m_{\psi}=m_{\chi_1}\sin^2\theta+m_{\chi_2}\cos^2\theta\,,\\
m_{\psi_1}=m_{\chi_1}\cos^2\theta+m_{\chi_2}\sin^2\theta\,.
\eea
$m_{\psi}$ denotes the mass of the charged component of vector like fermion doublet $\psi^{\pm}$.
The independent parameters of our model are $\{m_{\chi_1},~m_{\chi_2},~m_{\psi_2},~m_{\phi},~\sin\theta,~Y_2,\lambda_{\phi H}\}$. 
$\chi_1$ being the lightest neutral fermion, serves as the WIMP DM and the mass difference between $\chi_1$ and the second lightest neutral fermion 
$\chi_2$ is denoted as $\Delta m = m_{\chi_2}-m_{\chi_1}$. Mass difference between $\chi_1$ and $\phi$ is denoted by 
$\delta m = m_{\phi} - m_{\chi_1}$ and serves as an important parameter in the WIMP-pFIMP set up. 

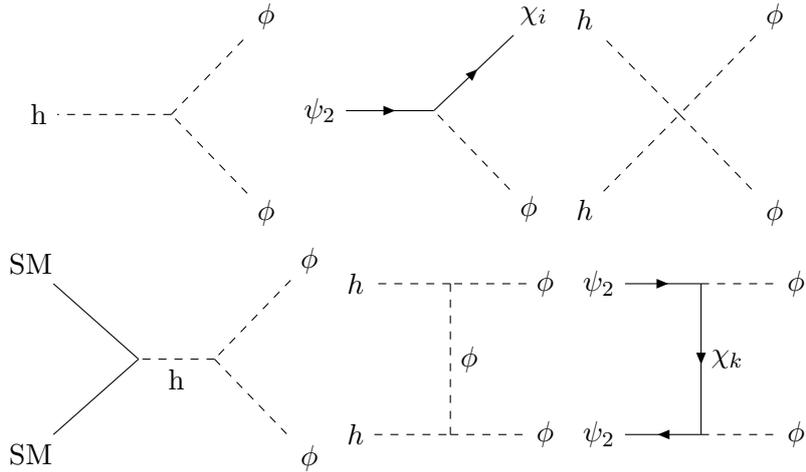
\begin{figure}[htb!]
	\centering	
	\begin{tikzpicture}[baseline={(current bounding box.center)}]
	\begin{feynman}
	\vertex (a);
	\vertex[left=1.5cm of a] (c){\(\rm{h}\)}; 
	\vertex[above right=1cm and 1cm of a] (a2){\(\phi\)};
	\vertex[below right=1cm and 1cm of a] (a1){\(\phi\)}; 
	\diagram* {
		(a1) -- [scalar] (a),(a) -- [scalar] (a2),(a) -- [scalar] (c) 
	};\end{feynman}
	\end{tikzpicture}
	\begin{tikzpicture}[baseline={(current bounding box.center)}]
	\begin{feynman}
	\vertex (a){\(\psi_2\)};
	\vertex[right=1.5cm of a] (b); 
	\vertex[above right=1cm and 1cm of b] (b1){\(\chi_i\)};
	\vertex[below right=1cm and 1cm of b] (b2){\(\phi\)}; 
	\diagram* {
		(a) -- [fermion ,arrow size=1pt] (b) -- [fermion ,arrow size=1pt] (b1),(b2) -- [scalar] (b) 
	};\end{feynman}
	\end{tikzpicture}
	 \begin{tikzpicture}[baseline={(current bounding box.center)}]
	\begin{feynman}
	\vertex (a);
	\vertex[above left=1cm and 1cm of a] (a2){\(h\)};
	\vertex[above right=1cm and 1cm of a] (a1){\(\phi\)}; 
	\vertex[below left=1cm and 1cm of a] (d1){\(h\)};
	\vertex[below right=1cm and 1cm of a] (d2){\(\phi\)};	
	\diagram* {
		(a1) -- [scalar] (a),(a) -- [scalar] (a2),(d1) --[scalar] (a)  --[scalar] (d2)
	};\end{feynman}
	\end{tikzpicture}
	
	\begin{tikzpicture}[baseline={(current bounding box.center)}]
	\begin{feynman}
	\vertex (a);
	\vertex[below right=1cm and 1cm of a] (a2){\(\phi\)};
	\vertex[above right=1cm and 1cm of a] (a1){\(\phi\)}; 
	\vertex[left=1cm of a] (d); 
	\vertex[above left=1cm and 1cm of d] (d1){\(\rm{SM}\)};
	\vertex[below left=1cm and 1cm of d] (d2){\(\rm{SM}\)};	
	\diagram* {
		(a1) -- [scalar] (a),(a) -- [scalar] (a2),(a) -- [scalar,edge label={\(\rm{h}\)}] (d) ,(d1) --[plain] (d)  --[plain] (d2)
	};\end{feynman}
	\end{tikzpicture}
	\begin{tikzpicture}[baseline={(current bounding box.center)}]
	\begin{feynman}
	\vertex (a);
	\vertex[left=1cm of a] (a2){\(h\)};
	\vertex[right=1cm of a] (a1){\(\phi\)}; 
        \vertex[below=2cm of a] (b); 
	\vertex[left=1cm of b] (d1){\(h\)};
	\vertex[right=1cm of b] (d2){\(\phi\)};	
	\diagram* {
		(a1) -- [scalar] (a),(a) -- [scalar] (a2),(a) -- [scalar,edge label={\(\phi\)}] (b),(d1) --[scalar] (b)  --[scalar] (d2)
	};\end{feynman}
	\end{tikzpicture}
	\begin{tikzpicture}[baseline={(current bounding box.center)}]
	\begin{feynman}
	\vertex (a);
	\vertex[left=1cm of a] (a1){\(\psi_2\)};
	\vertex[right=1cm of a] (a2){\(\phi\)}; 
        \vertex[below=2cm of a] (b); 
	\vertex[left=1cm of b] (d1){\(\psi_2\)};
	\vertex[right=1cm of b] (d2){\(\phi\)};	
	\diagram* {
		(a1) -- [fermion,arrow size=1pt] (a),(a) -- [scalar] (a2),(a) -- [fermion,arrow size=1pt,edge label={\(\chi_k\)}] (b),(b) --[fermion,arrow size=1pt] (d1),(b)  --[scalar] (d2)
	};\end{feynman}
	\end{tikzpicture}
%%%%%%%%%%%%%%%%%%%	
\begin{comment}	
        \begin{tikzpicture}[baseline={(current bounding box.center)}]
	\begin{feynman}
	\vertex (a){\(\psi_2\)};
	\vertex[right=1.5cm of a] (b); 
	\vertex[above right=2cm and 2cm of b] (b1){\(\chi_i\)};
	\vertex[above right=1cm and 1cm of b] (b11);
	\vertex[below right=1cm and 1cm of b] (b22); 
	\vertex[below right=2cm and 2cm of b] (b2){\(\phi\)}; 
	\diagram* {
		(a) -- [fermion ,arrow size=1pt] (b) -- [scalar,edge label={\(\phi\)}] (b11)-- [fermion ,arrow size=1pt] (b1),(b) -- [fermion ,arrow size=1pt,edge label'={\(\chi_j\)}] (b22) ,
		(b22) -- [fermion ,arrow size=1pt,edge label'={\(\psi_2\)}] (b11),(b22) -- [scalar] (b2)
	};\end{feynman}
	\end{tikzpicture}
\end{comment}
	\caption{Feynman diagrams for pFIMP $\phi$ production from thermal bath ($\{i,k=1,2\}$).}
        \label{fig:-fimp}
\end{figure}

\begin{figure}[htb!]
	\centering	
	\begin{tikzpicture}[baseline={(current bounding box.center)}]
	\begin{feynman}
	\vertex (a);
	\vertex[below right=1cm and 1cm of a] (a2){\(\rm{SM}\)};
	\vertex[above right=1cm and 1cm of a] (a1){\(\rm{SM}\)}; 
	\vertex[left=1cm of a] (d); 
	\vertex[above left=1cm and 1cm of d] (d1){\(\chi_i\)};
	\vertex[below left=1cm and 1cm of d] (d2){\(\chi_j\)};	
	\diagram* {
	(a1) -- [plain] (a),(a) -- [plain] (a2),(a) -- [scalar,edge label={\(\rm{h}\)}] (d) ,(d) --[fermion ,arrow size=1pt] (d1) ,(d2) --[fermion ,arrow size=1pt] (d)
	};\end{feynman}
	\end{tikzpicture}
	\begin{tikzpicture}[baseline={(current bounding box.center)}]
	\begin{feynman}
	\vertex (a);
	\vertex[below right=1cm and 1cm of a] (a2){\(\rm{SM}\)};
	\vertex[above right=1cm and 1cm of a] (a1){\(\rm{SM}\)}; 
	\vertex[left=1cm of a] (d); 
	\vertex[above left=1cm and 1cm of d] (d1){\(\chi_i\)};
	\vertex[below left=1cm and 1cm of d] (d2){\(\chi_j\)};	
	\diagram* {
	(a1) -- [plain] (a),(a) -- [plain] (a2),(a) -- [boson,edge label={\(\rm{Z}\)}] (d) ,(d) --[fermion ,arrow size=1pt] (d1) ,(d2) --[fermion ,arrow size=1pt] (d)
	};\end{feynman}
	\end{tikzpicture}
	\begin{tikzpicture}[baseline={(current bounding box.center)}]
	\begin{feynman}
	\vertex (a);
	\vertex[below right=1cm and 1cm of a] (a2){\(Z,\nu_{\ell},q_u\)};
	\vertex[above right=1cm and 1cm of a] (a1){\(W^{\pm},\ell,q_d\)}; 
	\vertex[left=1cm of a] (d); 
	\vertex[above left=1cm and 1cm of d] (d1){\(\chi_i\)};
	\vertex[below left=1cm and 1cm of d] (d2){\(\psi^{\pm}\)};	
	\diagram* {
		(a) -- [fermion ,arrow size=1pt] (a1),(a2) -- [fermion ,arrow size=1pt] (a),(a) -- [boson, edge label={\(W^{\pm}\)}] (d) ,(d) --[fermion ,arrow size=1pt] (d1) ,(d2) --[fermion ,arrow size=1pt] (d)
	};\end{feynman}
	\end{tikzpicture}
	\small
	\begin{tikzpicture}[baseline={(current bounding box.center)}]
	\begin{feynman}
	\vertex (a);
	\vertex[left=1cm of a] (a2){\(\chi_i\)};
	\vertex[right= 1cm of a] (a1){\(Z,W^{\pm}\)}; 
	\vertex[below=2cm of a] (d); 
	\vertex[left=1cm of d] (d1){\(\psi^{\pm}\)};
	\vertex[right=1cm of d] (d2){\(W^{\pm},Z\)};	
	\diagram* {
	(a2) -- [fermion ,arrow size=1pt] (a) -- [fermion ,arrow size=1pt] (a1),(d) -- [fermion,arrow size=1pt,edge label'={\(\chi_k,\psi^{\pm}\)}] (a) ,(d1) --[fermion ,arrow size=1pt] (d)  --[fermion ,arrow size=1pt] (d2)
	};\end{feynman}
	\end{tikzpicture}
	\begin{tikzpicture}[baseline={(current bounding box.center)}]
	\begin{feynman}
	\vertex (a);
	\vertex[left=1cm of a] (a2){\(\chi_i\)};
	\vertex[right= 1cm of a] (a1){\(\psi_2\)}; 
	\vertex[below=2cm of a] (d); 
	\vertex[left=1cm of d] (d1){\(\chi_j\)};
	\vertex[right=1cm of d] (d2){\(\psi_2\)};	
	\diagram* {
	(a2) -- [fermion ,arrow size=1pt] (a) -- [fermion ,arrow size=1pt] (a1),(a) -- [scalar,edge label={\(\phi\)}] (d) ,(d2) --[fermion ,arrow size=1pt] (d)  --[fermion ,arrow size=1pt] (d1)
	};\end{feynman}
	\end{tikzpicture}
	\begin{tikzpicture}[baseline={(current bounding box.center)}]
	\begin{feynman}
	\vertex (a);
	\vertex[left=1cm of a] (a2){\(\chi_i\)};
	\vertex[right= 1cm of a] (a1){\(W^+\)}; 
	\vertex[below=2cm of a] (d); 
	\vertex[left=1cm of d] (d1){\(\chi_j\)};
	\vertex[right=1cm of d] (d2){\(W^-\)};	
	\diagram* {
	(a2) -- [fermion ,arrow size=1pt] (a) -- [fermion ,arrow size=1pt] (a1),(a) -- [fermion,edge label={\(\psi^-\)},arrow size=1pt] (d) ,(d2) --[fermion ,arrow size=1pt] (d)  --[fermion ,arrow size=1pt] (d1)
	};\end{feynman}
	\end{tikzpicture}
	\begin{tikzpicture}[baseline={(current bounding box.center)}]
	\begin{feynman}
	\vertex (a);
	\vertex[left=1cm of a] (a2){\(\chi_i\)};
	\vertex[right= 1cm of a] (a1){\(h,Z,h\)}; 
	\vertex[below=2cm of a] (d); 
	\vertex[left=1cm of d] (d1){\(\chi_j\)};
	\vertex[right=1cm of d] (d2){\(Z,h,Z\)};	
	\diagram* {
	(a2) -- [fermion ,arrow size=1pt] (a) -- [fermion ,arrow size=1pt] (a1),(d) -- [fermion,edge label'={\(\chi_k\)},arrow size=1pt] (a) ,(d2) --[fermion ,arrow size=1pt] (d)  --[fermion ,arrow size=1pt] (d1)
	};\end{feynman}
	\end{tikzpicture}
	\caption{Feynman diagrams for the possible annihilation and co-annihilation channels of WIMP $\chi_1$ ($\{i,j,k=1,2\}$).}
        \label{fig:-wimp}
\end{figure}
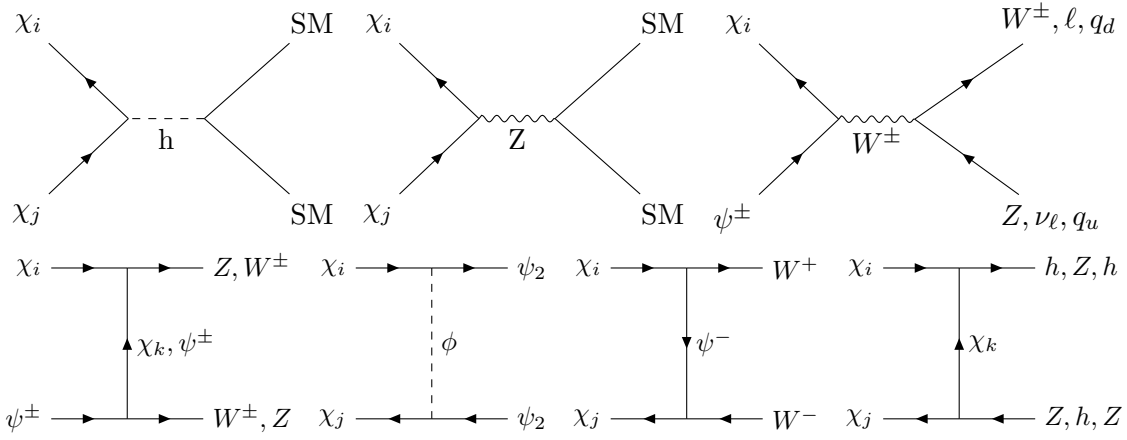

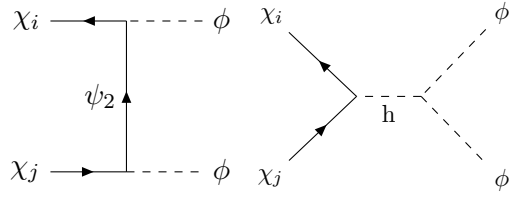
\begin{figure}[htb!]
	\centering	
	\begin{tikzpicture}[baseline={(current bounding box.center)}]
	\begin{feynman}
	\vertex (a);
	\vertex[left=1cm of a] (a2){\(\chi_i\)};
	\vertex[right=1cm of a] (a1){\(\phi\)}; 
        \vertex[below=2cm of a] (b); 
	\vertex[left=1cm of b] (d1){\(\chi_j\)};
	\vertex[right=1cm of b] (d2){\(\phi\)};	
	\diagram* {
		(a1) -- [scalar] (a),(a) -- [fermion ,arrow size=1pt] (a2),(b) -- [fermion ,arrow size=1pt ,edge label={\(\psi_2\)}] (a),(d1) --[fermion ,arrow size=1pt] (b)  --[scalar] (d2)
	};\end{feynman}
	\end{tikzpicture}
	\begin{tikzpicture}[baseline={(current bounding box.center)},style={scale=0.85, transform shape}]
	\begin{feynman}
	\vertex (a);
	\vertex[below right=1cm and 1cm of a] (a2){\(\phi\)};
	\vertex[above right=1cm and 1cm of a] (a1){\(\phi\)}; 
	\vertex[left=1cm of a] (d); 
	\vertex[above left=1cm and 1cm of d] (d1){\(\chi_i\)};
	\vertex[below left=1cm and 1cm of d] (d2){\(\chi_j\)};	
	\diagram* {
		(a1) -- [scalar] (a),(a) -- [scalar] (a2),(a) -- [scalar, edge label={\(\rm{h}\)}] (d) ,(d) --[fermion ,arrow size=1pt] (d1) ,(d2) --[fermion ,arrow size=1pt] (d)
	};\end{feynman}
	\end{tikzpicture}
	\caption{Feynman diagrams contributing to WIMP-pFIMP conversion ($\{i,j=1,2\}$).}
        \label{fig:-DM-DM}
\end{figure}

The interaction between WIMP and pFIMP occurs via the Yukawa term, $Y_2\overline{\psi}_1\psi_2\phi$, which is crucial for the WIMP-pFIMP conversion. 
We will be particularly interested in the region where $m_{\psi_2}>m_{\chi_1}+m_{\phi}$, so that $\psi_2$ can decay into $\chi_1$ and $\phi$. 
The relevant Feynman diagrams for pFIMP production, (co)-annihilation of WIMPs and 
WIMP-pFIMP conversions are shown in Figures.~\ref{fig:-fimp},~\ref{fig:-wimp} and ~\ref{fig:-DM-DM}.
We would like to further mention that a tiny $\lambda_{\phi H}\sim 10^{-10}$ as required for pFIMP realisation, helps evading the upper bound on Higgs invisible branching ratio 
BR$(h\rightarrow invisible) < 19\%$ at 2$\sigma$~\cite{CMS:2018yfx} as well as the direct detection constraints. It allows us to explore the mass range below the Higgs resonance 
$m_{\phi}<(m_h/2)$. We would like to mention, we have checked the constraints from limits on Higgs and $Z$ invisible decay width(see Appendix~\ref{appendixe}, when DM masses are below such resonances.

%%%%%%%%%%%%%%%%%%%%%%%%%%%%%%%%%%%%%%%%%%%%%%%%%%%%%%%%%

%%%%%%%%%%%%%%%%%%%
%%%%%%%%%%%%%%%%%%%

%%%%%%%%%%%%%%%%%%%%%%%%%%%%%%%%%%%%%%%%%%%%%%%%%%%%%%%%%
\section{Dark matter phenomenology}
\label{sec:phenomenology}
%%%%%%%%%%%%%%%%%%%%%%%%%%%%%%%%%%%%%%%%%%%%%%%%%%%%%%%%%
Having discussed the model, we will focus on the DM phenomenology highlighting the pFIMP behaviour.
%We will indeed draw parallels with the model-independent inferences from section 2. 

%%%%%%%%%%%%%%%%%%%%%%%%%%%%%%%%%%%%%%%%%%%%%%
\subsection{Coupled Boltzmann Equations and Relic allowed parameter space}
%%%%%%%%%%%%%%%%%%%%%%%%%%%%%%%%%%%%%%%%%%%%%

We begin with the cBEQ's, specific to our model after considering all the relevant processes 
\cite{Griest:1990kh,PhysRevLett.119.061102,Belanger:2014vza,DAgnolo:2018wcn,Chua:2013zpa},

%%%%%%%%%%%%%%%%%%%%%%%%%%%%%%%%%%%%%%%%%%%%%%%%%%%%%%%%%
\bea\begin{split}
\dfrac{dY_{\chi}}{dx}=&-\frac{2\pi^2\rm{M_{pl}}}{45\times 1.67}\frac{g_{\star}^s}{ \sqrt{g_{\star}^{\rho}}}\frac{\mu_{\chi_1\phi}}{x^2}\Biggl[\langle\sigma v\rangle_{\rm SM}^{\rm eff} \left(Y_{\chi}^2-Y_{\chi}^{\rm eq^2}\right)+\langle\sigma v\rangle_{\phi}^{\rm eff}\left(Y_{\chi}^2-Y_{\chi}^{\rm{eq^2}}\frac{Y_{\phi}^2}{Y_{\phi}^{\rm{eq^2}}}\right)
\\&-\langle\sigma v\rangle_{\psi_2}^{\rm eff}\left(Y_{\chi}^2-Y_{\chi}^{\rm{eq^2}}\frac{Y_{\psi_2}^2}{Y_{\psi_2}^{\rm{eq^2}}}\right)+\langle\sigma v\rangle_{\chi_1\overline{\psi}_2\to h\phi}^{\rm eff}\left(Y_{\psi_2}Y_{\chi}-Y_{\psi_2}^{\rm{eq}}Y_{\chi}^{\rm{eq}}\frac{Y_{\phi}}{Y_{\phi}^{\rm{eq}}}\right)\Biggr]
\\&+\frac{\rm{M_{pl}}}{1.67 \sqrt{g_{\star}^{\rho}}}\frac{x}{\mu_{\chi_1\phi}^2}\langle\Gamma\rangle^{\rm eff}_{\psi_2\to\chi_1\phi} \left(Y_{\psi_2}-Y_{\psi_2}^{\rm{eq}}\frac{Y_{\phi}}{Y_{\phi}^{\rm{eq}}}\frac{Y_{\chi}}{Y_{\chi}^{\rm{eq}}}\right),
\end{split}
\label{wimpeq}
\eea
%%%%%%%%%%%%%%%%%%%%%%%%%%%%%%%%%%%%%%%%%%%%%%%%%%%%%%%%%
\bea\begin{split}
\dfrac{dY_{\phi}}{dx}=&\frac{\rm{M_{pl}}}{1.67\times \sqrt{g_{\star }^{\rho}}}\frac{x}{\mu_{\chi_1\phi}^2}\left[2\langle\Gamma\rangle_{h\to\phi\phi} \left(Y_{h}^{\rm{eq}}-Y_h^{\rm{eq}}\frac{Y_{\phi}^2}{Y_{\phi}^{\rm{eq}^2}}\right)+\langle\Gamma\rangle^{\rm eff}_{\psi_2\to\chi_1\phi} \left(Y_{\psi_2}-Y_{\psi_2}^{\rm{eq}}\frac{Y_{\phi}}{Y_{\phi}^{\rm{eq}}}\frac{Y_{\chi}}{Y_{\chi}^{\rm{eq}}}\right)\right]
\\&+\frac{2\pi^2\rm{M_{pl}}}{45\times 1.67}\frac{g_{\star}^s}{ \sqrt{g_{\star}^\rho}}\frac{\mu_{\chi_1\phi}}{x^2}\Biggl[2\langle\sigma v\rangle_{\rm SM~SM \to \phi\phi}\left(Y_{\rm{SM}}^{\rm{eq}^2}-Y_{\rm{SM}}^{\rm{eq}^2}\frac{Y_{\phi}^2}{Y_{\phi}^{\rm{eq}^2}}\right)\\&+2\langle\sigma v\rangle_{\psi_2\overline{\psi}_2\to \phi\phi}\left(Y_{\psi_2}^2-Y_{\psi_2}^{\rm{eq^2}}\frac{Y_{\phi}^2}{Y_{\phi}^{\rm{eq^2}}}\right)+\langle\sigma v\rangle_{\chi_1\overline{\psi}_2\to h\phi}^{\rm eff}\left(Y_{\psi_2}Y_{\chi}-Y_{\psi_2}^{\rm{eq}}Y_{\chi}^{\rm{eq}}\frac{Y_{\phi}}{Y_{\phi}^{\rm{eq}}}\right)
\\&+2\langle\sigma v\rangle_{\phi}^{\rm eff}\left(Y_{\chi}^2-Y_{\chi}^{\rm{eq^2}}\frac{Y_{\phi}^2}{Y_{\phi}^{\rm{eq^2}}}\right)\Biggr].
\end{split}
\label{fimpeq}
\eea
%%%%%%%%%%%%%%%%%%%%%%%%%%%%%%%%%%%%%%%%%%%%%%%%%%%%%%%%%
Eq.~\ref{wimpeq}, and ~\ref{fimpeq}, are the cBEQ's of the WIMP($\chi_1$) and pFIMP($\phi$) respectively.
In Eq.~\ref{wimpeq}, $Y_{\chi}$ is the total WIMP DM yield and $Y_{\phi}$ is the pFIMP yield, the two crucial quantities for our analysis. 
%%%%%%%%%%%%%%%%%%%%%%%%%%%%%%%%%%%%%%%%%%%%%%%%%%%%%%%%%
In writing the Equations, we have used the following ansatz~\cite{Griest:1990kh},
 \bea\frac{n_i}{n}\approx\frac{n_i^{\rm eq}}{n^{\rm eq}}\,,\eea 
where \bea n^{\rm eq}=\sum_i n_i^{\rm eq}=\frac{T}{2\pi^2}\sum_ig_im_i^2K_2\left(\frac{m_i}{T}\right)\,.\eea

\noindent
We would like to mention, since in our model, we have an extended dark sector, where over and above the stable DM states $\chi_1$ and $\phi$, all the unstable heavy 
states such as $\psi^{\pm}$, $\chi_2$ will also take part in co-annihilation as well as decay processes when in equilibrium. After freeze-out they will eventually decay into 
the stable lightest particle of the spectrum, namely $\chi_1$. In order take this effect in account, we have considered the `effective' thermal average $\langle \sigma v\rangle^{\rm eff}$ 
of annihilation cross-section and decay width $\langle\Gamma\rangle^{\rm eff}$, see \cite{Edsjo:1997bg} and \autoref{coannihilation} for details.
We emphasize that the total WIMP yield $Y_{\chi}$ will be the sum of the yields of all the particles which transform under the same $Z_2$ symmetry as $\chi_1$, 

\begin{equation}
Y_{\chi} = \sum_i Y_{\chi_i}, ~~\chi_i=\{\chi_1,\overline{\chi}_1,\chi_2,\overline{\chi}_2,\psi^\pm\}\,.
\label{totyield}
\end{equation}

%%%%%%%%%%%%%%%%%%%
\noindent

The SM final states in Eq.~\ref{wimpeq} and \ref{fimpeq} includes all possible final states such as $h,W^{\pm},Z,\ell,q$. 
The common variable $x=\frac{\mu_{\chi_1 \phi}}{T}$ written in terms of the reduced mass $\mu_{\chi_1 \phi}=(\frac{1}{m_{\chi_1}}+\frac{1}{m_{\phi}})^{-1}$ 
caters to the two component DM system. A symmetry factor of $2$ applies in case of the scalar DM $\phi$ (see Eq.~\ref{fimpeq}).
The dark sector particles follow the non-relativistic equilibrium distribution given by,
\begin{align}
&Y^{\rm{eq}}_{\chi}=\frac{45}{4\pi^4}\sum_i\frac{g_i}{g_{\star}^s}\left(x\frac{m_{i}}{\mu_{\chi_1 \phi}}\right)^{2}K_2\left(x\frac{m_i}{\mu_{\chi_1 \phi}}\right)\,,
\\&Y^{\rm{eq}}_{\phi}=\frac{45}{4\pi^4}\frac{g_{\phi}}{g_{\star}^s}\left(x\frac{m_{\phi}}{\mu_{\chi_1 \phi}}\right)^{2}K_2\left(x\frac{m_{\phi}}{\mu_{\chi_1 \phi}}\right)\,.
\end{align}
In the above equation, $\rm M_{\rm pl}=1.22091\times 10^{19}$ GeV and $g_{\star}^{ s}\simeq g_{\star}^{ \rho}\approx106.7$.
We have assumed that in our model $\chi_i,\psi^{\pm}$ are in equilibrium by rapid annihilations into bath particles. 
$\psi_2$ will also remain in equilibrium by virtue of the sizeable $Y_2$, and we can assume $Y_{\psi_2} \approx Y_{\psi_2}^{\rm eq}$ 
and neglect the evolution of $\psi_2$ separately. The scalar DM $\phi$ is assumed out-of-equilibrium initially, due to its tiny coupling with the SM 
particles, while $\phi$ reaches thermal equilibrium and becomes pFIMP, aided by large conversion between $\chi\chi \rightarrow \phi\phi$ mediated by $\psi_2$, when $Y_2$ is large. 
Subsequent solution of the cBEQ provides relic density of the DM species by the following formula,
\bea\begin{split}
\Omega_{\rm{DM}}\rm{h}^2&=2.74385\times 10^8 \left(m_{\chi_1}Y_{\chi}\Biggl[{\frac{m_{\chi_1}}{\mu_{\chi_1 \phi}}x_{\infty}}\Biggr]+m_{\phi}Y_{\phi}\Biggl[{\frac{m_{\phi}}{\mu_{\chi_1 \phi}}x_{\infty}}\Biggr]\right)\,,
\end{split}\eea   
where $x_{\infty}$ corresponds to the present time.
%%%%%%%%%%%%%%%%%%%%%%%%%%%%%%%%%%%%%%%%%%%%%%%%%%%%%%%%%
 \begin{figure}[htb!]
\centering
\subfloat[]{\includegraphics[width=0.5\linewidth]{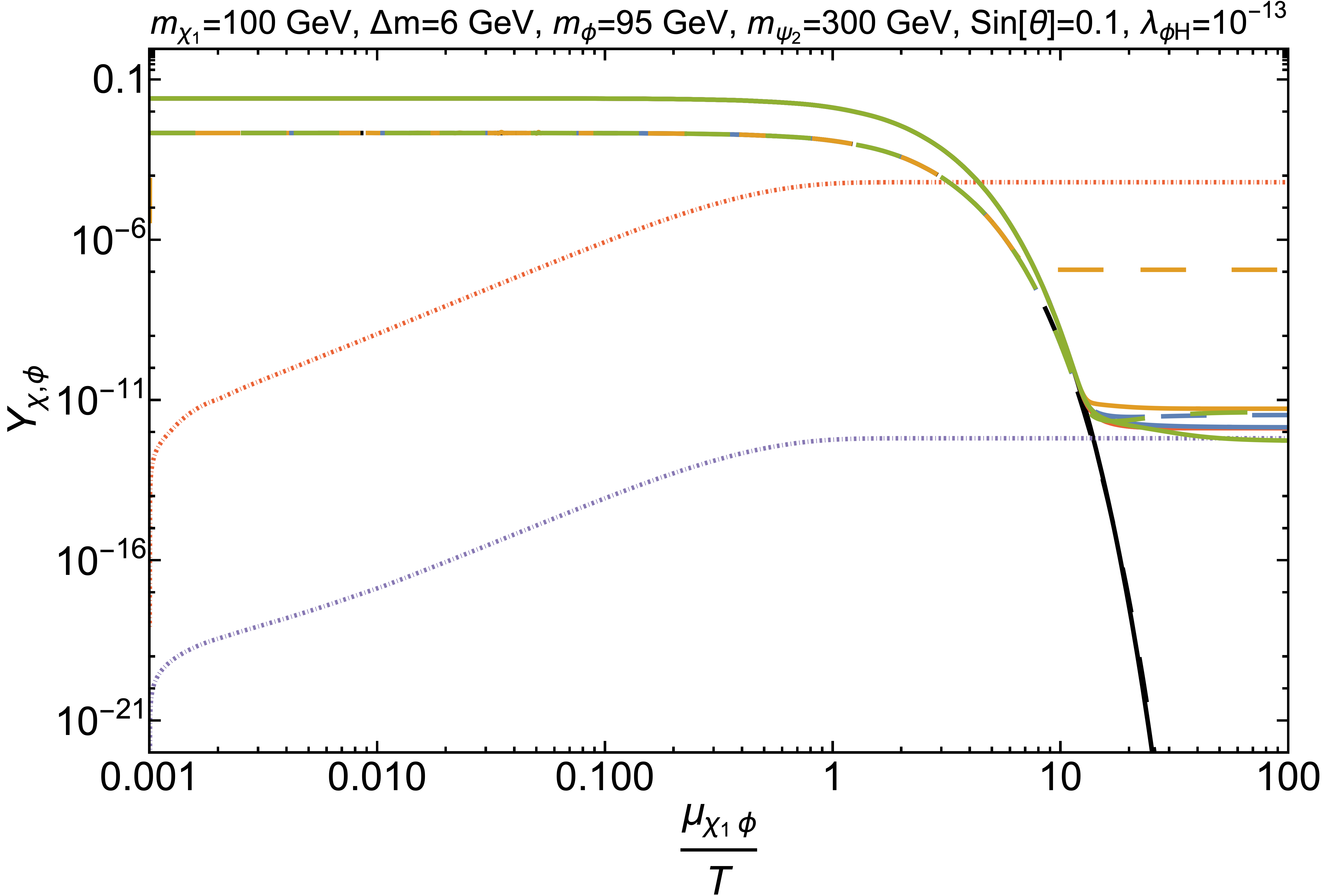}\label{fig:-yield_m1gmphi}}~~
\subfloat[]{\includegraphics[width=0.5\linewidth]{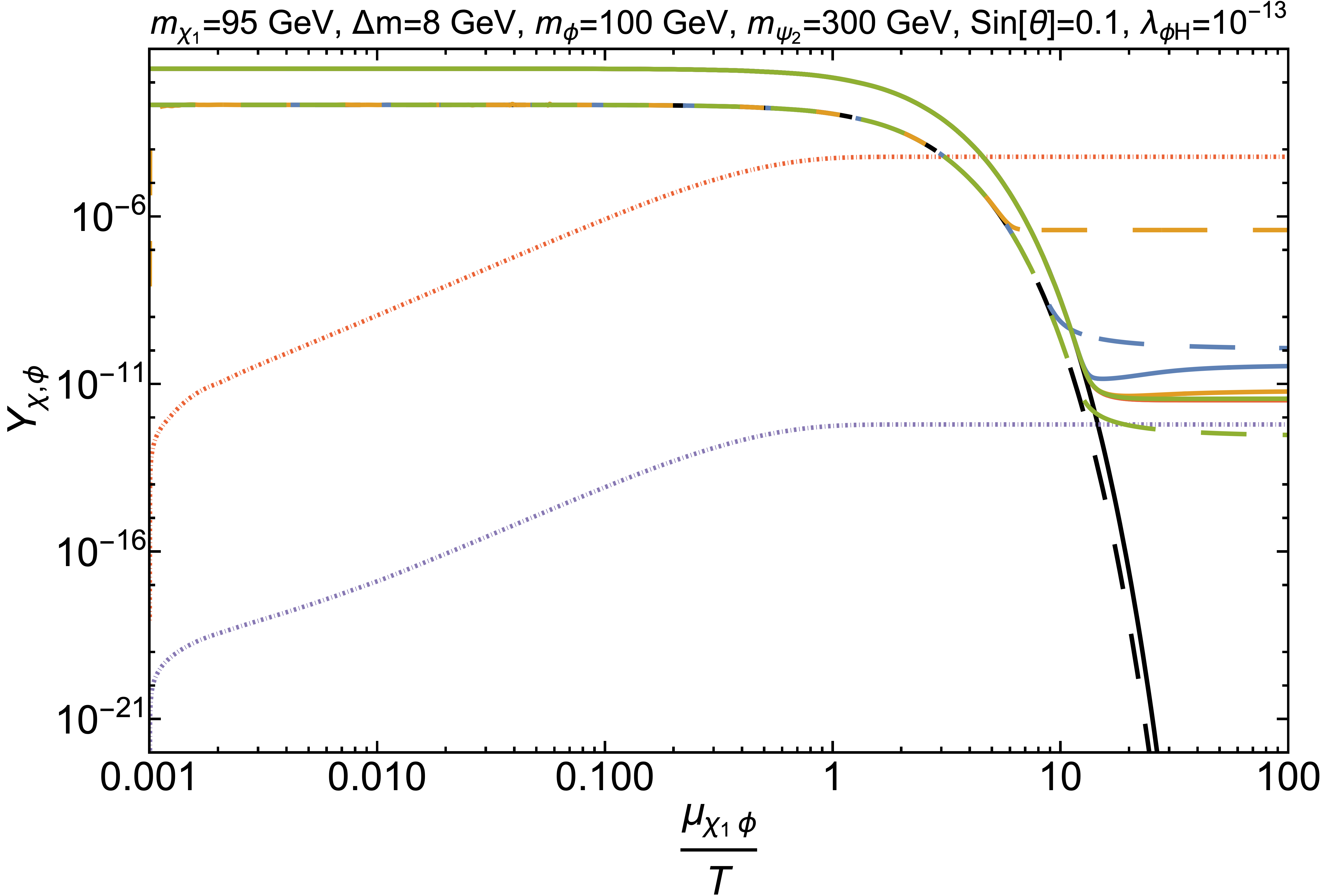}\label{fig:-yield_m1lmphi}}\\
%\subfloat[]{\includegraphics[width=0.5\linewidth]{figure/relic_m1gmphi}\label{fig:-relic_m1gmphi}}~~
%\subfloat[]{\includegraphics[width=0.5\linewidth]{figure/relic_m1lmphi}\label{fig:-relic_m1lmphi}}
  \caption{Figure \ref{fig:-mchimphi}(a) shows the variation of yield $(Y)$ of WIMP (thick lines), pure FIMP (dotted lines) and pFIMP (dashed lines) as a function of $x$ 
  where violet, red, yellow, blue and green lines correspond to different values of $Y_2$ $\{10^{-12},~10^{-8},~10^{-2},~1,~2\}$ respectively for $m_{\chi_1}>m_{\phi}$.  
  Figure \ref{fig:-mchimphi}(b) shows the same for $m_{\chi_1}<m_{\phi}$ with different values of Yukawa coupling $Y_2$ $\{10^{-12},~10^{-8},~10^{-3},~0.1,~0.5\}$ represented by 
  violet, red, yellow, blue and green lines respectively. The black thick and dashed lines show the equilibrium distribution of WIMP and pFIMP respectively.}
  \label{fig:-mchimphi}
\end{figure}
%%%%%%%%%%%%%%%%%%%%%%%%%%%%%%%%%%%%%%%%%%%%%%%%%%%%%%%%%%
The solutions of cBEQ's are presented in terms of DM yield as a function of $x$ in Figures.~\ref{fig:-mchimphi} (a) and (b) for two different mass hierarchies. 
Red and violet dotted lines represent the pure FIMP scenario when $Y_2=\{ 10^{-12},10^{-8}\}$ respectively. With larger $Y_2$, conversion from WIMP to FIMP 
and consequently the FIMP yield increases. However, in this region, the FIMP still freezes-in. With $Y_2$ increasing further, the FIMP yield thermalises to
equilibrium number density and enters into the pFIMP regime (yellow, blue and green dashed lines) to freeze-out subsequently. 
The pFIMP dynamics has been discussed in details in ~\cite{Bhattacharya:2022dco}.   
%%%%%%%%%%%%%%%%%%%%%%%%%%%%%%%%%%%%%%%%%%%%%%%%%%%%%%%%%%
 \begin{figure}[htb!]
\centering
\subfloat[]{\includegraphics[width=0.5\linewidth]{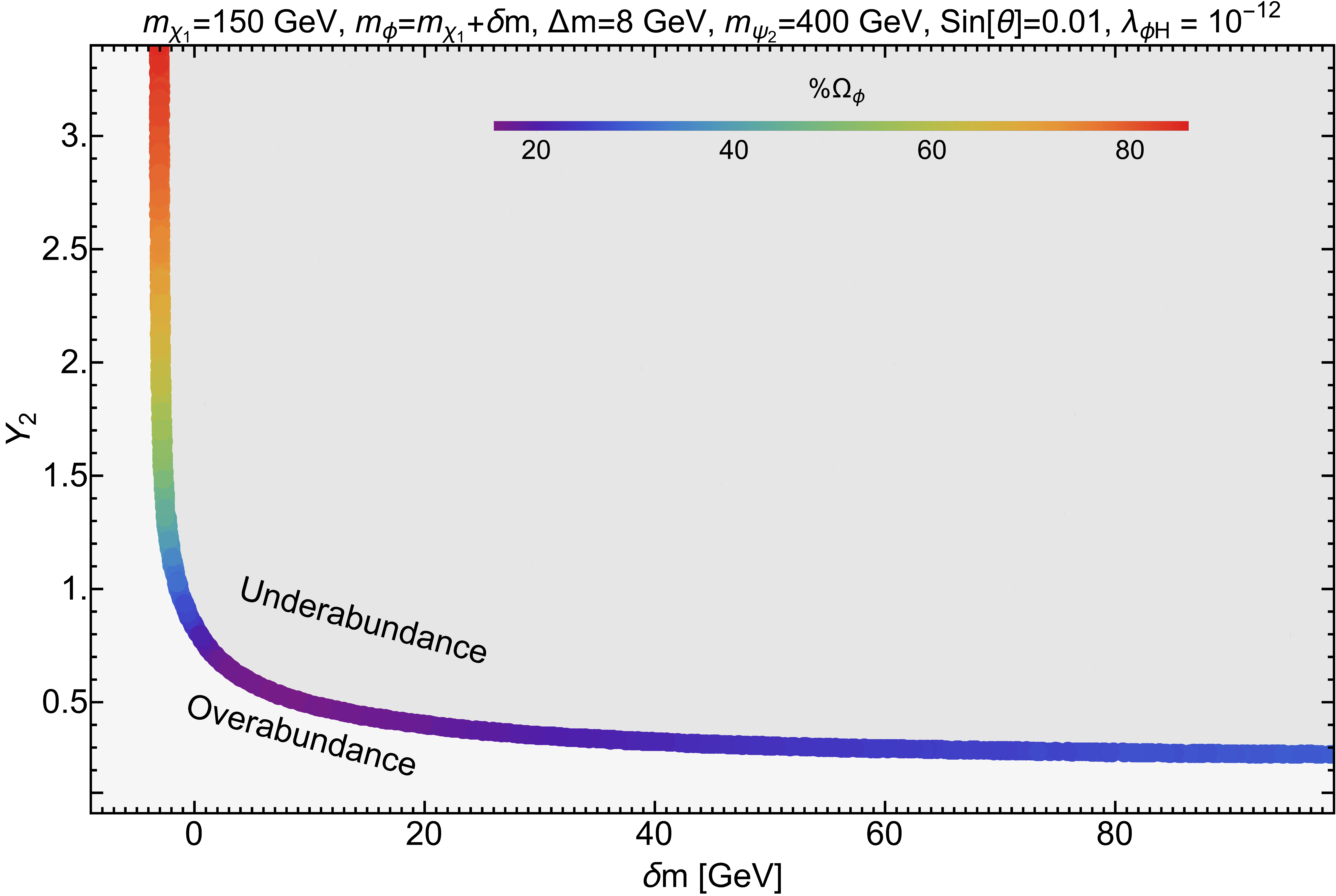}}~~
\subfloat[]{\includegraphics[width=0.5\linewidth]{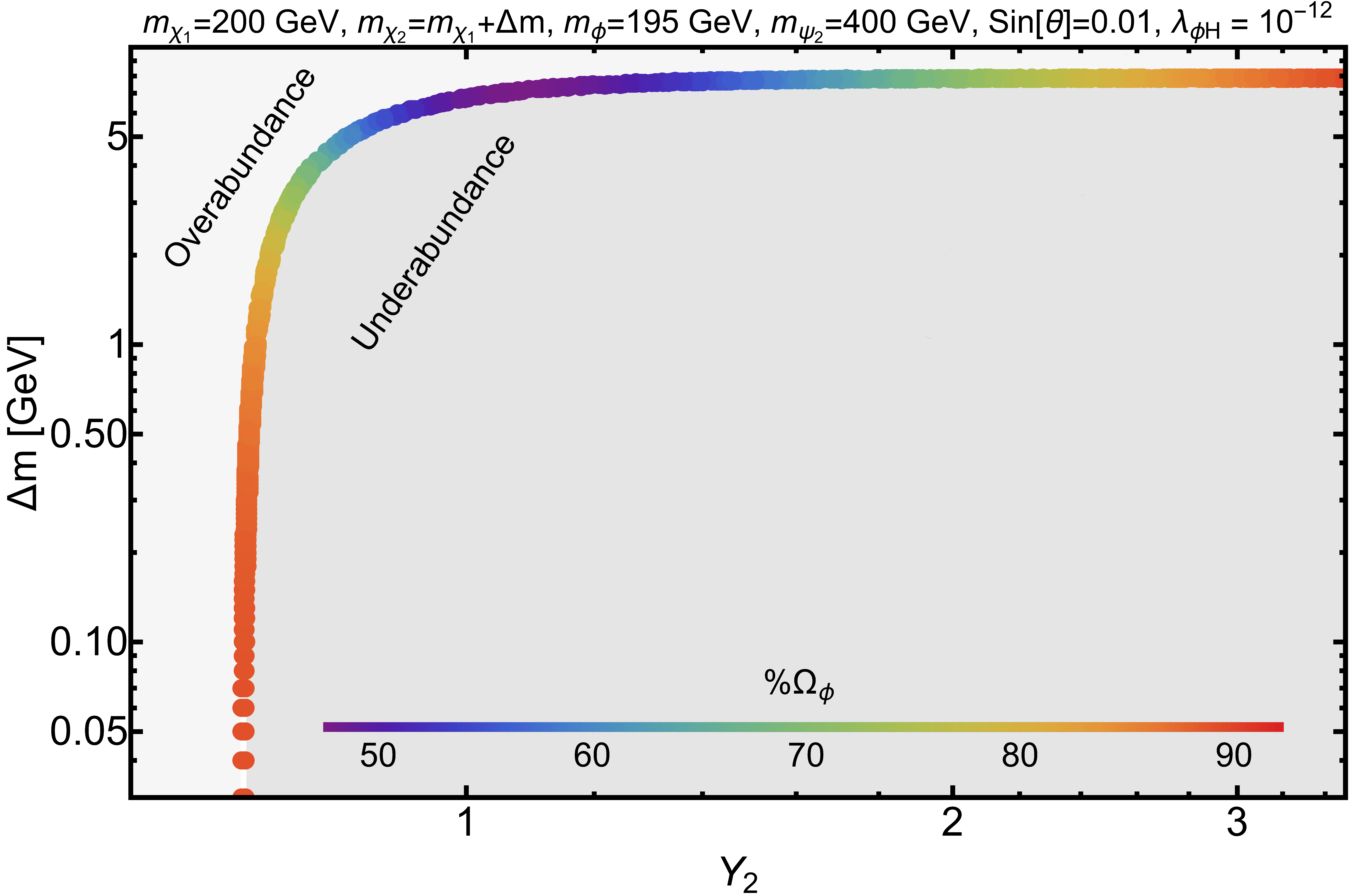}}\\
%\subfloat[]{\includegraphics[width=0.5\linewidth]{figure/y2-dm_phidd}}
\caption{Parameter space allowed by the observed relic ($0.1188\leq\Omega_{\rm DM}h^2\leq 0.1212$) in 
(a) $\delta m - Y_2$ plane (b) $Y_2 - \Delta m$ plane. In both the figures percentage contribution of pFIMP
$(\Omega_{\phi})$ is shown in the color bar. Parameters kept fixed are mentioned in the figure heading.}
\label{fig:deltam_y2_dd_relic}
\end{figure}
%%%%%%%%%%%%%%%%%%%%
In Fig.~\ref{fig:deltam_y2_dd_relic}(a), we show the allowed parameter space in $\delta m - Y_2$ plane, where the total relic adds to the observed one. The percentage contribution of the 
pFIMP ($\phi$) is shown in the color axis. Note that, $\delta m$ here is the mass difference between the two DM's. One can see, when $\delta m > 0$, i.e. the 
pFIMP is heavier than the WIMP, the decrease in $\delta m$ increase the pFIMP-WIMP conversion significantly and therefore, the relative contribution of pFIMP 
to the total relic decreases. On the other hand, for the opposite hierarchy, i.e. $m_\phi < m_{\chi_1}$, increasing $Y_2$ will increase WIMP-pFIMP conversion thereby increasing 
pFIMP contribution to the total relic density. 

In Fig.~\ref{fig:deltam_y2_dd_relic}(b), we show the relic density allowed parameter space in $Y_2 - \Delta m$ plane. Here, $\Delta m$ is the mass difference between the second 
lightest fermionic dark sector particle $\chi_2$ and the WIMP DM $\chi_1$. We see that $\Delta m \lsim 10$ GeV is required so that co-annihilation reduces $\chi_1$ relic to the correct 
ballpark for pFIMP $\phi$ to saturate the rest of it. With larger $Y_2$, the relative contribution of $\phi$ to the total relic density increases with more $\chi_1 \to \phi$ conversion.

%One can see that the decrease in $\Delta m$ opens up co-annihilation channels thereby decreasing $\chi_1$
%relic density contribution. 
%
%and correct relic one has to keep low $\Delta m \lsim 10$ GeV. In this case, $m_{\chi_1}$ has been chosen to be heavier than $m_{\phi}$ and therefore, with increasing $Y_2$, the relative contribution of $\phi$ to the total relic increases. On the other hand, if we decrease $\Delta m$ for a specific choice of $Y_2$, large co-annihilation will lead to less contribution of WIMP to relic, and consequently, percentage contribution of $\phi$ increases. 

%%%%%%%%%%%%%%%%%%%%%%%%%%%%%%%%%%%%%%%%%%%%%%%%%%%%%%%%%
\subsection{Direct detection prospect}
%%%%%%%%%%%%%%%%%%%%%%%%%%%%%%%%%%%%%%%%%%%%%%%%%%%%%%%%%

Now we delve into the direct search prospect of the two component DM's, which is our key focus in this study. 
First we will briefly discuss the direct detection of the WIMP and then explore the pFIMP case in detail. 

%%%%%%%%%%%%%%%%%%%%%%%%%%%%%
\subsubsection{Direct detection of WIMP}
%%%%%%%%%%%%%%%%%%%%%%%%%%%%%

\begin{figure}[htb!]
	\centering	
	\begin{tikzpicture}[baseline={(current bounding box.center)},style={scale=0.7, transform shape}]
	\begin{feynman}
	\vertex (a);
	\vertex[left=2cm of a] (a1){\(\chi_1\)};
	\vertex[right=2cm of a] (a2){\(\chi_1\)}; 
	\vertex[below=2cm of a] (b); 
	\vertex[left=2cm of b] (c1){\(N\)};
	\vertex[right=2cm of b] (c2){\(N\)};
	
	\diagram* {
		(a1) -- [fermion, arrow size=1pt,style=red] (a) -- [fermion, arrow size=1pt,style=red] (a2),(a) -- [scalar, edge label={\(h\)},style=blue] (b) ,(c1) --[fermion, arrow size=1pt] (b)  --[fermion, arrow size=1pt] (c2),
	};\end{feynman}
	\end{tikzpicture}
	\begin{tikzpicture}[baseline={(current bounding box.center)},style={scale=0.7, transform shape}]
	\begin{feynman}
	\vertex (a);
	\vertex[ left=2cm of a] (a1){\(\chi_1\)};
	\vertex[ right= 2cm of a] (a2){\(\chi_1\)}; 
	\vertex[below= 2cm of a] (b); 
	\vertex[ left= 2cm of b] (c1){\(N\)};
	\vertex[ right= 2cm of b] (c2){\(N\)};
	
	\diagram* {
		(a1) -- [fermion, arrow size=1pt,style=red] (a) -- [fermion, arrow size=1pt,style=red] (a2),(a) -- [boson, edge label={\(Z\)}, style=blue] (b) ,(c1) --[fermion, arrow size=1pt] (b)  --[fermion, arrow size=1pt] (c2),
	};\end{feynman}
	\end{tikzpicture}
	\caption{The Feynman diagrams for the direct detection of WIMP ($\chi_1$).}
    \label{fig:wimp-dd}
\end{figure}
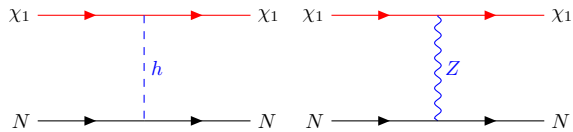

\noindent
In our model, the spin-independent direct detection cross section for WIMP $\chi_1$ ($\rm \sigma^{ SI}_{\chi_1 N}$) gets major contribution from $Z$ and Higgs-mediated 
$t$-channel diagrams (Fig.~\ref{fig:wimp-dd}) and therefore the singlet-doublet mixing parameter $\sin\theta$ plays an important role, apart from the DM mass $m_{\chi_1}$. 
The $Z$ mediated contribution is required to be small to abide by the non-observation of a spin independent direct search of the DM. This is possible when 
the singlet doublet mixing ($\sin\theta$) is small, since the effective coupling involved in the $Z$ mediated vertex is 
$\lambda_{Z \overline{\chi}_1\chi_1}=\frac{m_Z}{v}\sin^2{\theta}$, whereas the effective Higgs coupling is $\lambda_{h\overline{\chi}_1\chi_1}=-\frac{Y_1}{\sqrt{2}}\sin2{\theta}$.
%two cases are $\lambda_{h\overline{\chi}_1\chi_1}=-\frac{Y_1}{\sqrt{2}}\sin2{\theta}$ and  respectively.
%The contribution from Higgs-mediated diagram dominates over the $Z$-mediated one when the singlet doublet mixing ($\sin\theta$) is small, since the couplings involved in the 
%two cases are  and $\lambda_{Z \overline{\chi}_1\chi_1}=\frac{m_Z}{v}\sin^2{\theta}$ respectively. 

 \begin{figure}[htb!]
\centering
\subfloat[]{\includegraphics[width=0.5\linewidth]{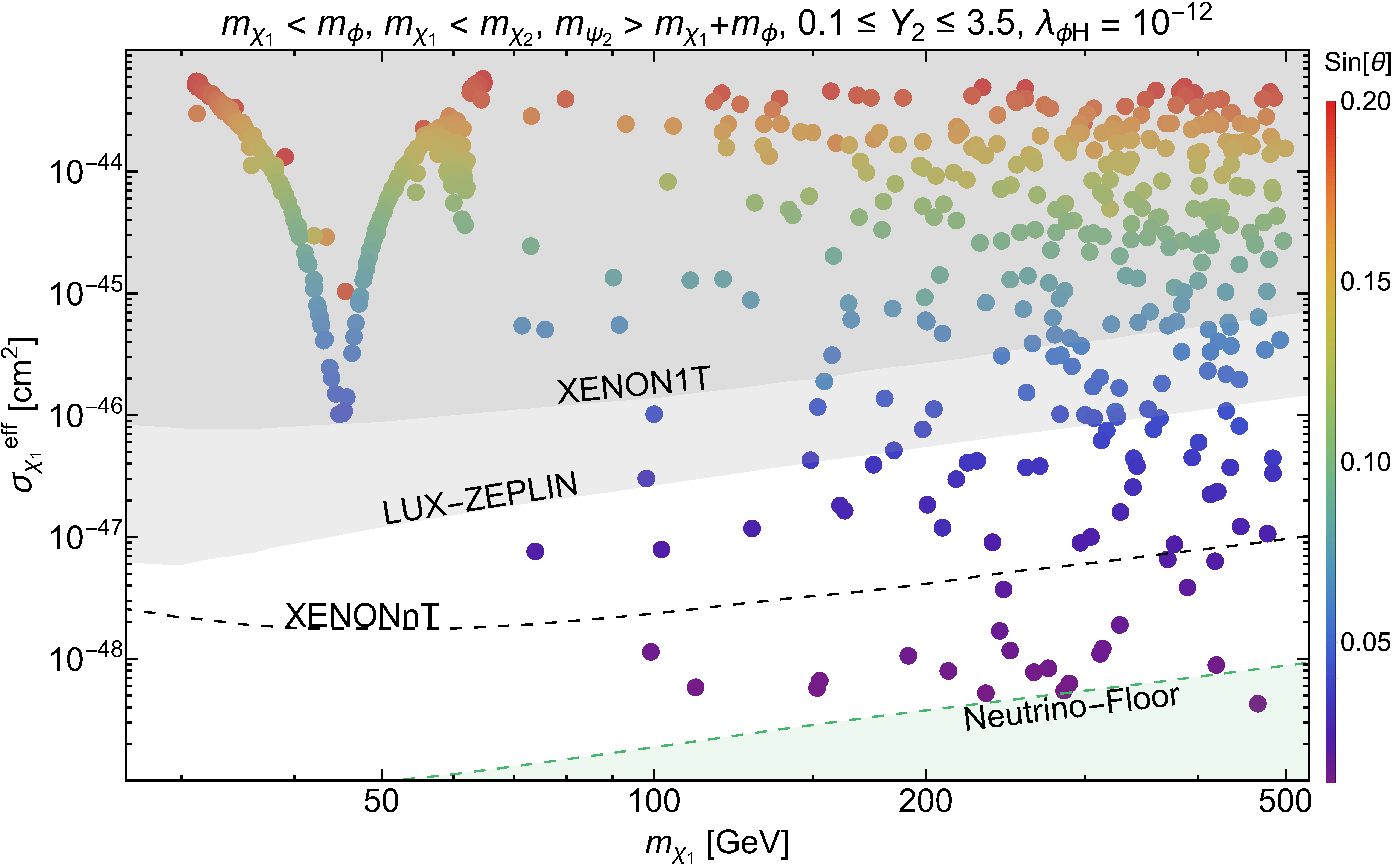}}~~
\subfloat[]{\includegraphics[width=0.5\linewidth]{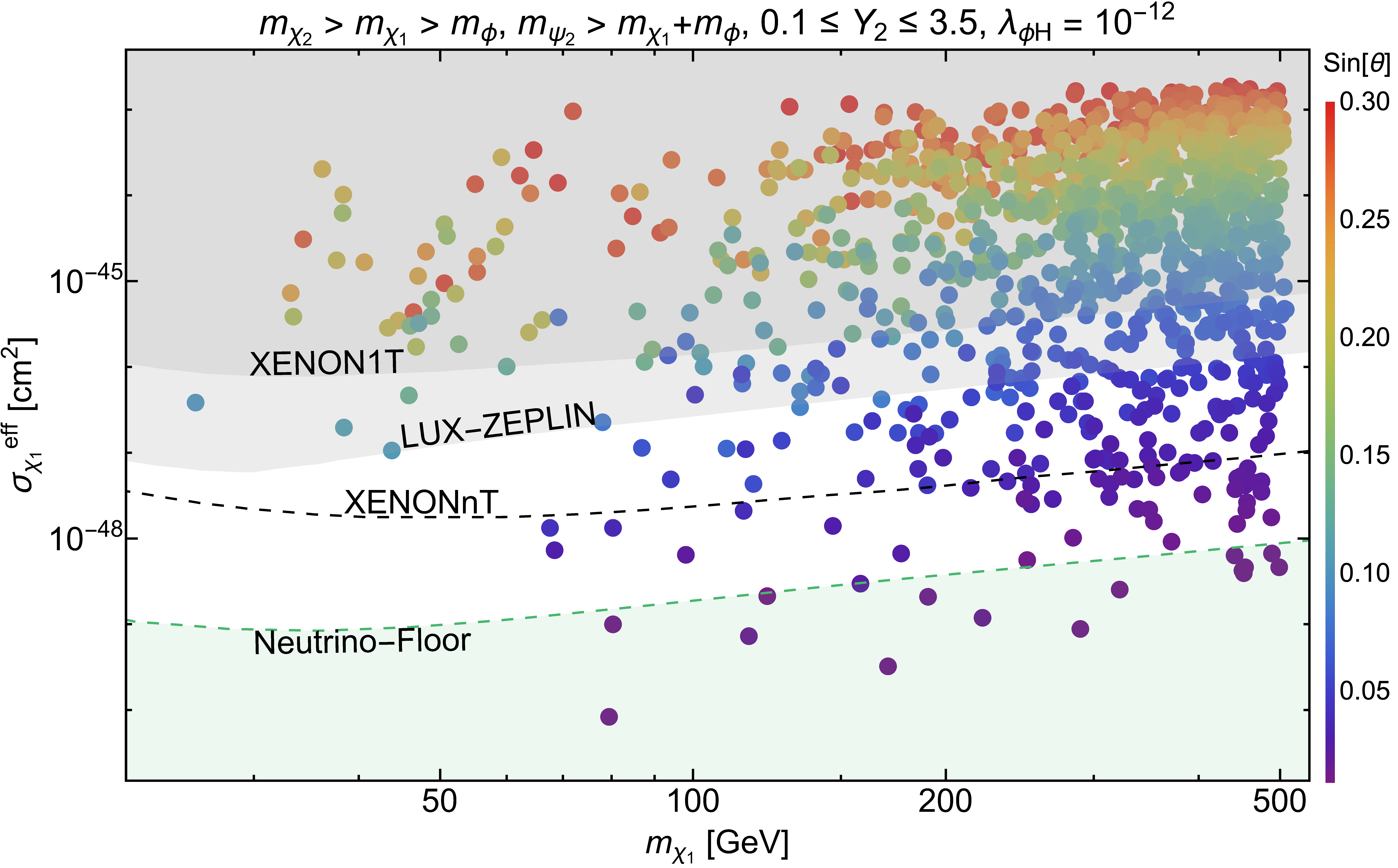}}
  \caption{Effective spin-independent direct detection cross-section ($\sigma^{\rm eff}_{\chi_1}$) for WIMP $\chi_1$ for (a) $m_{\chi_1} < m_{\phi}$ and (b) $m_{\chi_1} > m_{\phi}$. 
  All the points satisfy the present DM relic density bound $0.1188\leq\Omega_{\chi_1}h^2+\Omega_{\phi}h^2\leq 0.1212$ via the combined contribution of both DM's. 
  $\sin\theta$ is shown as color axis in both the figures. Other parameters kept fixed are shown in figure heading. The limits from XENON1T, LUX-ZEPLIN data and future sensitivities 
  from XENONnT and Neutrino floor are shown.}
  \label{fig:-mchi1mphisinth}
\end{figure}

In Fig.~\ref{fig:-mchi1mphisinth}(a) and (b) we show the effective spin-independent direct detection cross-section ($\sigma^{\rm eff}_{\chi_1}$) of the 
WIMP-like fermion DM $\chi_1$ as a function of its mass ($m_{\chi_1}$) for two different mass hierarchies. $\sigma^{\rm eff}_{\chi_1}$ is defined \cite{Duda:2002hf} as follows,

\begin{equation}
\rm \sigma^{\rm eff}_{\chi_1}=\frac{\Omega_{\chi_1}}{\Omega_{\chi_1}+\Omega_{\phi}}\sigma^{\rm{SI}}_{\chi_1 N}\,.
\end{equation}

In both figures~\ref{fig:-mchi1mphisinth} (a) and (b) $\sin\theta$ is shown as the color axis. It is clear that with increasing $\sin\theta$, with more doublet contribution, 
the direct detection cross-section for $\chi_1$ increases. One finds in Fig.~\ref{fig:-mchi1mphisinth}(a), when $m_{\chi_1} \lsim 100$ GeV, only in the vicinity of $Z$-resonance, 
we get points allowed by relic density, but disfavoured from direct search data. But in the reverse hierarchy (Fig.~\ref{fig:-mchi1mphisinth}(b)), the $Z$-resonance region is not particularly distinct. 
The reason behind this is the following. When $m_{\chi_1} > m_{\phi}$ the conversion channel from $\chi_1$ to $\phi$ is open which helps $\chi_1$ deplete considerably and become 
under-abundant, whereas when $m_{\chi_1} < m_{\phi}$, this conversion is kinematically disfavoured and therefore the under-abundance for $\chi_1$ is achieved primarily near 
$Z$-resonance. Here too, small mass difference between $\chi_1$ and $\chi_2$ can facilitate co-annihilation and there is possibility of under-abundance with appropriate choice of 
$\Delta m$. The detailed calculation of direct detection cross-section of WIMP can be found in Appendix~\ref{appendixd}. Broadly we see that WIMP mass is required to be larger than 
$\sim$ 100 GeV with $\sin\theta \lesssim 0.1$.

%%%%%%%%%%%%%%%%%%%
\subsubsection{Direct detection of pFIMP}
%%%%%%%%%%%%%%%%%%%

The FIMP having negligible coupling with SM states is difficult to probe in direct search experiments. pFIMP on the other hand, despite having negligible couplings to SM, 
has a prospect of being detected at direct search experiments via substantial coupling to the WIMP.  As discussed in Section~\ref{sec:model}, the pFIMP coupling to SM occurs 
via the WIMP-loop, which can have a non-negligible contribution to the elastic scattering between pFIMP and detector nucleon.

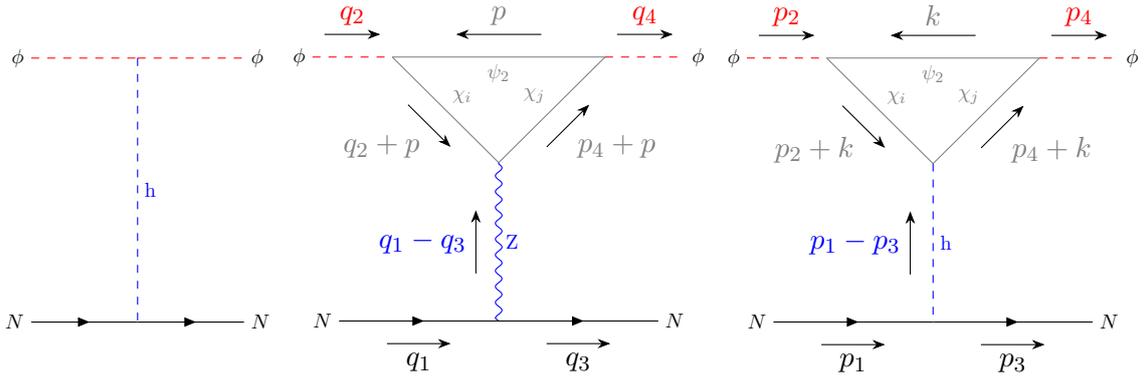
\begin{figure}[htb!]
	\centering	
	\begin{tikzpicture}[baseline={(current bounding box.center)},style={scale=0.7, transform shape}]
	\begin{feynman}
	\vertex (a);
	\vertex[left=2cm of a] (a1){\(\phi\)};
	\vertex[right=2cm of a] (a2){\(\phi\)}; 
	\vertex[below=5cm of a] (b); 
	\vertex[left=2cm of b] (c1){\(N\)};
	\vertex[right=2cm of b] (c2){\(N\)};
	\diagram* {
		(a1) -- [scalar,style=red] (a) -- [scalar,style=red] (a2),(a) -- [scalar, edge label={\(\rm h\)},,style=blue] (b) ,(c1) --[fermion, arrow size=1pt] (b)  --[fermion, arrow size=1pt] (c2)
	};\end{feynman}
	\end{tikzpicture}
	\begin{tikzpicture}[baseline={(current bounding box.center)},style={scale=0.7, transform shape}]
	\begin{feynman}
	\vertex (a);
	\vertex[above left=2cm and 2cm of a] (d1);
	\vertex[above right=2cm and 2cm of a] (d2); 
	\vertex[left=1.5cm of d1] (a1){\(\phi\)};
	\vertex[right=1.5cm of d2] (a2){\(\phi\)}; 
	\vertex[below=3cm of a] (b); 
	\vertex[left=3cm of b] (c1){\(N\)};
	\vertex[right=3cm of b] (c2){\(N\)};	
	\diagram* {
		(a1) -- [scalar, momentum ={\(q_2\)},style=red] (d1),(d2) -- [scalar, momentum ={\(q_4\)},style=red] (a2),(d1) --[plain, edge label={\(\chi_i\)},  momentum' ={[arrow shorten=0.3]\(q_2+p\)},style=gray](a),(a)--[plain, edge label={\(\chi_j\)},  momentum' ={[arrow shorten=0.3]\(p_4+p\)},style=gray](d2),(a) -- [boson, edge label={\(\rm Z\)},style=blue,reversed momentum' ={[arrow shorten=0.3]\(q_1-q_3\)}] (b) ,(c1) --[fermion, arrow size=1pt, momentum' ={[arrow shorten=0.3]\(q_1\)}] (b)  --[fermion, arrow size=1pt, momentum' ={[arrow shorten=0.3]\(q_3\)}] (c2),(d2)--[plain, edge label={\(\psi_2\)}, momentum' ={[arrow shorten=0.3]\(p\)},style=gray](d1)
	};\end{feynman}
	\end{tikzpicture}
	\begin{tikzpicture}[baseline={(current bounding box.center)},style={scale=0.7, transform shape}]
	\begin{feynman}
	\vertex (a);
	\vertex[above left=2cm and 2cm of a] (d1);
	\vertex[above right=2cm and 2cm of a] (d2); 
	\vertex[left=1.5cm of d1] (a1){\(\phi\)};
	\vertex[right=1.5cm of d2] (a2){\(\phi\)}; 
	\vertex[below=3cm of a] (b); 
	\vertex[left=3cm of b] (c1){\(N\)};
	\vertex[right=3cm of b] (c2){\(N\)};	
	\diagram* {
		(a1) -- [scalar, momentum ={\(p_2\)},style=red] (d1),(d2) -- [scalar, momentum ={\(p_4\)},style=red] (a2),(d1) --[plain, edge label={\(\chi_i\)},  momentum' ={[arrow shorten=0.3]\(p_2+k\)},style= gray](a),(a)--[plain, edge label={\(\chi_j\)},  momentum' ={[arrow shorten=0.3]\(p_4+k\)},style= gray](d2),(a) -- [scalar, edge label={\(\rm h\)},style=blue, reversed momentum' ={[arrow shorten=0.3]\(p_1-p_3\)}] (b) ,(c1) --[fermion, arrow size=1pt, momentum' ={[arrow shorten=0.3]\(p_1\)}] (b)  --[fermion, arrow size=1pt, momentum' ={[arrow shorten=0.3]\(p_3\)}] (c2),(d2)--[plain, edge label={\(\psi_2\)}, momentum' ={[arrow shorten=0.3]\(k\)},style= gray](d1)
	};\end{feynman}
	\end{tikzpicture}
	\caption{The (left) tree-level and (middle) and (right) one-loop Feynman diagram for the direct detection of pFIMP $\phi$ .}
    \label{fig:loopdd}
\end{figure}

In Figure~\ref{fig:loopdd}, we show the diagrams which contribute to the direct search cross-section of pFIMP ($\phi$) in our model. 
The diagram (Figure~\ref{fig:loopdd} (left)) involving the Higgs portal coupling of pFIMP, contributes negligibly to the total amplitude. 
Figure~\ref{fig:loopdd} (middle) shows the WIMP-loop induced contribution with a $t$-channel $Z$ mediator, whereas Figure~\ref{fig:loopdd} (right) 
shows the same with $t$-channel Higgs mediation. 

The coupling that plays crucial role in determining the loop amplitude is the WIMP-pFIMP coupling (Yukawa coupling $Y_2$), which was also 
a key parameter in governing the pFIMP dynamics. On the other hand, one has to also rememeber that the loop contributions are also a direct 
consequence of the singlet-doublet mixing in our model. In the absence of mixing, the loop contribution to pFIMP-nucleon interaction vanishes. 
Therefore, not only the Yukawa coupling $Y_2$, but also the Yukawa coupling $Y_1$ is crucial in this context. In addition, the mixing term is 
directly proportional to the mass difference between the WIMP and the second lightest dark sector particle $(m_{\chi_2}-m_{\chi_1})$, as we 
have seen in Equation~\eqref{eq:massdiff}. The smaller the mass difference, the weaker is the detectability of the pFIMP at direct detection experiments. 
Similar to the WIMP case, here too the Higgs mediated diagram contributes much more compared to the $Z$-mediated case, thanks to the small $\sin\theta$. 
For an order-of-magnitude estimate of the Higgs and $Z$ mediated contributions, one can see Figure~\ref{fig:-lh-dmy2} and \ref{fig:-lz-dmy2} in Appendix~\ref{appendixb}. 
We would like to emphasize here that, we get an advantage by choosing a fermionic WIMP candidate over a scalar. It was shown in \cite{DiazSaez:2021pfw}, 
that the pFIMP in a two-component scalar DM model, will have negligible contribution to the direct detection as the scalar loop-amplitude vanishes at the low transfer 
momentum limit (unless the WIMP is a warm DM in the keV mass-range). This is certainly not the case with fermion WIMP loop, which is our case. One can see the detailed calculation 
involving the fermion loop in Appendix~\ref{appendixb}. Herein lies another very important motivation behind choosing our model. 
The detailed calculation of direct detection cross-section of pFIMP can be found in Appendix~\ref{appendixc}, which we use for the 
parameter space scan discussed next.

 \begin{figure}[htb!]
\centering
\subfloat[]{\includegraphics[width=0.5\linewidth]{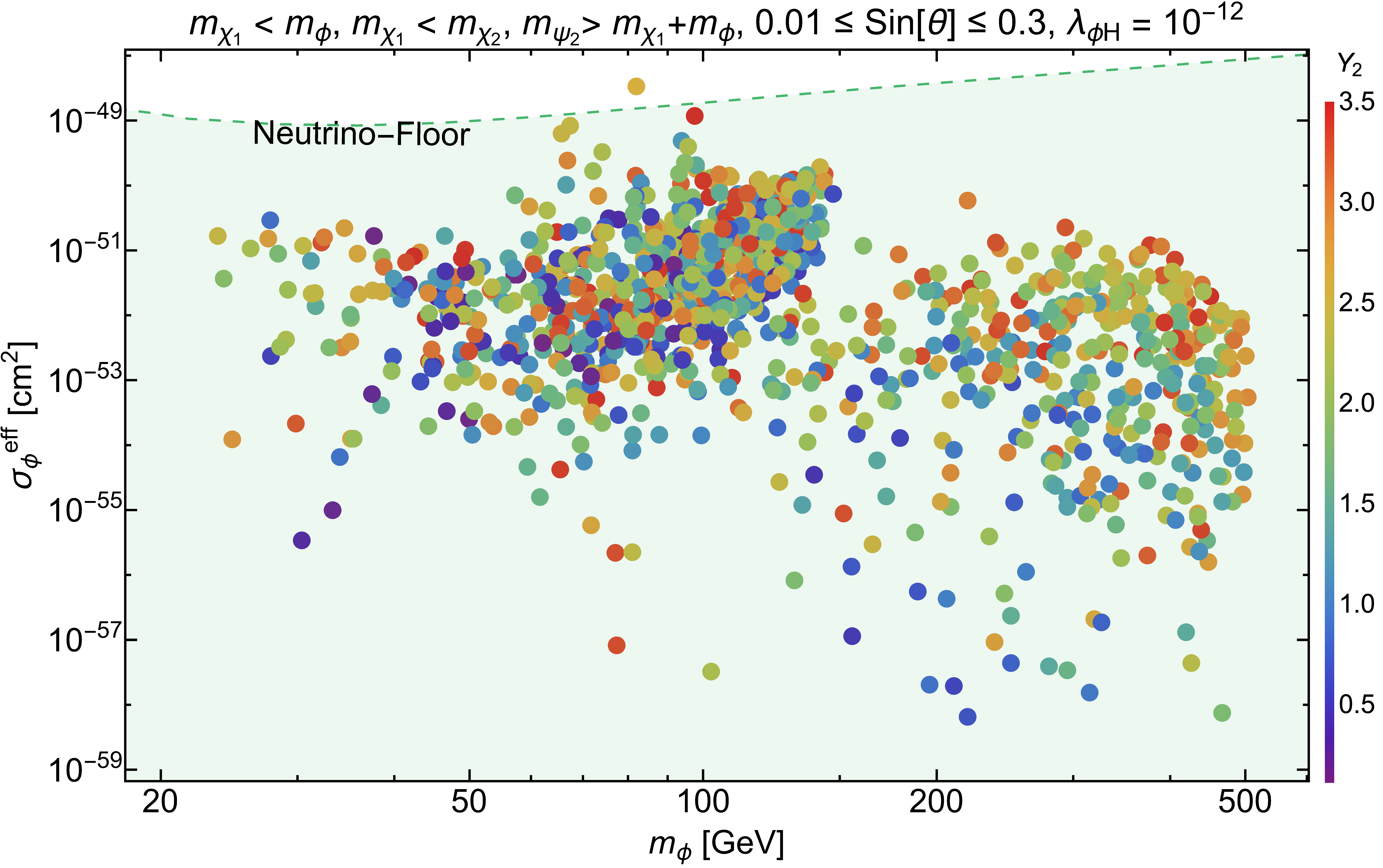}\label{fig:-mchi1lmphi_mphi-dd-y2}}~~
\subfloat[]{\includegraphics[width=0.5\linewidth]{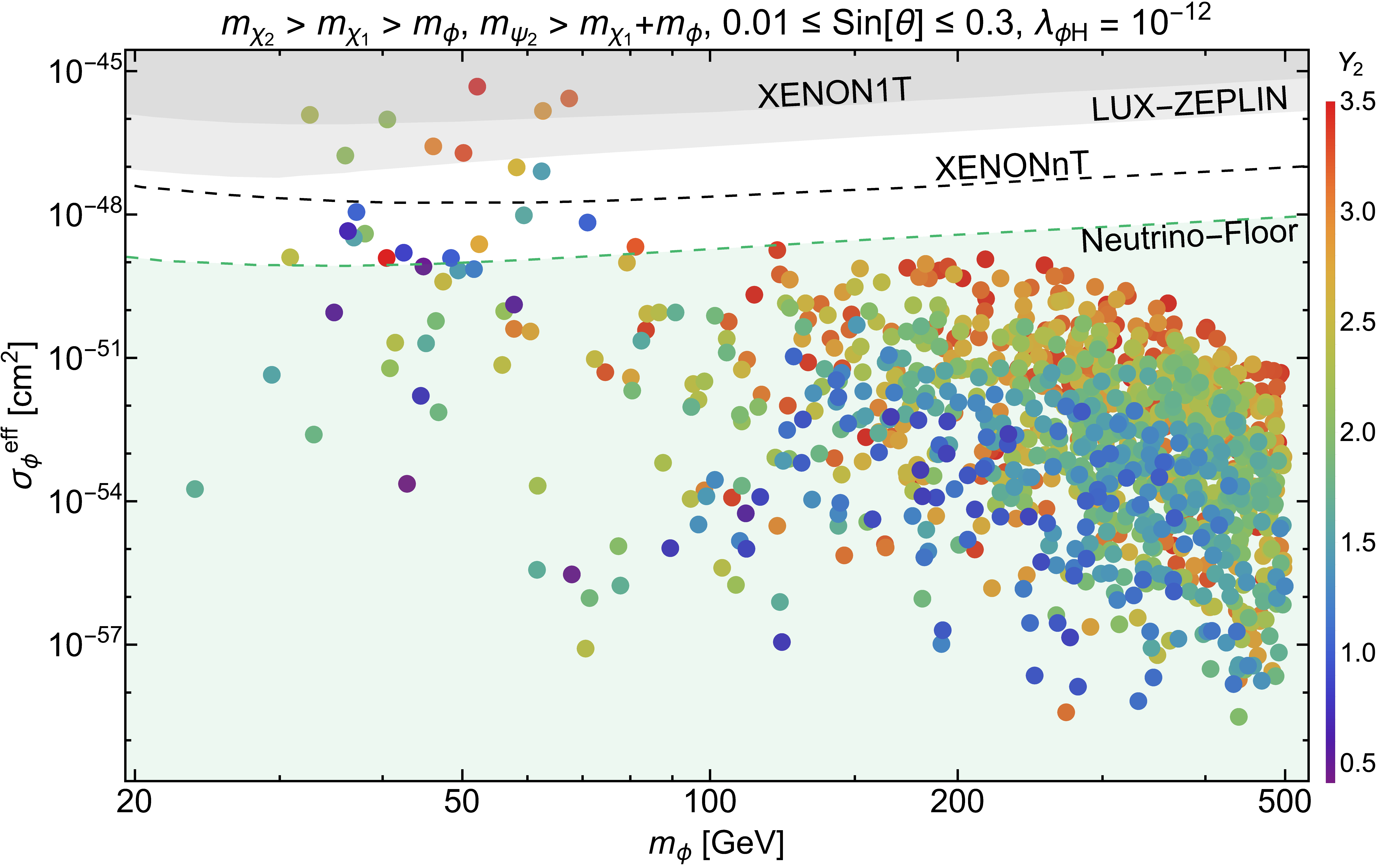}\label{fig:-mchi1gmphi_mphi-dd-y2}}
  \caption{Effective spin-independent direct detection cross-section ($\sigma^{\rm eff}_{\phi}$) for pFIMP $\phi$ for (a) $m_{\chi_1} < m_{\phi}$ and (b) $m_{\chi_1} > m_{\phi}$. 
  All the points satisfy the present DM relic density bound $0.1188\leq\Omega_{\chi_1}h^2+\Omega_{\phi}h^2\leq 0.1212$ via the combined contribution of both DM's. 
  $Y_2$ is shown as the color axis in both figures. Other parameters kept fixed are shown in figure heading. The limits from XENON1T, LUX-ZEPLIN data and future sensitivities 
  from XENONnT and Neutrino floor are shown.}
  \label{fig:-mchi1mphiy2}
\end{figure}
%%%%%%%%%%%%%%%%%%%

In Fig.~\ref{fig:-mchi1mphiy2}, we present the effective spin-independent direct detection cross-section ($\sigma^{\rm eff}_{\phi}$) of the pFIMP $\phi$ as a function of its mass 
($m_{\phi}$) for two different mass hierarchies. The definition of $\sigma^{\rm eff}_{\phi}$ follows as before,

\bea\begin{split}
\sigma^{\rm eff}_{\phi}=\frac{\Omega_{\phi}}{\Omega_{\chi_1}+\Omega_{\phi}}\sigma^{\rm{SI}}_{\phi N}\,.
\end{split}
\eea

\noindent
Since the tree-level coupling of $\phi$ with Higgs is extremely small $\lambda_{\phi H} \sim 10^{-12}$, the major contribution to $\sigma^{\rm eff}_{\phi}$ comes from 
the fermion-loop induced diagrams (Fig.~\ref{fig:loopdd}). Therefore, the parameters $\Delta m$, $Y_2$ and $\sin\theta$ play a crucial role. In order to achieve considerable 
direct detection cross-section a large $\Delta m$ is desirable, as already pointed out. On the other hand, large $\Delta m$ in turn means the absence of co-annihilation and therefore, 
over-abundance of $\chi_1$. This situation is evident in Figure~\ref{fig:-mchi1mphiy2}(a), where the parameter space allowed by observed relic density immediately implies 
direct detection cross-section for the pFIMP below the neutrino floor. This tension is relaxed when $m_{\chi_1} > m_{\phi}$, since in this case the conversion channel from 
$\chi_1$ to $\phi$ becomes kinematically favoured and under-abundance of $\chi_1$ is possible even with large $\Delta m$. This in turn ensures moderate direct search 
cross-section ($10^{-49} - 10^{47}$) for pFIMP $\phi$ with mass $\lsim 100$ GeV for the next generation direct detection experiments like Xenon-nT (projected limit $10^{-49}$ cm$^2$) 
to probe such cases, as shown in Figure~\ref{fig:-mchi1mphiy2}(b). Hereby, we draw a crucial inference that $m_{\chi_1} > m_{\phi}$ is more favorable scenario for the detection of pFIMP, 
as compared to the inverse hierarchy.

%\subsubsection{Direct detection of WIMP-pFIMP scenario}

 \begin{figure}[htb!]
\centering
\includegraphics[width=0.5\linewidth]{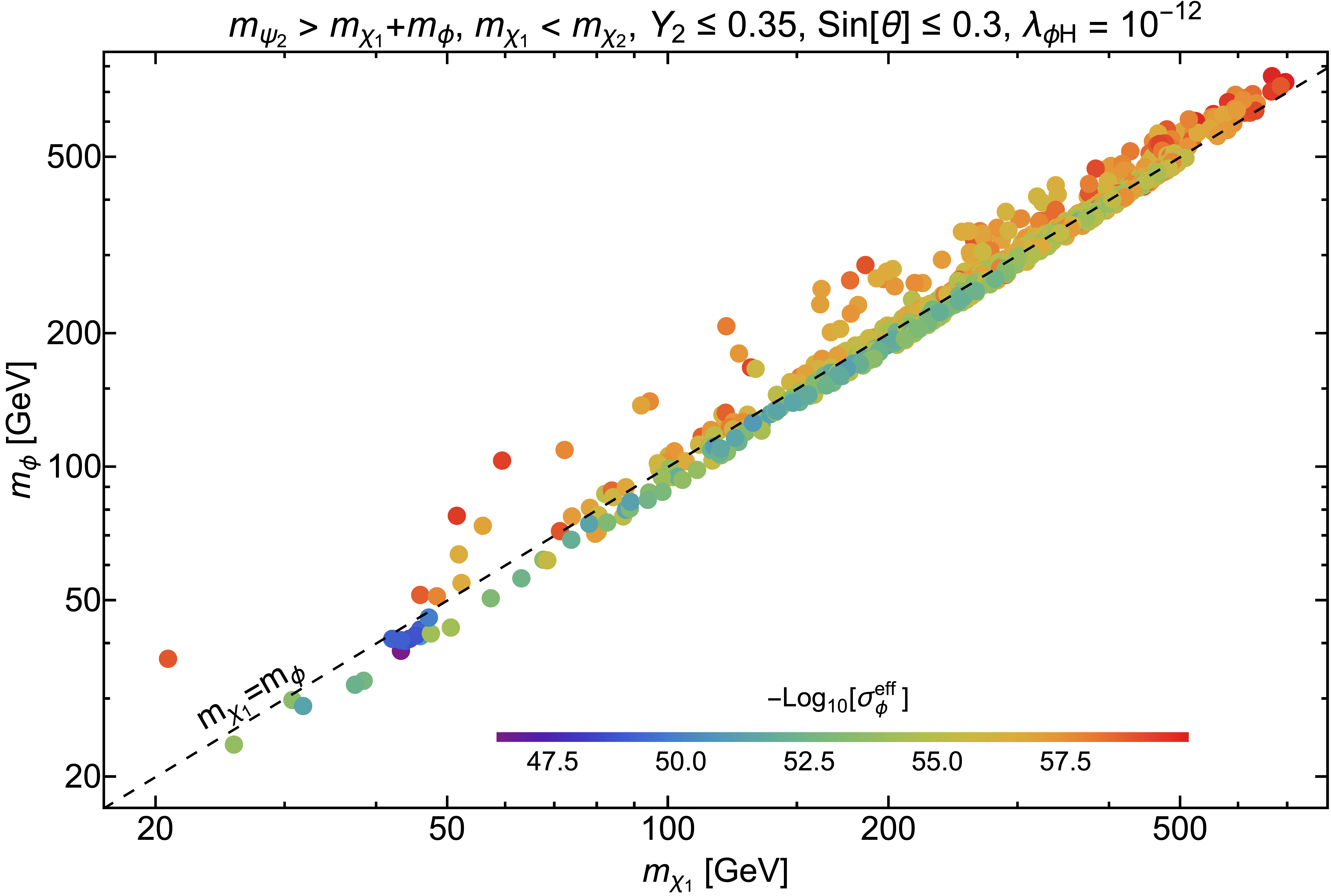}
 \caption{Allowed parameter space in $m_{\chi_1}-m_{\phi}$ plane, which respect the present observed relic density as well as have sensitivities for future 
 direct detection experiments beyond LUX-ZEPLIN bound. The color axis represents the effective spin-independent pFIMP-nucleon scattering cross-section, 
 $\sigma_{\phi}^{\rm eff}$ in $\rm cm^2$ in Log scale.}
\label{fig:mchi1mphi-dd}
\end{figure}

Having discussed the individual aspects of WIMP and pFIMP direct detection, we would also like to make a connection between the two.  
In Figure \ref{fig:mchi1mphi-dd}, we have shown the allowed parameter space which respect the present relic density and direct detection 
(LUX-ZEPLIN) bound in $m_{\chi_1}-m_{\phi}$ plane. The color axis represents the effective spin-independent pFIMP-nucleon scattering 
cross-section, $\sigma_{\phi}^{\rm eff}$, in $\rm cm^2$ in Log scale. It has been shown in \cite{Bhattacharya:2022dco}, 
when $m_{\chi_1} > m_{\phi}$, WIMP-pFIMP conversion is significant via large $Y_2$ as well as via small mass difference between the WIMP 
and pFIMP. Furthermore, the contribution of $\phi$ to total relic in this case increases due to $\chi_1$ to $\phi$ conversion, and therefore the 
effective direct detection too cross-section increases. In addition, if $\chi_1$ is in the $Z$ resonance, under-abundance of $\chi_1$ becomes 
further enhanced and it becomes easier to achieve large $\Delta m$ and consequently large direct detection cross-section for the pFIMP 
(blue points in the vicinity of $Z$ resonance in Fig.~\ref{fig:mchi1mphi-dd}). For the inverse hierarchy on the other hand, the under-abundance of 
WIMP is solely dependent on its co-annihilation and therefore, large $\Delta m$ values are disfavoured, resulting in small direct detection cross-section 
for the pFIMP. We have checked that even with $\chi_1$ in the vicinity $Z$ resonance, the dependence on co-annihilation is not relaxed and therefore, 
direct detection cross-section for the pFIMP remains below the neutrino floor for almost the entire parameter space.

%It is clear that when $m_{\chi_1} > m_{\phi}$, 
%$\sigma_{\phi}^{\rm eff}$ can be compared to the inverse hierarchy as discussed earlier. 
% and Fig.~\ref{fig:dmdm-bb}%(a) and (b). 

 \begin{figure}[htb!]
\centering
\includegraphics[width=0.5\linewidth]{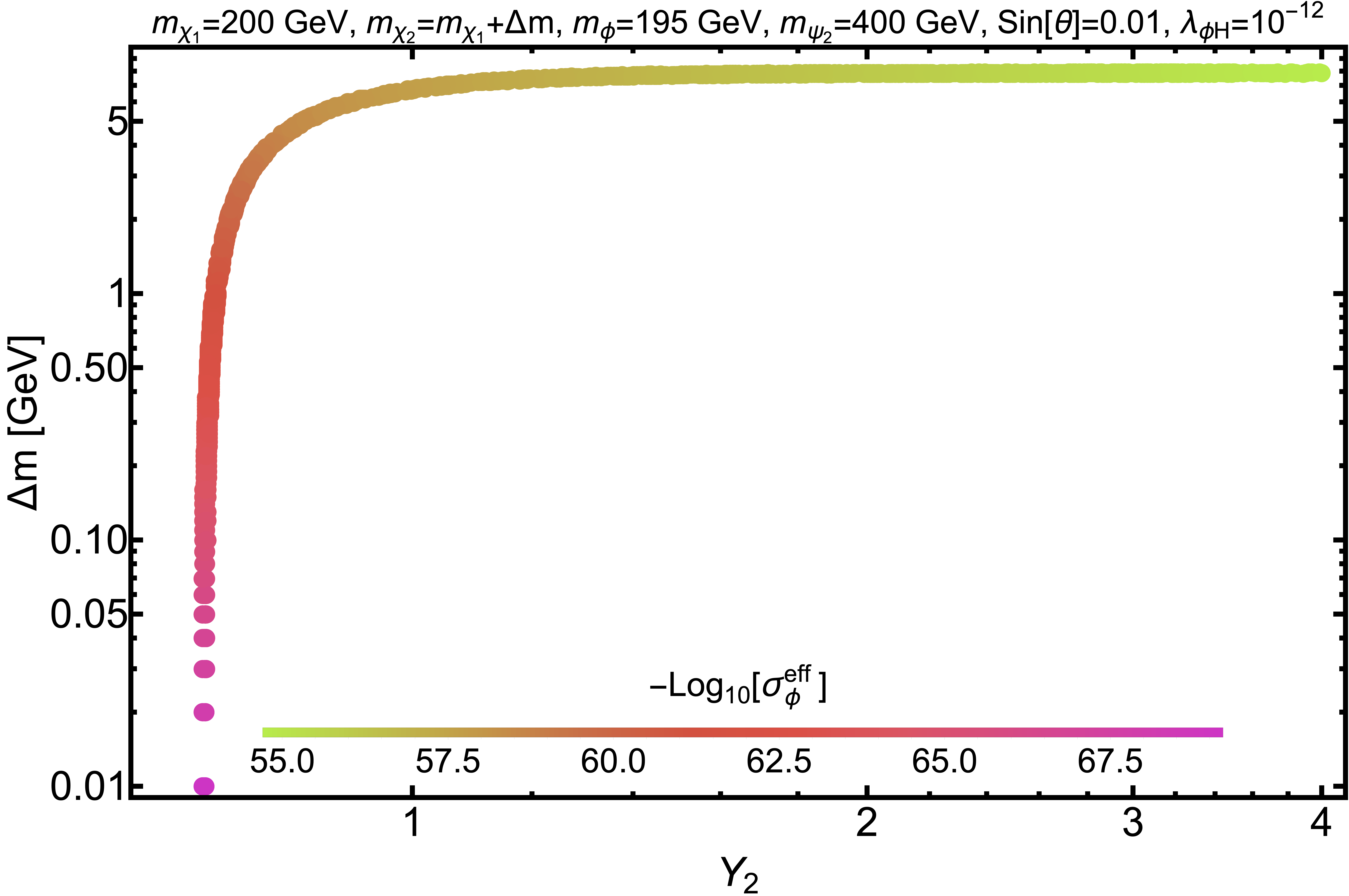}
\caption{Parameter space allowed by observed relic and direct search in $Y_2 - \Delta m$ plane, where the spin-independent effective DM-nucleon elastic scattering cross-section of scalar DM ($\sigma_\phi^{\rm eff}$ in $\rm cm^2$) is shown as the color axis in Log scale.}
\label{fig:deltam_y2_dd}
\end{figure}

\noindent
In Fig.~\ref{fig:deltam_y2_dd}, we show the relic density and direct search allowed points in $Y_2-\Delta m$ plane, with the color axis denoting the direct detection cross-section of 
the pFIMP $\phi$. As discussed before large $\Delta m$ will imply enhanced contribution from the fermion loop (see Figure~\ref{fig:loopdd}). Therefore the spin-independent 
pFIMP-nucleon cross-section will increase with increasing $\Delta m$, which is evident from the transition in color in Figure~\ref{fig:deltam_y2_dd}. Needless to say, increasing 
$Y_2$ is also crucial in obtaining enhanced contribution of $\phi$ to $\sigma_{\phi}^{\rm eff}$. Increasing $\Delta m$ can also enhance the direct detection contribution of pFIMP. 
However, the requirement from observed relic forbids us to go beyond $\Delta m \sim 10$ GeV, a restriction which can be relaxed in the $Z$-resonance region, as pointed out earlier.

\subsection{Indirect detection possibility}
%%%%%%%%%%%%%%%%%%%%%%%%%%%%%%%%%%%%%%%%%%%%%%%

Similar to direct detection of pFIMP, one may also look for indirect signal evidence of pFIMP, analysing the photon flux \cite{Ullio:2002pj} 
in the existing and future indirect detection experiments such as Fermi-LAT \cite{PhysRevD.104.083026}, SK \cite{Super-Kamiokande:2020sgt}, 
H.E.S.S \cite{HESS:2022ygk}, IceCube \cite{IceCube:2016umi,IceCube:2022vtr,IceCube:2021xzo} etc. We have considered the recent data for 
DM annihilation channels to $b\overline{b}$, $\tau^+\tau^-$ and $W^+W^-$ from various experiments and study their effect on our model parameter space. 
The strongest bounds come from the $b\overline{b}$ annihilation channel. The effective annihilation cross-section 
of a DM pair to $b \bar b$ final state is given as follows \cite{DiazSaez:2021pmg,Reinert:2017aga}, 

\begin{equation}
\langle\sigma v\rangle_{\rm{DM~DM}\to b\overline{b}}^{\rm ID}=\frac{\Omega_{\rm DM}^2}{(\Omega_{\rm \chi_1}+\Omega_{\phi})^2}\langle\sigma v\rangle_{\rm{DM ~DM}\to b\overline{b}}\,.
\end{equation}

% and are at the level of $7\times 10^{-26}\rm cm^3 s^{-1}$ at $\rm6~GeV$ WIMP mass \cite{PhysRevD.104.083026}. 

%The cosmic particle flux for dark matter annihilation at the centre of our galaxy \cite{Gaskins:2016cha},
%\bea
%&\frac{d\Phi_{i}}{dE_{i }}\bigg|_{ann}=\frac{<\sigma v>}{\kappa~ m_{DM}^2}\frac{dN_{i}}{dE_{i }}\frac{1}{4\pi}J_{ann}(\psi)\\&
%\frac{d\Phi_{i}}{dE_{i }}\bigg|_{dec}=\frac{<\sigma v>}{ m_{DM}\tau}\frac{dN_{i}}{dE_{i }}\frac{1}{4\pi}J_{decay}(\psi)
%\eea
%Where $i$ denotes the observed secondary particle (gamma-rays, neutrinos), $J_{\rm ann}(\psi)=\int_{los}\rho^2(\psi,l)dl$ and $J_{\rm dec}(\psi)=\int_{los}\rho(\psi,l)dl$, $\psi $ is the direction from which $i$ particle produced by dark matter annihilation, $l$ is the distance along the line-of-sight (loc) and $\rho$ is the dark matter density. $\frac{dN_{i}}{dE_{i}}$ is the differential spectral of $i$ particle emitted per annihilation/decay. $\tau $ is the life time of dark matter particle. $\kappa $ is 2 for the dark matter which is its own anti-particle, and becomes 4 if the dark matter is not its own anti-particle. The total spectrum of $i$ particles emitted per annihilation/decay $(dN_{i}/dE_i)$ can be written as the sum of the
%spectra produced for all possible final states $f$ can be SM particle which is kinematically accessible, $\frac{dN_{i}}{dE_i}=\sum\limits_{f}B_f\frac{dN_{i ,f}}{dE_i}$ where $B_f$ is the branching ratio to final state $f$.

%\subsubsection{DM annihilation to pair of charge particle}
 
%%%%%%%%%%%%%%%%%%%%%%%%%%%%%%%%%%%%%%%%%%%%%%%%
\begin{figure}[htb!]
	\centering	
	\begin{tikzpicture}[baseline={(current bounding box.center)}]
	\begin{feynman}
	\vertex (a);
	\vertex[below right=1cm and 1cm of a] (a2){\(\rm{SM}\)};
	\vertex[above right=1cm and 1cm of a] (a1){\(\rm{SM}\)}; 
	\vertex[left=1cm of a] (d); 
	\vertex[above left=1cm and 1cm of d] (d1){\(\chi_1\)};
	\vertex[below left=1cm and 1cm of d] (d2){\(\chi_1\)};	
	\diagram* {
	(a1) -- [plain] (a),(a) -- [plain] (a2),(a) -- [scalar,style=blue,edge label={\(\rm{h}\)}] (d) ,(d) --[fermion ,arrow size=1pt] (d1) ,(d2) --[fermion ,arrow size=1pt] (d)
	};\end{feynman}
	\end{tikzpicture}
	\begin{tikzpicture}[baseline={(current bounding box.center)}]
	\begin{feynman}
	\vertex (a);
	\vertex[below right=1cm and 1cm of a] (a2){\(\rm{SM}\)};
	\vertex[above right=1cm and 1cm of a] (a1){\(\rm{SM}\)}; 
	\vertex[left=1cm of a] (d); 
	\vertex[above left=1cm and 1cm of d] (d1){\(\chi_1\)};
	\vertex[below left=1cm and 1cm of d] (d2){\(\chi_1\)};	
	\diagram* {
	(a1) -- [plain] (a),(a) -- [plain] (a2),(a) -- [boson,style=blue,edge label={\(\rm Z \)}] (d) ,(d) --[fermion ,arrow size=1pt] (d1) ,(d2) --[fermion ,arrow size=1pt] (d)
	};\end{feynman}
	\end{tikzpicture}
	\begin{tikzpicture}[baseline={(current bounding box.center)}]
	\begin{feynman}
	\vertex (a);
	\vertex[below right=1cm and 1cm of a] (a2){\(\rm{SM}\)};
	\vertex[above right=1cm and 1cm of a] (a1){\(\rm{SM}\)}; 
	\vertex[left=1cm of a] (d); 
	\vertex[above left=1cm and 1cm of d] (d1){\(\phi\)};
	\vertex[below left=1cm and 1cm of d] (d2){\(\phi\)};	
	\diagram* {
	(a1) -- [plain] (a),(a) -- [plain] (a2),(a) -- [scalar,style=blue,edge label={\(\rm{h}\)}] (d) ,(d) --[scalar ,arrow size=1pt] (d1) ,(d2) --[scalar ,arrow size=1pt] (d)
	};\end{feynman}
	\end{tikzpicture}
	
	\begin{tikzpicture}[baseline={(current bounding box.center)},style={scale=0.7, transform shape}]
	\begin{feynman}
	\vertex (a);
	\vertex[above left=2cm and 2cm of a] (d1);
	\vertex[below left=2cm and 2cm of a] (d2); 
	\vertex[left=2cm of d1] (a1){\(\phi\)};
	\vertex[left=2cm of d2] (a2){\(\phi\)}; 
	\vertex[right=2cm of a] (b); 
	\vertex[above right=2.83cm of b] (c1){\(\rm SM\)};
	\vertex[below right=2.83cm of b] (c2){\(\rm SM\)};	
	\diagram* {
		(a1) -- [scalar,style=black] (d1),(d2) -- [scalar,style=black] (a2),(d1) --[plain, edge label={\(\chi_i\)},  style=gray](a),(a)--[plain, edge label={\(\chi_j\)},  style=gray](d2),(b) -- [scalar, edge label={\(\rm h\)},style=blue] (a) ,(c1) --[plain] (b)  --[plain] (c2),(d2)--[plain, edge label={\(\psi_2\)},style=gray](d1)
	};\end{feynman}
	\end{tikzpicture}
	\begin{tikzpicture}[baseline={(current bounding box.center)},style={scale=0.7, transform shape}]
	\begin{feynman}
	\vertex (a);
	\vertex[above left=2cm and 2cm of a] (d1);
	\vertex[below left=2cm and 2cm of a] (d2); 
	\vertex[left=2cm of d1] (a1){\(\phi\)};
	\vertex[left=2cm of d2] (a2){\(\phi\)}; 
	\vertex[right=2cm of a] (b); 
	\vertex[above right=2.83cm of b] (c1){\(\rm SM\)};
	\vertex[below right=2.83cm of b] (c2){\(\rm SM\)};	
	\diagram* {
		(a1) -- [scalar,style=black] (d1),(d2) -- [scalar,style=black] (a2),(d1) --[plain, edge label={\(\chi_i\)},  style=gray](a),(a)--[plain, edge label={\(\chi_j\)},  style=gray](d2),(b) -- [boson, edge label={\(\rm Z\)},style=blue] (a) ,(c1) --[plain] (b)  --[plain] (c2),(d2)--[plain, edge label={\(\psi_2\)},style=gray](d1)
	};\end{feynman}
	\end{tikzpicture}
	\caption{The Tree-level (left) and one-loop (right) Feynman diagrams for the Indirect detection of WIMP $(\chi_1)$ and pFIMP $(\phi)$.}
    \label{fig:loopindirect}
\end{figure}
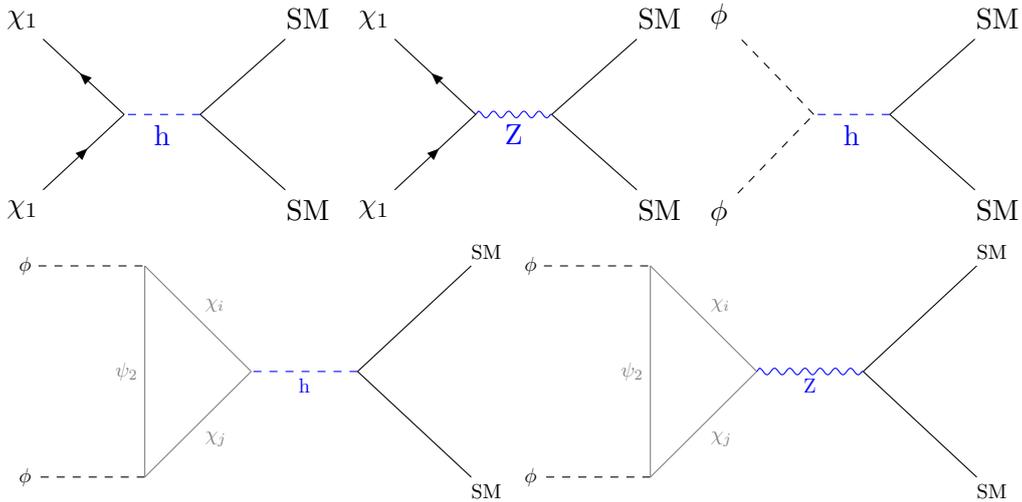
%%%%%%%%%%%%%%%%%%%%%%%%%%%%%%%%%%%%%%%%%%%%%%%%

 \begin{figure}[htb!]
 \centering
\subfloat[]{\includegraphics[width=0.475\linewidth]{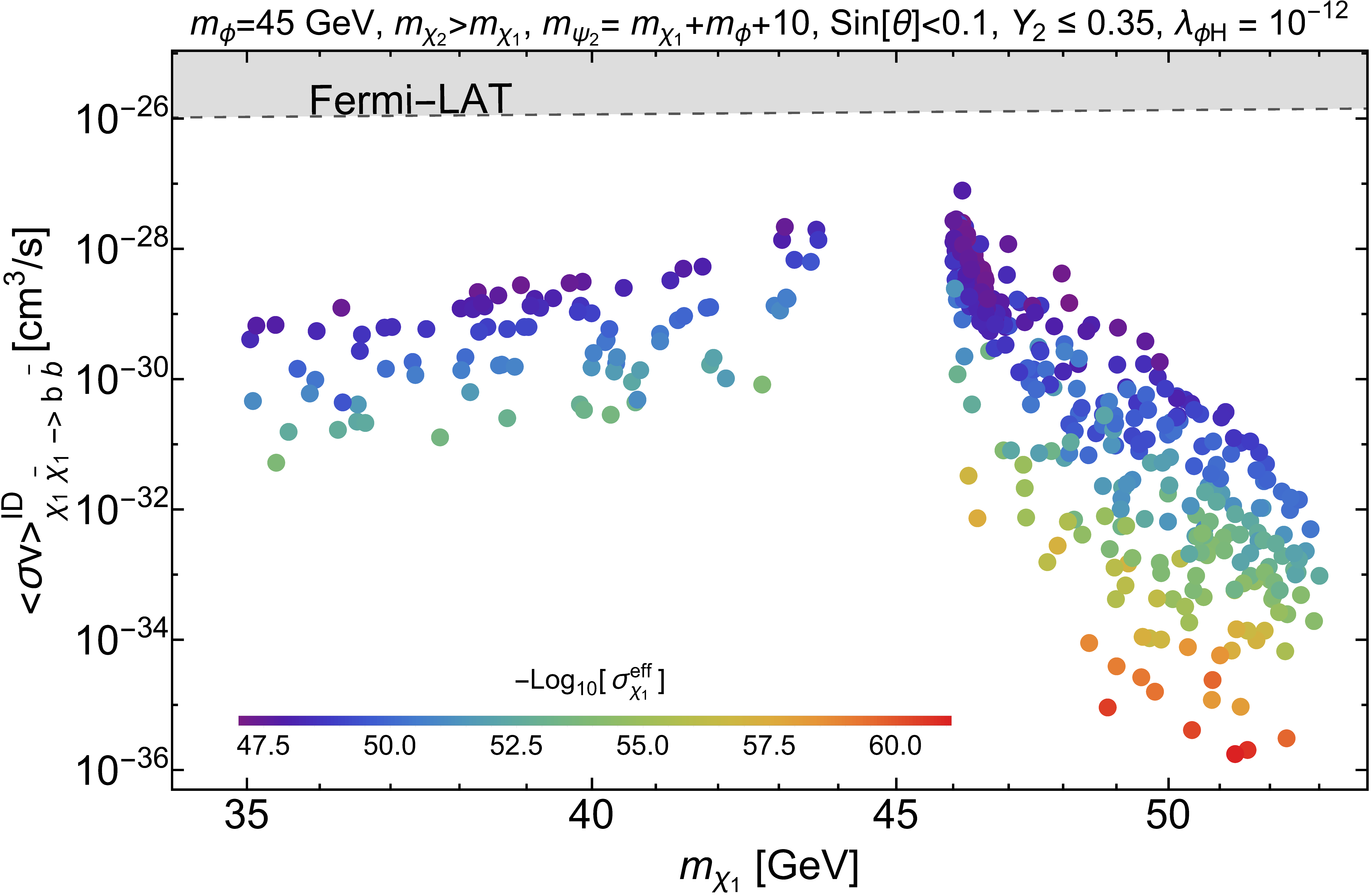}}~~
\subfloat[]{\includegraphics[width=0.475\linewidth]{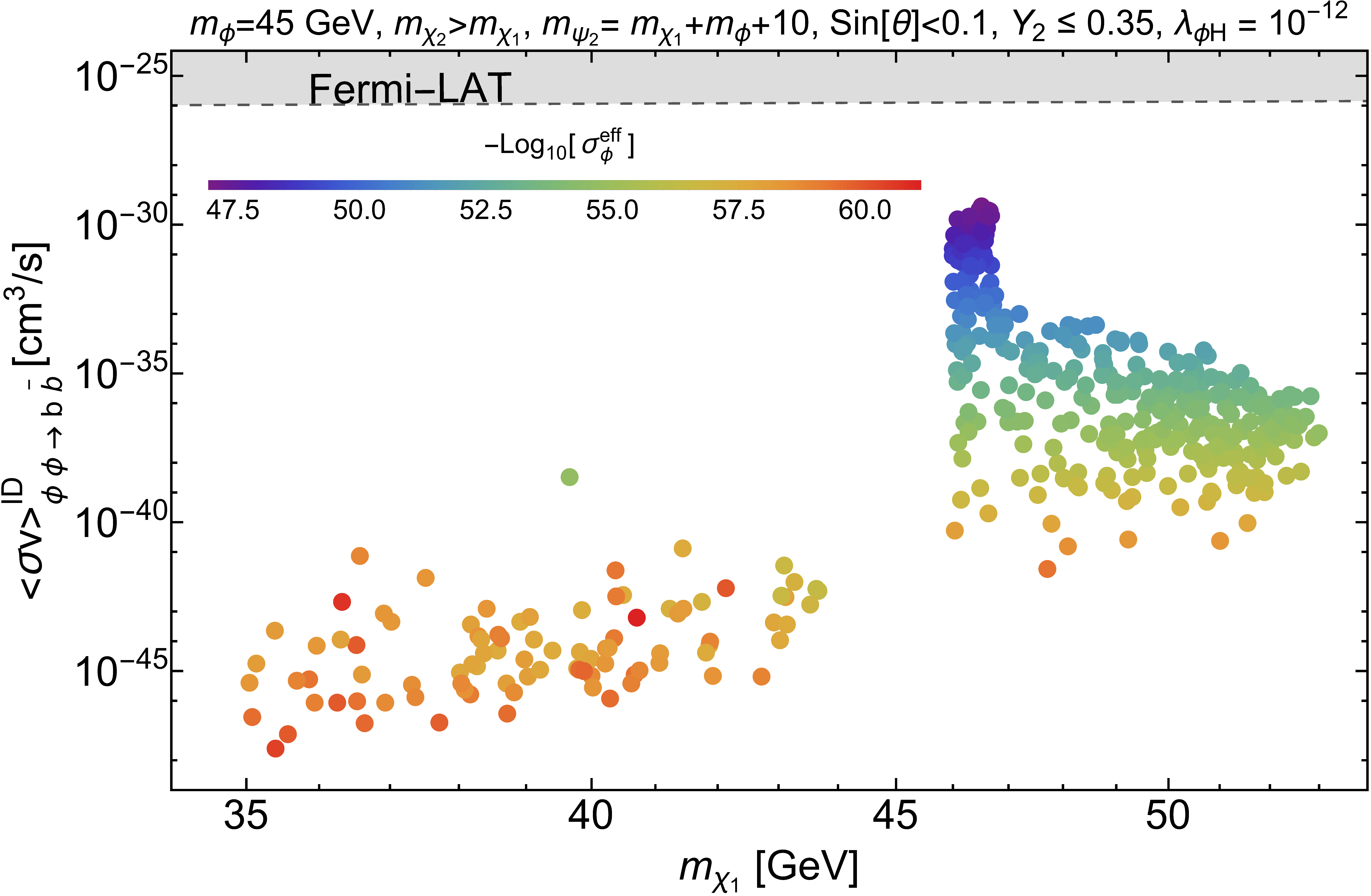}}
 \caption{DM annihilation cross-section to $b\overline{b}$, $\langle\sigma v\rangle^{\rm ID}_{\chi_1 \overline{\chi}_1\to b\overline{b}}$ and $\langle\sigma v\rangle^{\rm ID}_{\phi \phi\to b\overline{b}}$ 
 as a function of DM mass. The color axis denotes effective direct detection cross-section for pFIMP($\sigma_\phi^{\rm eff}$). All points in $\sigma^{\rm ID}-m_{\chi_1}$ plane satisfy 
 observed relic and LUX-ZEPLIN bound for both WIMP and pFIMP. Grey shaded region is excluded by DM annihilation to $b\overline{b}$ search at Fermi-LAT \cite{Fermi-LAT:2015att,PhysRevD.104.083026} and H.E.S.S \cite{HESS:2022ygk} data.}
 \label{fig:dmdm-bb}
\end{figure}

\noindent
In Figure~\ref{fig:loopindirect}, we show the processes that contribute to the aforementioned annihilation channels. 
We can see that the WIMP-loop induced diagrams take part in pFIMP annihilation, similar to the direct search case.  

In Figure~\ref{fig:dmdm-bb}, we plot the quantity $\langle\sigma v\rangle_{\rm{DM~DM}\to b\overline{b}}^{\rm ID}$ as a function of DM mass for WIMP 
(Figure~\ref{fig:dmdm-bb}(a)) and pFIMP (Figure~\ref{fig:dmdm-bb}(b)). The color axis denotes effective direct detection cross-section for pFIMP ($\sigma_\phi^{\rm eff}$). 
All the points in both the plots satisfy the observed relic and LUX-ZEPLIN bound. It is evident from the figures that the entire parameter space is allowed by 
indirect bound from Fermi-LAT (black dashed line) and can be probed at future experiments. Interestingly, the region that is most sensitive to the future indirect 
detection experiments are in the vicinity of $Z$-resonance. The same points produce maximum direct search cross-section for the pFIMP (dark blue points in Figure~\ref{fig:dmdm-bb}(b)) 
and thereby yield best discovery potential.

%During analysis, for simplicity, we have neglected the $Z$ mediated and tree level $\phi\phi\to b\overline{b}$ annihilation of $\phi$ which are suppressed compared to Higgs-mediated processes has shown in Fig. \ref{fig:loopindirect}. Near $Z$ resonance $\chi_1$ annihilation to $b\overline{b}$ cross-section enhanced due to $Z-$mediated process but in case of $\phi$, sufficient relaxation has come on $\Delta m$ near Z-resonance which effect $\mathbb{L}_h$ and finally cross-section. So, near Z-resonance annihilation of $\phi$ to $b\overline{b}$ is enhanced to get the relic allowed parameter space. The most of this allowed parameter space are in neutrino floor which are a big challenge to distinguish out the dark matter signal from neutrino but may have a possibility \cite{OHare:2015utx,Aalbers:2022dzr}.

%%%%%%%%%%%%%%%%%%%%%%%%%%%%%%%%%%%%%%%%%%%%%%%%

%\subsection{{\color{red}{Collider signature of WIMP and pFIMP???}}}

%Collider signal analysis for WIMP $(\chi_1)$ already has in \cite{Bhattacharya:2018cgx} while for pFIMP signal in collider, channels are bottom two figure inFig. \ref{fig:loopindirect} ,is only missing energy (dark matter $\phi$) plus mono-photon (lepton collider) is duly suppressed by background. So, in presence of pFIMP total signal is dominantly contributed mostly by WIMP $(\chi_1)$.

%%%%%%%%%%%%%%%%%%%%%%%%%
\section{Conclusion}
\label{conclusion}
%%%%%%%%%%%%%%%%%%%%%%%%%

Our focus is on a two-component DM scenario involving a thermal WIMP and a non-thermal FIMP having negligible interaction with SM. 
When the interaction between WIMP and FIMP becomes of weak interaction strength, the FIMP thermalises and freezes-out just like the WIMP. 
We categorized such DM candidates as pseudo-FIMP (pFIMP) and explored various aspects of this mechanism in our earlier work~\cite{Bhattacharya:2022dco} 
in a model independent manner. In this work, we focus on a specific two component DM model to analyze the parameter space on the pFIMP region.
The two different mass hierarchies present interesting dynamics as well as detection possibilities.
  
A crucial aspect of this work is to explore the possibility of detection of pFIMP in direct and indirect search experiments. It is well-known that a FIMP 
having tiny coupling to SM states, evades both present and future (projected) direct detection bounds. However, the pFIMP, aided by its significant interaction 
with the WIMP, can have considerable DM-nucleon cross-section at the direct search experiments. Although, the pFIMP has no direct connection to the SM, 
it can produce WIMP-loop induced amplitudes, which can bring the pFIMP under future experimental sensitivities. Having identified all such one-loop possibilities
with scalar, fermion and vector boson particles as WIMP and pFIMP, we choose a specific model which is likely to provide a better pFIMP direct and indirect search sensitivity. 

Our model consists of a fermion DM as WIMP which is an admixture of a singlet and a doublet. The pFIMP is a scalar singlet having negligible Higgs portal interaction and a 
substantial WIMP-pFIMP conversion via Yukawa interaction. We have scanned the parameter space of our model and identified the region which is most sensitive 
to both direct and indirect search experiments. We found that, one of the mass hierarchies, where WIMP is heavier than pFIMP, is more favored in terms of its direct detection. 
On the other hand, in our model, a substantial mass difference between the WIMP and the second lightest dark sector particle, which is directly related to the singlet-doublet mixing, 
is required for a better detection prospect. Interestingly, the WIMP mass range in the $Z$-resonance region, turns out to be most sensitive in both direct and indirect searches.
Importantly, the WIMP being a fermion in this model, also helps to generate a significant loop-induced amplitude. However, we do not claim that our model is the only scenario 
that is conducive for detecting a pFIMP. There is a plethora of possibilities, already outlined briefly in the beginning of this work, where similar analyses can be performed 
to study the rich phenomenology therein.  

One of the crucial features of WIMP-pFIMP set up to address the correct relic density is to have the mass difference between the two around $\lesssim$ 10 GeV. Therefore, we expect 
two DM signals in the same mass range. While this is very predictive in one hand, to disentangle them in such cases provide another important challenge ahead. 
The collider search prospect for pFIMP can similarly be studied involving a WIMP loop. This will be taken up in a future analysis.    
  
{\bf Acknowledgments:} SB and JL acknowledge the grant CRG/2019/004078 from SERB, Govt. of India.

\section*{Appendix}
%===============================================
\begin{appendix}
%%%%%%%%%%%%%%%%%%%%%%%%%%%%%%%%%%%%%%%%%%%%%%%%
\section{BEQ with Coannihilation of dark matter in Wimp-Fimp framework}\label{coannihilation}
Let us consider, $n_i$ dark sector particles have the same $Z_2$ symmetry, and their masses are $m_i$ ($m_1$ being the mass of the stable DM), internal d.o.f $g_i$. The evolution of the number density $n_i$ of particle $i$ can be written as

\bea
\dot{n}_i+3Hn_i=-\sum_j\langle\sigma v\rangle_{ij\to \rm SM}\left(n_in_j-n_i^{\rm eq}n_j^{\rm eq}\right)
\eea

\noindent
 Since all these dark sector particles with $i>1$, will eventually decay into the stable DM candidate after their respective freeze-out. Therefore the total DM density will be result of the combined yield of all the dark sector particles. Therefore, its final abundance($n$) can be described
by the sum of the density of all dark sector particle that transform under same $Z_2$ symmetry as the DM.

\bea
n=\sum_i n_i
\eea
The corresponding evolution equation for $n$ therefore can be written as follows~\cite{Griest:1990kh,Edsjo:1997bg}, without solving $n_i$ BEQ simultaneously.

\bea
\dot{n} = - 3Hn -\sum_{i,j}\langle\sigma v\rangle_{ij\to \rm SM}\left(n_in_j-n_i^{\rm eq}n_j^{\rm eq}\right)
\label{eq:10}
\eea
We have assumed that $n_i$ dark sector particles are initially in thermal bath with SM via annihilation process.
As the $n_i$ remains in thermal equilibrium,  and in particular
their ratios follow their equilibrium value, we use the relation $\frac{n_i}{n}=\frac{n_i^{\rm eq}}{n^{\rm eq}}$ and the Eq. \ref{eq:10} becomes,

%{\color{blue}{We have assumed that decay rate is larger than the scattering rates and the freeze-out of dark matter is determined only by the annihilation.}} 

\bea\begin{split}
\dot{n} = - 3Hn &=-\sum_{ij}\langle\sigma v\rangle_{ij\to \rm SM}\left(n_i^{\rm eq}\frac{n}{n^{\rm eq}}n_j^{\rm eq}\frac{n}{n^{\rm eq}}-n_i^{\rm eq}n_j^{\rm eq}\right)\\&
=-\sum_{ij}\langle\sigma v\rangle_{ij\to \rm SM}\frac{n_i^{\rm eq}}{n^{\rm eq}}\frac{n_j^{\rm eq}}{n^{\rm eq}}\left(n^2-n^{\rm eq^2}\right)\\&
=-<\sigma v>_{\rm SM}^{\rm eff}\left(n^2-n^{\rm eq^2}\right)
\label{eq:11}
\end{split}\eea
where,
\bea
<\sigma v>^{\rm eff}=\sum_{i,j} <\sigma v>_{ij}\frac{n_i^{\rm eq}n_j^{\rm eq}}{n_{\rm eq}^2}~~\text{ and }~n^{\rm eq}=\sum_in_i^{\rm eq}
\label{sigmaeff}
\eea
%%%%%%%%%%%%%%%%%%%%%%%%%%%%%%%%%%%%%%%%%%%%%%%%%%%%%%%%%
Following Eq.~\ref{sigmaeff}, it is straightforward to calculate the $<\sigma v>^{\rm eff}$ for all possible channels in Eqs.~\ref{wimpeq} and \ref{fimpeq}. 

\small
\begin{align}
&<\Gamma >_{\psi_2\to\chi_1\phi}^{\rm eff}=\sum_{i}<\Gamma>_{\psi_2\to i\phi}\\
&<\sigma v>_{\chi_1\overline{\psi}_2\to h\phi}^{\rm eff}=\left[\sum_ig_im_i^2K_2\left(\frac{m_i}{T}\right)\right]^{-1}\sum_{i}{\sb 2}<\sigma v>_{i\overline{\psi}_2\to h\phi} g_im_i^2K_2\left(\frac{m_i}{T}\right)\\&
<\sigma v>_{\phi}^{\rm eff}=\left[\sum_ig_im_i^2K_2\left(\frac{m_i}{T}\right)\right]^{-2}\sum_{i,j}{\sb 2}<\sigma v>_{ij\to\phi\phi} g_ig_jm_i^2m_j^2K_2\left(\frac{m_i}{T}\right)K_2\left(\frac{m_j}{T}\right)\\&
<\sigma v>_{\psi_2}^{\rm eff}=\left[\sum_ig_im_i^2K_2\left(\frac{m_i}{T}\right)\right]^{-2}\sum_{i,j}{\sb 2}<\sigma v>_{ij\to\psi_2\overline{\psi}_2} g_ig_jm_i^2m_j^2K_2\left(\frac{m_i}{T}\right)K_2\left(\frac{m_j}{T}\right)\\&
<\sigma v>_{\rm SM}^{\rm eff}=\left[\sum_ig_im_i^2K_2\left(\frac{m_i}{T}\right)\right]^{-2}\sum_{i,j} {\sb 2}<\sigma v>_{ij\to \rm SM~ SM}g_ig_jm_i^2m_j^2K_2\left(\frac{m_i}{T}\right)K_2\left(\frac{m_j}{T}\right)
\end{align}
%%%%%%%%%%%%%%%%%%%%%%%%%%%%%%%%%%%%%%%%%%%%%%%%%%%%%%%%%

%%%%%%%%%%%%%%%%%%%%%%%%%%%%%%%%%%%%%%%%%%%%%%%%%%%%%%%%%
\section{Relevant Fermion Loop Calculations for direct search}
\label{appendixb}
%%%%%%%%%%%%%%%%%%%%%%%%%%%%%%%%%%%%%%%%%%%%%%%%%%%%%%%%%
\begin{figure}[htb!]
\centering
	\begin{tikzpicture}[baseline={(current bounding box.center)},style={scale=0.575, transform shape}]
	\begin{feynman}
	\vertex (a);
	\vertex[ right=3cm of a] (a2){\(\phi\)};
	\vertex[ left=3cm of a] (a1){\(\phi\)}; 
	\vertex[below=4cm of a] (c); 
	\diagram* {
		(a1) -- [scalar,style=red] (a),(a) -- [scalar,style=red] (a2),(a) -- [scalar,edge label={\(\rm h\)},style=blue] (c) 
	};\end{feynman}
	\end{tikzpicture} + 
	\begin{tikzpicture}[baseline={(current bounding box.center)},style={scale=0.75, transform shape}]
	\begin{feynman}
	\vertex (a);
	\vertex[above left=2cm and 2cm of a] (d1);
	\vertex[above right=2cm and 2cm of a] (d2); 
	\vertex[ left=1.5cm of d1] (a1){\(\phi\)};
	\vertex[ right=1.5cm of d2] (a2){\(\phi\)}; 
	\vertex[below=2cm of a] (b); 
	\diagram* {
		(a1) -- [scalar, momentum ={\(p_2\)},style=red] (d1),(d2) -- [scalar, momentum ={\(p_4\)},style=red] (a2),(d1) --[fermion, edge label={\(\chi_i\)}, arrow size=1pt,  momentum' ={[arrow shorten=0.3]\(p_2+k\)},style=gray](a),(a)--[fermion, edge label={\(\chi_j\)}, arrow size=1pt,  momentum' ={[arrow shorten=0.3]\(p_4+k\)},style=gray](d2),(a) -- [scalar,style=blue, edge label={\(\rm h\)},reversed momentum' ={[arrow shorten=0.3]\(p_4-p_2\)}] (b) ,(d2)--[fermion, edge label={\(\psi_2\)}, arrow size=1pt, momentum' ={[arrow shorten=0.3]\(k\)},style=gray](d1)
	};\end{feynman}
	\end{tikzpicture}  + 
	\begin{tikzpicture}[baseline={(current bounding box.center)},style={scale=0.575, transform shape}]
	\begin{feynman}
	\vertex (c);
	\node[above=4cm of c, crossed dot, style=black](a);
	\vertex[right=3cm and 3cm of a] (a2){\(\phi\)};
	\vertex[left=3cm and 3cm of a] (a1){\(\phi\)}; 
	\diagram* {
		(a1) -- [scalar,style=red] (a),(a) -- [scalar,style=red] (a2),(a) -- [scalar,edge label={\(\rm h\)},style=blue] (c) 
	};\end{feynman}
	\end{tikzpicture}
	\caption{Tree-level, one-loop and counter-term interaction vertex for the interaction term $h\phi\phi$ where $\{i,j=1,2\}$\cite{Schwartz:2014sze, Peskin:1995ev}.}
        \label{fig:-higgs_loop}
\end{figure}
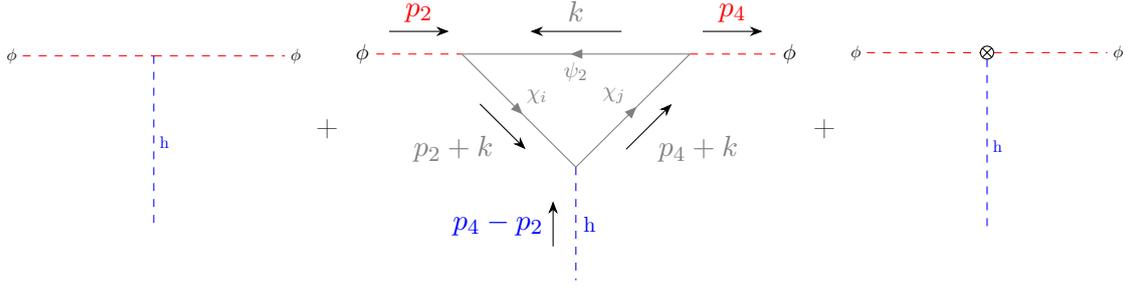
\noindent
We consider first the three-point vertex and its one loop contribution, for the $\phi\phi h$ interaction, which plays crucial role in the direct detection of the pFIMP. The 1-loop  amplitude part can be written as,
\bea
\mathbb{L}_h=-i\lambda_{h\phi\phi}+\sum_{i,j=1,2}\Gamma^{\rm{1-loop}}_{ij}+\delta^{\lambda}_{ij}
\eea
We have assumed that at tree level $\lambda_{\phi H}$ is very small w.r.t other couplings. The one-loop diagram contributes to the $h\phi\phi$ vertex is given in the middle of Fig. \ref{fig:-higgs_loop} and the amplitude is given by,
%%%%%%%%%%%%%%%%%%%%%%%%%%%%%%%%%%%%%%%%%%%%%%%%%%%%%%%%%
\bea\footnotesize\begin{split}
\Gamma^{1-\rm{loop}}_{ij}&=\int \frac{d^4k}{(2\pi)^4}(-1)\rm{Tr}\biggl[\frac{(-i\lambda_{\phi \psi_2\chi_j})i(\slashed k+m_{\psi_2})}{\left[k^2-m_{\psi_2}^2+i\epsilon\right]}\frac{(-i\lambda_{\phi \psi_2\chi_i})i(\slashed p_2 +\slashed k+m_{\chi_i})}{\left[(p_2+k)^2-m_{\chi_i}^2+i\epsilon\right]}\frac{(-i\lambda_{h\chi_i\chi_j})i(\slashed p_4 +\slashed k+m_{\chi_j})}{\left[(p_4+k)^2-m_{\chi_j}^2+i\epsilon\right]}\biggr]
\\&=-\lambda_{h\chi_i\chi_j}\lambda_{\phi \psi_2\chi_j}\lambda_{\phi \psi_2\chi_i}\int \frac{d^4k}{(2\pi)^4}\frac{\rm{Tr}\left[(\slashed k+m_{\psi_2})(\slashed p_2 +\slashed k+m_{\chi_i})(\slashed p_4 +\slashed k+m_{\chi_j})\right]}{\left[k^2-m_{\psi_2}^2+i\epsilon\right]\left[(p_2+k)^2-m_{\chi_i}^2+i\epsilon\right]\left[(p_4+k)^2-m_{\chi_j}^2+i\epsilon\right]}\\&
=-4\int \frac{d^4k}{(2\pi)^4}\frac{m_{\psi_2}(p_2.p_4+k.p_4+k.p_2+k^2)+m_{\chi_i}(k.p_4+k^2)+m_{\chi_j}(k.p_2+k^2)+m_{\chi_i}m_{\chi_j}m_{\psi_2}}{\left[k^2-m_{\psi_2}^2+i\epsilon\right]\left[(k+p_2)^2-m_{\chi_i}^2+i\epsilon\right]\left[(k+p_4)^2-m_{\chi_j}^2+i\epsilon\right]}\\&\hspace{12cm}\times \lambda_{h\chi_i\chi_j}\lambda_{\phi \psi_2\chi_j}\lambda_{\phi \psi_2\chi_i}\\&
=-4\int\frac{d^4k}{(2\pi)^4}\frac{m_{\chi_i}m_{\chi_j}m_{\psi_2}+m_{\psi_2}(m_{\phi}^2-\frac{t}{2})+(m_{\chi_i}+m_{\chi_j}+m_{\psi_2})k^2+(m_{\psi_2}+m_{\chi_j})k.p_2+(m_{\psi_2}+m_{\chi_i})k.p_4}{\left[k^2-m_{\psi_2}^2+i\epsilon\right]\left[(k+p_2)^2-m_{\chi_i}^2+i\epsilon\right]\left[(k+p_4)^2-m_{\chi_j}^2+i\epsilon\right]}\\&\hspace{12cm}\times \lambda_{h\chi_i\chi_j}\lambda_{\phi \psi_2\chi_j}\lambda_{\phi \psi_2\chi_i}
\end{split}\eea
\begin{align*}
&\biggl[\text{Where, }l=k+yp_2+zp_4,\Delta_{ij}=(y+z)(y+z-1)m_{\phi}^2-t yz+xm_{\psi_2}^2+ym_{\chi_i}^2+zm_{\chi_j}^2,\\
&\delta m_{ij}=m_{\psi_2}\left(m_{\chi_i}m_{\chi_j}+m_{\phi}^2(1-y-z)^2-\frac{t}{2}(1-y-z+2yz)\right)+m_{\phi}^2(m_{\chi_i}+m_{\chi_j})(y+z)(y+z-1)+\\&\frac{t}{2}m_{\chi_i}y(1-2z)+\frac{t}{2}m_{\chi_j}z(1-2y),~c_{ij}=m_{\psi_2}+m_{\chi_i}+m_{\chi_j}\biggr].\\
&\text{As } l.p_{2,4} \text{ are odd under }l\to -l\text{ while the  rest of the integrand, is even.}
\end{align*}
\bea\footnotesize\begin{split}
\Gamma^{1-\rm{loop}}_{ij}&=-8\lambda_{h\chi_i\chi_j}\lambda_{\phi \psi_2\chi_j}\lambda_{\phi \psi_2\chi_i}\int \frac{d^4l}{(2\pi)^4}\int_0^1dx~dy~dz\frac{\delta m_{ij}+ c_{ij} ~l^2}{(l^2-\Delta_{ij}+i\epsilon)^3}\delta(x+y+z-1)\\&
=-8\lambda_{h\chi_i\chi_j}\lambda_{\phi \psi_2\chi_j}\lambda_{\phi \psi_2\chi_i}\mu^{4-d}\int \frac{d^dl}{(2\pi)^d}\int_0^1dx~dy~dz\frac{\delta m_{ij}+ c _{ij}~l^2}{(l^2-\Delta_{ij}+i\epsilon)^3}\delta(x+y+z-1)
\\&\text{Where $\mu$ is dimension regularization parameter (basically a mass scale) introduced to keep $\lambda$}\\& \text{dimensionless and $ d = 4-2\epsilon$ in the limit, $\epsilon\to 0_+$.}
\\&=-8\lambda_{h\chi_i\chi_j}\lambda_{\phi \psi_2\chi_j}\lambda_{\phi \psi_2\chi_i}\mu^{4-d}\int_0^1dx~dy~dz\biggl[\delta m_{ij}\frac{(-1)^{3}i}{(4\pi)^{d/2}}\frac{\Gamma(3-d/2)}{\Gamma(3)}\left(\frac{1}{\Delta_{ij}}\right)^{3-d/2}+\\&\hspace{7cm}c_{ij} \frac{(-1)^{3-1}i}{(4\pi)^{d/2}}\frac{d}{2}\frac{\Gamma(3-d/2-1)}{\Gamma(3)}\left(\frac{1}{\Delta_{ij}}\right)^{3-d/2-1}\biggr]\delta(x+y+z-1)
\\&=8i\lambda_{h\chi_i\chi_j}\lambda_{\phi \psi_2\chi_j}\lambda_{\phi \psi_2\chi_i}\int_0^1dx~dy~dz\left[\frac{\delta m_{ij}}{32\pi^{2}}\frac{\Gamma(1+\epsilon)}{\Delta_{ij}^{1+\epsilon}}\left(4\pi\mu^2\right)^{\epsilon}- \frac{c_{ij}}{32\pi^2}\left(2-\epsilon\right)\frac{\Gamma(\epsilon)}{\Delta_{ij}^{\epsilon}}\left(4\pi\mu^2\right)^{\epsilon}\right]\delta(x+y+z-1).
\end{split}\eea
%%%%%%%%%%%%%%%%%%%%%%%%%%%%%%%%%%%%%%%%%%%%%%%%%%%%%%%%%
Around $\epsilon\to 0_{+}$ limit we have the expansions,
\bea\begin{split}
&\Gamma(\epsilon)\simeq\frac{1}{\epsilon}-\gamma_E + \mathcal{O}(\epsilon),\\&
\frac{1}{\Delta^{\epsilon}}\simeq 1-\frac{\epsilon}{2}\ln \Delta^2+\mathcal{O}(\epsilon),\\&
\mu^{\epsilon}\simeq 1+\frac{\epsilon}{2}\ln\mu^2+\mathcal{O}(\epsilon).
\end{split}\eea
where $\gamma_E\approx 0.5772156649$ is the Euler-Mascheroni constant.
Then,
\bea
\nonumber\Gamma^{1-\rm{loop}}_{ij}&=\frac{i}{4\pi^2}\lambda_{h\chi_i\chi_j}\lambda_{\phi \psi_2\chi_j}\lambda_{\phi \psi_2\chi_i}\int_0^1dx~dy~dz\left[\frac{\delta m_{ij}}{\Delta_{ij}}- 2c_{ij}\left(\underbrace{\frac{1}{\epsilon}-\gamma_E+\ln[4\pi]}_{\text{counter term}}-\frac{1}{2}+\ln\frac{\mu^2}{\Delta_{ij}}\right)+\mathcal{O}(\epsilon)\right]\\&\hspace{10cm}\times\delta(x+y+z-1).\label{eq:-loop-div}
\eea
Clearly, in the $\overline{\rm{MS}}$ renormalization scheme the counter term $\delta^{\lambda}_{ij}$ has to be fixed at the following value to cancel the pole in the Eq. \ref{eq:-loop-div},
\bea\begin{split}\delta^{\lambda}_{ij}&=i\lambda_{h\chi_i\chi_j}\lambda_{\phi \psi_2\chi_j}\lambda_{\phi \psi_2\chi_i} \frac{c_{ij}}{2\pi^2}\left(\frac{1}{\epsilon}-\gamma_E+\ln 4\pi\right).\end{split}\eea 
and the total amplitude at $\epsilon\to 0$ becomes,
\begin{align}
\mathbb{L}_h&=-i\lambda_{h\phi\phi}+\frac{i}{4\pi^{2}}\sum_{i,j=1,2}\lambda_{h\chi_i\chi_j}\lambda_{\phi \psi_2\chi_j}\lambda_{\phi \psi_2\chi_i}\int_0^1dx~dy~dz\left[\frac{\delta m_{ij}}{\Delta_{ij}}+c_{ij}\left(1-2\ln\frac{\mu^2}{\Delta_{ij}}\right)\right]\delta(x+y+z-1)\label{loop-fermion}.
\end{align}

 \begin{figure}[htb!]
\centering
\includegraphics[width=0.6\linewidth]{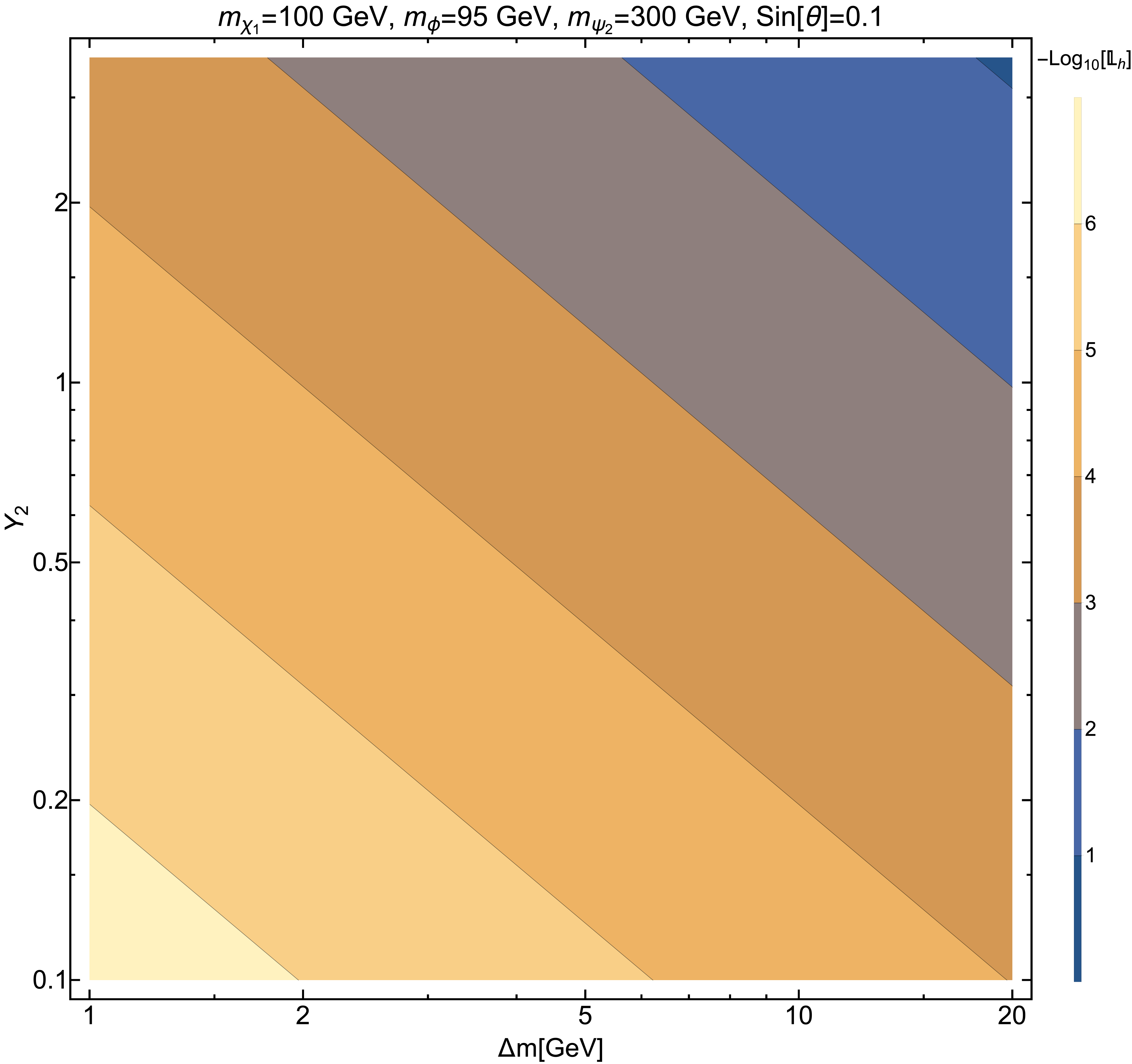}
  \caption{One-loop amplitude for the Higgs-mediated process.}
  \label{fig:-lh-dmy2}
\end{figure}
with,
\bea\small\begin{split}
&\lambda_{\phi \psi_2\overline{\chi}_1}=\lambda_{\phi \overline{\psi}_2\chi_1}=-Y_2\cos\theta ,\\
&\lambda_{\phi \psi_2\overline{\chi}_2}=\lambda_{\phi \overline{\psi}_2\chi_2}=Y_2\sin\theta ,\\
&\lambda_{h\overline{\chi}_1\chi_1}=-\frac{Y_1}{\sqrt{2}}\sin2{\theta} ,\\
&\lambda_{h\overline{\chi}_2\chi_2}= \frac{Y_1}{\sqrt{2}}\sin2{\theta},\\
&\lambda_{h\overline{\chi}_1\chi_2}=\lambda_{h \overline{\chi}_2\chi_1}=-\frac{Y_1}{\sqrt{2}}\cos2\theta .\\
\end{split}\eea
%%%%%%%%%%%%%%%%%%%%%%%%%%%%%%%%%%%%%%%%%%%%%%%%%%%%%%%%%

%\begin{itemize}
%\item{\textbf{Renormalization group equations:}}
%%%%%%%%%%%%%%%%%%%%%%%%%%%%%%%%%%%%%%%%%%%%%%%%%%%%%%%%%
%\end{itemize}

We present next the 1-loop contribution of the $Z$-mediated diagram to the pFIMP-nucleon cross-section. 
%%%%%%%%%%%%%%%%%%%%%%%%%%%%%%%%%%%%%%%%%%%%%%%%%%%%%%%%%
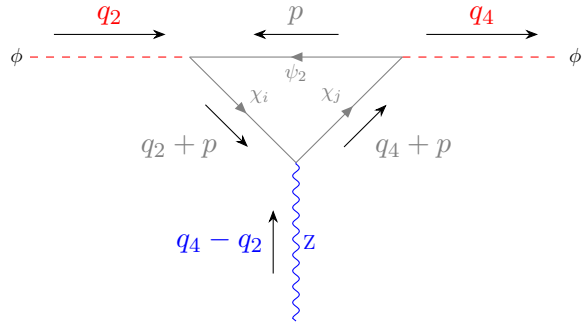
\begin{figure}[htb!]
\centering
	\begin{tikzpicture}[baseline={(current bounding box.center)},style={scale=0.7, transform shape}]
	\begin{feynman}
	\vertex (a);
	\vertex[above left=2cm and 2cm of a] (d1);
	\vertex[above right=2cm and 2cm of a] (d2); 
	\vertex[ left=3cm of d1] (a1){\(\phi\)};
	\vertex[ right= 3cm of d2] (a2){\(\phi\)}; 
	\vertex[below=3cm of a] (b); 
	\diagram* {
		(a1) -- [scalar, momentum ={\(q_2\)},style=red] (d1),(d2) -- [scalar, momentum ={\(q_4\)},style=red] (a2),(d1) --[fermion, edge label={\(\chi_i\)}, arrow size=1pt,  momentum' ={[arrow shorten=0.3]\(q_2+p\)},style=gray](a),(a)--[fermion, edge label={\(\chi_j\)}, arrow size=1pt,  momentum' ={[arrow shorten=0.3]\(q_4+p\)},style=gray](d2),(a) -- [boson,style=blue, edge label={\(\rm Z\)},reversed momentum' ={[arrow shorten=0.3]\(q_4-q_2\)}] (b) ,(d2)--[fermion, edge label={\(\psi_2\)}, arrow size=1pt, momentum' ={[arrow shorten=0.3]\(p\)},style=gray](d1)
	};\end{feynman}
	\end{tikzpicture}  
	\caption{One-loop Feynman diagram for the interaction term $Z\phi\phi$ where $\{i,j=1,2\}$.}
        \label{fig:-z_loop}
\end{figure}
\bea\footnotesize\begin{split}
\mathbb{L}^{ij}_{\mu}&=(-1)\int \frac{d^4p}{(2\pi)^4}\rm{Tr}\biggl[\frac{(-i\lambda_{\phi \psi_2\chi_j})i(\slashed p+m_{\psi_2})}{\left[p^2-m_{\psi_2}^2+i\epsilon\right]}\frac{(-i\lambda_{\phi \psi_2\chi_i})i(\slashed q_2+\slashed p+m_{\chi_i})}{\left[(q_2+p)^2-m_{\chi_i}^2+i\epsilon\right]}\frac{(-i\gamma_{\mu}\lambda_{Z\chi_i\chi_j})i(\slashed q_4+\slashed p+m_{\chi_j})}{\left[(q_4+p)^2-m_{\chi_j}^2+i\epsilon\right]}\biggl]\\&
=-4\lambda_{\phi \psi_2\chi_j}\lambda_{\phi \psi_2\chi_i}\lambda_{Z\chi_i\chi_j}\int \frac{d^4p}{(2\pi)^4}\biggl[\frac{(m_{\psi_2}m_{\chi_j}+p.p+p.q_{4})q_{2_{\mu}}+(m_{\psi_2}m_{\chi_i}+p.p+p.q_2)q_{4_{\mu}}}{\left[p^2-m_{\psi_2}^2+i\epsilon\right]\left[(q_2+p)^2-m_{\chi_i}^2+i\epsilon\right]\left[(q_4+p)^2-m_{\chi_j}^2+i\epsilon\right]}\\&\hspace{4.9cm}+\frac{(m_{\chi_i}m_{\chi_j}+m_{\psi_2}m_{\chi_i}+m_{\psi_2}m_{\chi_j}+p.p-q_2.q_4)p_{\mu}}{\left[p^2-m_{\psi_2}^2+i\epsilon\right]\left[(q_2+p)^2-m_{\chi_i}^2+i\epsilon\right]\left[(q_4+p)^2-m_{\chi_j}^2+i\epsilon\right]}\biggr] \\&
\biggl[\text{Using Feynman parametrization $l=p+yq_2+zq_4$, we define,}\\&
 \Delta^{ij}=(y+z)(y+z-1)m_{\phi}^2-tyz+xm_{\psi_2}^2+ym_{\chi_i}^2+zm_{\chi_j}^2,\\&
 \delta m^{ij}_{\mu}=\left[m_{\psi_2}m_{\chi_j}-(m_{\psi_2}m_{\chi_j}+m_{\psi_2}m_{\chi_i}+m_{\chi_i}m_{\chi_j})y+tyz(y-1)-m_{\phi}^2(z+(y-1)(y+z)^2)\right]q_{2_{\mu}}+\\&
 \hspace{1.15cm}\left[m_{\psi_2}m_{\chi_i}-(m_{\psi_2}m_{\chi_i}+m_{\psi_2}m_{\chi_j}+m_{\chi_i}m_{\chi_j})z+tyz(z-1)-m_{\phi}^2(y+(z-1)(y+z)^2)\right]q_{4_{\mu}},\\&
 c_{\mu}=(1-y)q_{2_{\mu}}+(1-z)q_{4_{\mu}}\biggr].\\&
 =-8 \lambda_{\phi \psi_2\chi_j}\lambda_{\phi \psi_2\chi_i}\lambda_{Z\chi_i\chi_j}\int_0^1~dx~dy~dz~\int \frac{d^4l}{(2\pi)^4}\frac{\delta m^{ij}_{\mu}+c_{\mu} l^2-2yq_2^{\alpha}l_{\alpha}l_{\mu}-2zq_4^{\beta}l_{\beta}l_{\mu}}{\left[l^2-\Delta^{ij}+i\epsilon\right]^3}\delta(x+y+z-1)\\&
 \text{using dimensional regularisation method we may write,}\\&
\equiv -8 \lambda_{\phi \psi_2\chi_j}\lambda_{\phi \psi_2\chi_i}\lambda_{Z\chi_i\chi_j}\mu^{4-d}\int_0^1~dx~dy~dz~\int \frac{d^dl}{(2\pi)^d}\frac{\delta m^{ij}_{\mu}+c_{\mu} l^2-\frac{2}{d}yq_2^{\alpha}l^2g_{\alpha\mu}-\frac{2}{d}zq_4^{\beta}l^2g_{\beta\mu}}{\left[l^2-\Delta^{ij}+i\epsilon\right]^3}\\&\hspace{13cm}\times\delta(x+y+z-1)\\&
=8i\lambda_{\phi \psi_2\chi_j}\lambda_{\phi \psi_2\chi_i}\lambda_{Z\chi_i\chi_j}\int_0^1dx~dy~dz\biggl[\frac{\delta m^{ij}_{\mu}}{32\pi^{2}}\frac{\Gamma(1+\epsilon)}{\Delta^{{ij}^{1+\epsilon}}}\left(4\pi\mu^2\right)^{\epsilon}\\&\hspace{5cm}- \left(2-\epsilon\right)\frac{\left(4\pi\mu^2\right)^{\epsilon}\Gamma(\epsilon)}{32\pi^2\Delta^{{ij}^{\epsilon}}}c_{\mu}+\frac{(4\pi\mu^2)^{\epsilon}\Gamma(\epsilon)}{32\pi^2\Delta^{ij^{\epsilon}}}\left(yq_{2_{\mu}}+zq_{4_{\mu}}\right)\biggr]\delta(x+y+z-1)\\&
=\frac{i}{4\pi^2}\lambda_{\phi \psi_2\chi_j}\lambda_{\phi \psi_2\chi_i}\lambda_{Z\chi_i\chi_j}\int_0^1dx~dy~dz\biggl[\delta m^{ij}_{\mu}\frac{\Gamma(1+\epsilon)}{\Delta^{{ij}^{1+\epsilon}}}\left(4\pi\mu^2\right)^{\epsilon}\\&\hspace{5cm}- (2c_{\mu}-yq_{2_\mu}-zq_{4_\mu})\frac{\Gamma(\epsilon)}{\Delta^{{ij}^{\epsilon}}}\left(4\pi\mu^2\right)^{\epsilon}+c_{\mu}\epsilon\frac{\Gamma(\epsilon)}{\Delta^{ij^{\epsilon}}}(4\pi\mu^2)^{\epsilon}\biggr]\delta(x+y+z-1)\\&
=\frac{i}{4\pi^{2}}\lambda_{\phi \psi_2\chi_j}\lambda_{\phi \psi_2\chi_i}\lambda_{Z\chi_i\chi_j}\int_0^1dx~dy~dz\biggl[\frac{\delta m^{ij}_{\mu}}{\Delta^{ij}}-(2c_{\mu}-yq_{2_\mu}-zq_{4_\mu})\left(\frac{1}{\epsilon}-\gamma_E+\ln[4\pi\mu^2]-\ln\Delta^{ij}\right)\\&\hspace{11.5cm}+c_{\mu}+\mathcal{O}(\epsilon)\biggr]\delta(x+y+z-1).
\end{split}\eea
Where, 
\bea\small\begin{split}
&\lambda_{\phi \psi_2\overline{\chi}_1}=\lambda_{\phi \overline{\psi_2}\chi_1}=-Y_2\cos\theta ,\\
&\lambda_{\phi \psi_2\overline{\chi}_2}=\lambda_{\phi \overline{\psi_2}\chi_2}=Y_2\sin\theta ,\\
&\lambda_{Z \overline{\chi}_1\chi_1}=\frac{g}{2\cos\theta_w}\sin^2{\theta} =\frac{m_Z}{v}\sin^2{\theta} ,\\
&\lambda_{Z \overline{\chi}_2\chi_2}=\frac{g}{2\cos\theta_w}\cos^2{\theta}=\frac{m_Z}{v}\cos^2{\theta} ,\\
&\lambda_{Z \overline{\chi}_1\chi_2}=\lambda_{Z \overline{\chi}_2\chi_1}=\frac{g}{2\cos\theta_w}\sin\theta\cos\theta=\frac{m_Z}{v}\sin\theta\cos\theta.\\
\end{split}\eea
and the total loop amplitude becomes,
\begin{align*}
\mathbb{L}^Z_{\mu}&=\sum_{i,j=1,2}\mathbb{L}^{ij}_{\mu}\\&
=\frac{i}{4\pi^{2}}\sum_{i,j=1,2}\lambda_{\phi \psi_2\chi_j}\lambda_{\phi \psi_2\chi_i}\lambda_{Z\chi_i\chi_j}\int_0^1dx~dy~dz\biggl[\frac{\delta m^{ij}_{\mu}}{\Delta^{ij}}\\&\hspace{3cm}-(2c_{\mu}-yq_{2_\mu}-zq_{4_\mu})\left(\frac{1}{\epsilon}-\gamma_E+\ln[4\pi\mu^2]-\ln\Delta^{ij}\right)+c_{\mu}+\mathcal{O}(\epsilon)\biggr]\delta(x+y+z-1)\\&
\underset{\epsilon\to 0}{=}\frac{i}{4\pi^{2}}\sum_{i,j=1,2}\lambda_{\phi \psi_2\chi_j}\lambda_{\phi \psi_2\chi_i}\lambda_{Z\chi_i\chi_j}\int_0^1dx~dy~dz\left[\frac{\delta m^{ij}_{\mu}}{\Delta^{ij}}+(2c_{\mu}-yq_{2_\mu}-zq_{4_\mu})\ln\Delta^{ij}\right]\delta(x+y+z-1)\\&
\xrightarrow[m_{\chi_1}=m_{\chi_2}]{}0\\&
\xrightarrow[m_{\chi_1}\neq m_{\chi_2}]{} \mathbb{L}_Z (q_{2_{\mu}}+q_{4_{\mu}})
\end{align*}
We have cross-checked our analytical solution with Package‑X \cite{Patel:2016fam} and FeynCalc \cite{Shtabovenko:2020gxv}.

 \begin{figure}[htb!]
\centering
\includegraphics[width=0.6\linewidth]{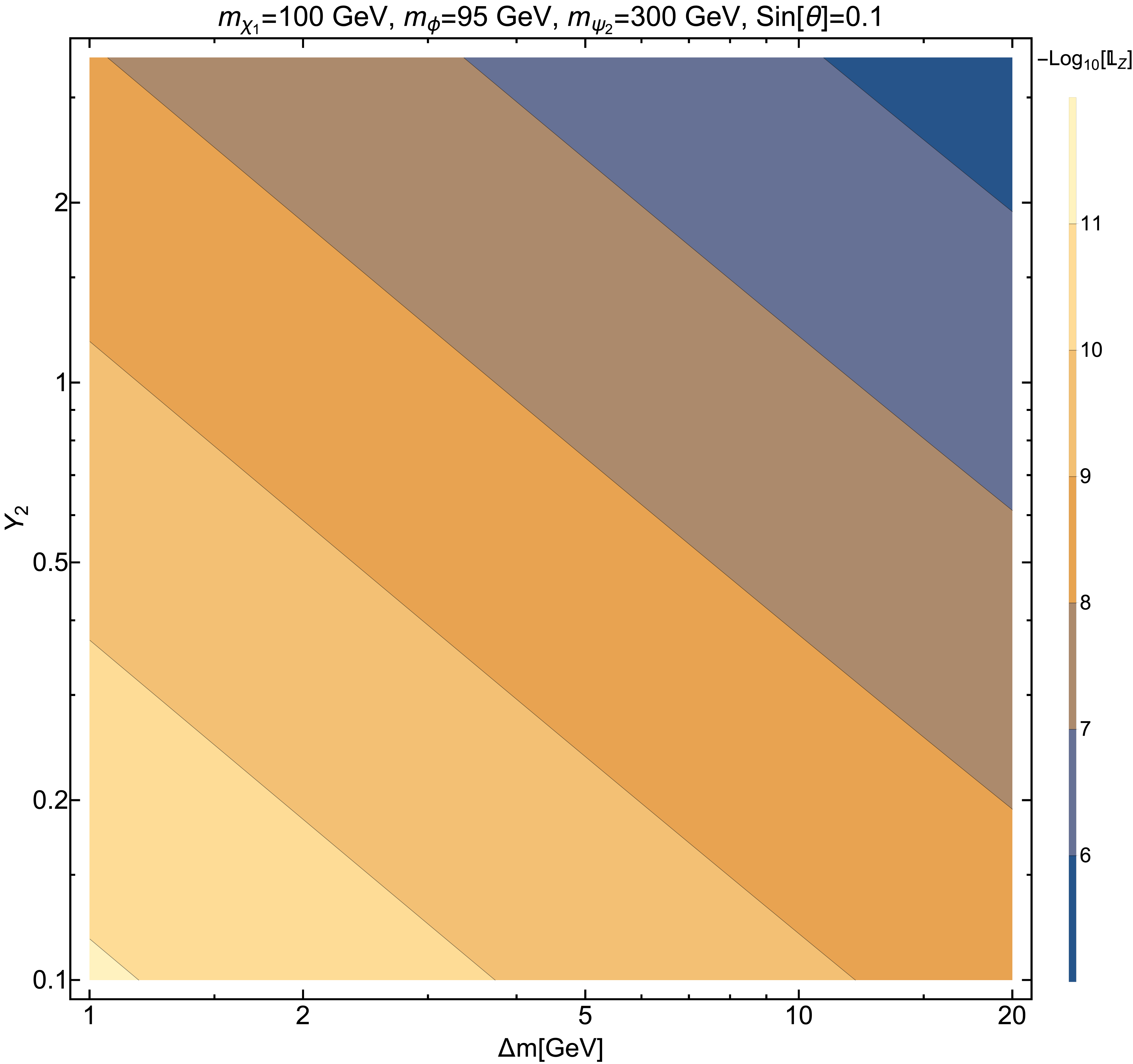}
  \caption{One-loop amplitude for the $Z$-mediated process.}
  \label{fig:-lz-dmy2}
\end{figure}

\section{Calculation of direct detection cross-section of pFIMP}
%%%%%%%%%%%%%%%%%%%%%%%%%%%%%%%%%%%%%%%%%%%%%%%%%%%%%%%%%
\label{appendixc}

The Feynman diagrams for DM $\phi$ scattering off a nucleon at tree-level and one-loop level are shown in Fig. \ref{fig:loopdd}.
%%%%%%%%%%%%%%%%%%%%%%%%%%%%%%%%%%%%%%%%%%%%%%%%%%%%%%%%%
\paragraph{Higgs Mediated pFIMP-Nucleon Scattering:\\}

\noindent
For higgs mediator, the tree-level contribution will be negligible because of tiny $h\phi\phi$ coupling that we have assumed, justifiably, in the pFIMP scenario. The dominant Higgs-mediated contribution therefore comes from 1-loop diagram. The two interaction vertices involved in the loop-induced Higgs-mediated process, are $\mathbb{L}_hh\phi\phi$ and $\frac{m_q}{v} hq\bar{q}$, the Effective Lagrangian at the parton level can be written as,

 \bea\mathcal{L}_{eff}^h=\mathbb{L}_h\frac{m_q}{v}\frac{1}{m_h^2}q\bar{q}\phi\phi=f_q^h q\bar{q}\phi\phi.\eea
The matrix element for a scattering $N~\phi\to N~\phi$ via Higgs-mediation, where $N$ stands for nucleon, is the following:
\begin{align}
i\mathcal{M}^h_{N\phi}&=\alpha_N^h\biggl[\overline{u_N}(q_3)u_N(q_1)\biggr]1
\end{align}
where $\alpha_N^h$ is the effective DM-nucleon coupling and the relation with quark level coupling is,
 \bea
 \frac{\alpha_N^{h}}{m_N}=\sum_{q=u,d,s}f^{N}_{T_q}\frac{f_q^{\rm h}}{m_q}+\frac{2}{27}(1-\sum_{q=u,d,s}f^N_{T_q})\sum_{q=c,b,t}\frac{f_q^{\rm h}}{m_q}.
 \label{eq:alph-h}\eea
 Where the nuclear form-factors are defined by~\cite{Bertone:2004pz,Ellis:2018dmb} 
\bea
<N|m_q\overline{q}q|N>\equiv m_Nf^N_{T_q}<N|N>(q=u,d,s)
\eea

 \begin{table}[htb!]
\vspace{-0.3cm}
\begin{center}
%\scriptsize
\begin{tabular}{|c||c|c|c|c||c|c|c|}
\hline
Nucleon & $f^N_{T_u}$ & $f^N_{T_d}$ & $f^N_{T_s}$ & $f^N_{T_G}$ & $f^N_{T_c}$ & $f^N_{T_b}$ & $f^N_{T_t}$\\
\hline
Proton & 0.018(5) & 0.027(7) & 0.037(17) & 0.917(19) &0.078(2) & 0.072(2) & 0.069(1) \\
Neutron & 0.013(3) & 0.040(10) & 0.037(17) & 0.910(20) & 0.078(2) & 0.071(2) & 0.068(2) \\
\hline
\end{tabular}\caption{Values of the $f^{N}_{T_{q,G}}$. }\label{tab:fTvalues}
\end{center}
\end{table}
 %Finally,
\bea\begin{split}
\overline{|\mathcal{M}_{N\phi}^h|^2}&=\frac{1}{2}\sum_{\rm{all~spin}}|\mathcal{M}^h_{N\phi}|^2\\&=4m_N^2|\alpha_N^h|^2
\end{split}\eea
%%%%%%%%%%%%%%%%%%%%%%%%%%%%%%%%%%%%%%%%%%%%%%%%%%%%%%%%%
\paragraph{Z Mediated pFIMP-Nucleon Scattering:\\}
%%%%%%%%%%%%%%%%%%%%%%%%%%%%%%%%%%%%%%%%%%%%%%%%%%%%%%%%%
The Effective Lagrangian for Z mediator direct search process can be written as, 
\begin{align}
\mathcal{L}_{eff}^Z&\nonumber=\bar{q}\left[\frac{g}{\cos\theta_W}\gamma^{\mu}\frac{1}{2}\left(c_V^q-c_A^q\gamma^5\right)\right]q\frac{\mathbb{L}_Z}{m_Z^2}\phi(q_{2_{\mu}}+q_{4_{\nu}})\phi\rm \hspace{3cm}SI+SD\\\nonumber&\to\frac{m_Z}{v}c_V^q\frac{\mathbb{L}_Z}{m_Z^2}\bar{q}\gamma^{\mu}q\phi(q_{2_{\mu}}+q_{4_{\mu}})\phi\rm \hspace{6cm}SI\\\nonumber&
=\frac{c_V^q}{v}\frac{\mathbb{L}_Z}{m_Z}\bar{q}\gamma^{\mu}q\phi(q_{2_{\mu}}+q_{4_{\mu}})\phi\\\nonumber&
=\frac{c_V^q}{v}\frac{\mathbb{L}_Z}{m_Z}\bar{q}\gamma^{\mu}q\phi(q_{2_{\mu}}+q_{4_{\mu}})\phi\\&
=b_q^Z\bar{q}\gamma^{\mu}q\phi(q_{2_{\mu}}+q_{4_{\mu}})\phi
\end{align}
%The Feynman diagram for DM $\phi$ scattering off a nucleon at 1-loop level is shown in the right of Fig. \ref{fig:loopdd} (Z-boson mediator) and
The matrix element for DM nucleon scattering $N~\phi\to N~\phi$, is the following by assuming $\rm Let,~\mathbb{L}_{\mu}^Z=\mathbb{L}_Z(q_{2_{\mu}}+q_{4_{\mu}})$:
\begin{align}
i\mathcal{M}^Z_{N\phi}&=b_N^Z\biggl[\overline{u_N}(q_3)\gamma^{\mu}u_N(q_1)\biggr](q_{2_{\mu}}+q_{4_{\mu}})
\end{align}
where $b_N^Z$ is the DM-nucleon effective coupling. As the sea-quarks and the gluons do not contribute to the vector current. Only valence quark contributions all add up due to the conservation of the vector current which gives us, $b_p = 2b_u + b_d$ and $b_n = b_u + 2b_d$. So the relation effective DM-nucleon couplings with quark level coupling are \cite{Agrawal:2010fh}, 
\bea\begin{split}
&b_p^Z=2b_u^Z+b_d^Z\\&
b_n^Z=b_u^Z+2b_d^Z
\label{eq:vec-quark-nucleon}
\end{split}\eea
with $b_q^Z=\frac{c_V^q}{v}\frac{\mathbb{L}_Z}{m_Z}$.
\begin{align}
\overline{|\mathcal{M}^Z_{N\phi}|^2}&=\frac{|b_N^Z|^2}{2}\rm Tr\left[(\slashed q_3+m_N)\gamma^{\mu}(\slashed q_1+m_{N})\gamma^{\nu}\right](q_{2_{\mu}}+q_{4_{\mu}})(q_{2_{\nu}}+q_{4_{\nu}})\\\nonumber
&=\frac{|b_N^Z|^2}{2} 4(q_1^{\mu}q_3^{\nu}+q_3^{\mu}q_1^{\nu})(q_{2_{\mu}}+q_{4_{\mu}})(q_{2_{\nu}}+q_{4_{\nu}})\\\nonumber
&=4|b_N^Z|^2 \left(q_1.q_2+q_1.q_4)~(q_3.q_2+q_3.q_4\right) \hspace{1.7cm}\\
&=4|b_N^Z|^2 4m_{\phi}^2m_N^2\hspace{1cm}\text{  as initially the nucleus is in rest, }q_1\sim \{m_N,\vec{0}\}
\end{align}

\paragraph{The interference:\\}
Let us now calculate the cross-term,
\begin{align}
\nonumber\overline{|\mathcal{M}^Z_{N\phi}|^{\dagger}|\mathcal{M}^h_{N\phi}|}
&=\frac{1}{2}\sum_{\rm{all~spin}}\left[b_N^Z\left(\overline{u}_N(q_3)\gamma^{\mu}u_N(q_1)\right)(q_{2_{\mu}}+q_{4_{\mu}})\right]^{\dagger}\left[\alpha_N^h\left(\overline{u}_N(q_3)u_N(q_1)\right)1\right]
\\\nonumber&=\frac{b_N^{Z^{\dagger}}\alpha_N^h}{2}\rm Tr\left[\gamma^{\mu}(\slashed q_3+m_N)(\slashed q_1+m_{N})\right](q_{2_{\mu}}+q_{4_{\mu}})
\\\nonumber&=2m_Nb_N^{Z^{\dagger}}\alpha_N^h \left(q_1.q_2+q_1.q_4+q_2.q_3+q_3.q_4\right) \hspace{1.7cm}\\\nonumber
&=2m_Nb_N^{Z^{\dagger}}\alpha_N^h 4m_{\phi}m_N\hspace{1cm}\text{  as initially the nucleus is in rest, }q_1\sim \{m_N,\vec{0}\}\\&=8m_N^2m_{\phi}b_N^{Z^{\dagger}}\alpha_N^h
\end{align}
In a similar way,
\begin{align}
\nonumber\overline{|\mathcal{M}^Z_{N\phi}||\mathcal{M}^h_{N\phi}|^{\dagger}}
&=\frac{1}{2}\sum_{\rm{all~spin}}\left[b_N^Z\left(\overline{u}_N(q_3)\gamma^{\mu}u_N(q_1)\right)(q_{2_{\mu}}+q_{4_{\mu}})\right]\left[\alpha_N^h\left(\overline{u}_N(q_3)u_N(q_1)\right)1\right]^{\dagger}
\\\nonumber&=\frac{b_N^{Z}\alpha_N^{h^{\dagger}}}{2}\rm Tr\left[\gamma^{\mu}(\slashed q_1+m_N)(\slashed q_3+m_{N})\right](q_{2_{\mu}}+q_{4_{\mu}})
\\\nonumber&=2m_Nb_N^{Z}\alpha_N^{h^{\dagger}} \left(q_1.q_2+q_1.q_4+q_2.q_3+q_3.q_4\right) \hspace{1.7cm}\\\nonumber
&=2m_Nb_N^{Z}\alpha_N^{h^{\dagger}} 4m_{\phi}m_N\hspace{1cm}\text{  as initially the nucleus is in rest, }q_1\sim \{m_N,\vec{0}\}\\&=8m_N^2m_{\phi}b_N^{Z}\alpha_N^{h^{\dagger}}
\end{align}
The spin-independent $\rm pFIMP-nucleon$ scattering cross-section for $Z$ mediator, taking the non-relativistic limit and assuming the initial nucleon is at rest, is given by \cite{Lin:2019uvt}, 
\begin{align}
\nonumber\rm \sigma_{\phi N}&=\frac{1}{4m_{\phi}m_N|w-v_N|}\int \frac{d^3q_3}{(2\pi)^32m_{N}}\frac{d^3q_4}{(2\pi)^32m_{\phi}}\overline{|\mathcal{M}^Z_{N\phi}+\mathcal{M}^h_{N\phi}|^2}(2\pi)^4\delta^4(q_1+q_2-q_3-q_4)\\&
=\int\frac{\overline{|\mathcal{M}^Z_{N\phi}+\mathcal{M}^h_{N\phi}|^2}}{4\pi^2(4m_{\phi}m_N)^2|w-v_N|} d^3q_3~d^3q_4\delta(E_1+E_2-E_3-E_4)\delta^3(\vec{q}_1+\vec{q}_2-\vec{q}_3-\vec{q}_4)\label{eq:nr-sz}
\end{align}
From energy conservation,
 \begin{align}
 &\nonumber2\mu_{\phi N}\vec{q}_2.\vec{q}_3=m_{\phi}|\vec{q}_3|^2\\&
 |\vec{q}_3|=2\mu_{\phi N}w\cos\theta
\end{align}
where $w$, $v_N$ is the initial velocity of the dark matter and nucleus. We have assumed that in Lab-frame the nucleus initially is in rest i.e. $|\vec{v}_N|=0$ so the relative velocity between DM and nucleus becomes $w$. The angle between $\vec{q}_2$ and $\vec{q}_3$ is $\theta$.
Then Eq. \ref{eq:nr-sz} becomes,
\begin{align}
\nonumber\rm \sigma_{\phi N}&=\int\frac{\overline{|\mathcal{M}^Z_{N\phi}+\mathcal{M}^h_{N\phi}|^2}}{4\pi^2(4m_{\phi}m_N)^2w} (\pi |\vec{q}_3|d\cos\theta~ d|\vec{q}_3|^2)~d^3q_4\delta(E_1+E_2-E_3-E_4)\delta^3(\vec{q}_2-\vec{q}_3-\vec{q}_4)
\\\nonumber&=\frac{\left|2m_N\alpha_N^h+2m_Nb_N^{Z^{\dagger}}2m_{\phi}\right|^2}{4\pi(4m_{\phi}m_N)^2w}\int (|\vec{q}_3|d\cos\theta~ d|\vec{q}_3|^2)\delta(E_1+E_2-E_3-E_4)
\\\nonumber&=\frac{\left|2m_N\alpha_N^h+2m_Nb_N^{Z^{\dagger}}2m_{\phi}\right|^2}{4\pi(4m_{\phi}m_N)^2w}\int (|\vec{q}_3|d\cos\theta~ d|\vec{q}_3|^2)\frac{1}{|\vec{q}_3|w}\delta(\cos\theta-\frac{|\vec{q}_3|}{2\mu_{\phi N}w})
\\\nonumber&=\frac{\left|2m_N\alpha_N^h+2m_Nb_N^{Z^{\dagger}}2m_{\phi}\right|^2}{4\pi(4m_{\phi}m_N)^2w^2}\int_{-1}^1 \int_0^{4\mu_{\phi N}^2 w^2}d\cos\theta~ d|\vec{q}_3|^2\delta(\cos\theta-\frac{|\vec{q}_3|}{2\mu_{\phi N}w})
\\\nonumber&=\frac{\left|2m_N\alpha_N^h+2m_Nb_N^{Z^{\dagger}}2m_{\phi}\right|^2}{4\pi(4m_{\phi}m_N)^2w^2}4\mu_{\phi N}^2 w^2
\\&=\frac{\mu_{\phi N}^2}{4\pi m_{\phi}^2}\left|\alpha_N^h+2m_{\phi}b_N^{Z^{\dagger}}\right|^2
\end{align}

%\par The spin-independent $\phi$ (FIMP) DM-nucleon effective scattering cross-section in presence of another DM $\chi_1$ is given by,{\sb
%\bea\begin{split}
% \sigma_{{\phi N}_{\rm eff}}^{\rm{SI}} &=\frac{\Omega_{\phi}}{\Omega_{\chi_1}+\Omega_{\phi}} \frac{\mu_{\phi N}^2}{4\pi m_{\phi}^2}\left|\alpha_N^h+2m_{\phi}b_N^{Z^{\dagger}}\right|^2
%\end{split}\eea}
%where $\mu_{\phi N}=m_{\phi}m_N/(m_{\phi}+m_N)$ is the pFIMP-nucleon reduced mass.
%\par The standard spin-independent (SI), Scalar DM (pFIMP) Nucleus cross-section at zero transfer momentum is \cite{PhysRevD.89.123521},
%\begin{align}
% \nonumber\sigma_{\rm T}^{\rm SI} &=\frac{\mu_{\rm T}^2}{4\pi m_{\phi}^2}\left|Z(\alpha_p^h+2m_{\phi}b_p^{Z^{\dagger}})+(A-Z)(\alpha_n^h+2m_{\phi}b_n^{Z^{\dagger}})\right|^2
%\end{align}
%where $\rm \mu_T=\frac{m_Tm_{DM}}{m_T+m_{DM}}$ is DM-nucleus reduced mass with $\rm m_T,~m_{DM}$ is the mass of the target nucleus and dark matter respectively.
%%%%%%%%%%%%%%%%%%%%%%%%%%%%%%%%%%%%%%%%%%%%%%%%%%%%%%%%%
\section{Calculation of direct detection cross-section of WIMP}
%%%%%%%%%%%%%%%%%%%%%%%%%%%%%%%%%%%%%%%%%%%%%%%%%%%%%%%%%
\label{appendixd}

The Feynman diagrams corresponding to WIMP DM $\chi_1$ scattering off a nucleon at tree level are shown in Fig. \ref{fig:wimp-dd}. 
%%%%%%%%%%%%%%%%%%%%%%%%%%%%%%%%%%%%%%%%%%%%%%%%%%%%%%%%%

\paragraph{Higgs Mediated WIMP-Nucleon Scattering:\\\\}
%%%%%%%%%%%%%%%%%%%%%%%%%%%%%%%%%%%%%%%%%%%%%%%%%%%%%%%%%
The two relevant interaction vertices here are $\lambda_{h\chi_1\overline{\chi}_1}h\chi_1\overline{\chi}_1$ and $\frac{m_q}{v} hq\bar{q}$. Effective Lagrangian for spin-independent direct search process can be written for higgs mediator as,
\bea\mathcal{L}_{eff}^h=\frac{m_q}{v}\frac{1}{m_h^2}\lambda_{h\chi_1\overline{\chi}_1}q\bar{q}\chi_1\overline{\chi}_1=F_q^h q\bar{q}\chi_1\overline{\chi}_1\eea
Where $\lambda_{h\chi_1\overline{\chi}_1}=-\frac{Y_1}{\sqrt{2}}\sin 2\theta$ in our model.

The matrix element for a scattering $N\chi_1\to N\chi_1$ via higgs mediation where $N$ stands for nucleon, is the following:
\begin{align}
i\mathcal{M}^h_{N\chi_1}&=\beta_N^h\biggl[\overline{u}_N(q_3)u_N(q_1)\biggr]\biggl[\overline{u}_{\chi_1}(q_4)u_{\chi_1}(q_2)\biggr]
\end{align}
where $\beta_N^h$ is the DM-nucleon coupling is related with the quark level coupling $F_q^h$ following Eq.~\ref{eq:alph-h}. 
And,
\bea\begin{split}
\overline{|\mathcal{M}_{N\chi_1}^h|^2}&=\frac{1}{4}\sum_{\rm{all~spin}}|\mathcal{M}^h_{N\chi_1}|^2
\\&=16m_{\chi_1}^2m_N^2|\beta_N^h|^2
\end{split}\eea
%%%%%%%%%%%%%%%%%%%%%%%%%%%%%%%%%%%%%%%%%%%%%%%%%%%%%%%%%
\paragraph{Z Mediated WIMP-Nucleon Scattering:\\\\}
%%%%%%%%%%%%%%%%%%%%%%%%%%%%%%%%%%%%%%%%%%%%%%%%%%%%%%%%%
 In the case of the $Z$ mediator, similar to pFIMP, only vector term will contribute to SI cross-section and the effective Lagrangian,
  \begin{align}
  \nonumber\mathcal{L}_{eff}^Z&=\frac{m_Z}{v}\frac{1}{m_Z^2}\lambda_{Z\chi_1\overline{\chi}_1}\overline{q}\gamma^{\mu}\left(c_V^q-c_A^q\gamma^5\right)q\overline{\chi}_1\gamma_{\mu}\chi_1\hspace{4cm}\rm SI+SD\\\nonumber& \to 
  \frac{m_Z}{v}\frac{c_V^q}{m_Z^2}\lambda_{Z\chi_1\overline{\chi}_1}\overline{q}\gamma^{\mu}q\overline{\chi}_1\gamma_{\mu}\chi_1\hspace{6cm}\rm SI
  \\&=B_q \overline{\chi}_1\gamma^{\mu}\chi_1\bar{q}\gamma_{\mu}q
  \end{align}
  where, $\lambda_{Z\chi_1\overline{\chi}_1}=\frac{m_Z}{v}\sin^2\theta$.\par
The matrix element for a scattering $N~\chi_1\to N~\chi_1$ via $Z$ mediation is the following:
\begin{align}
i\mathcal{M}^Z_{N\chi_1}&=B_N^Z\biggl[\overline{u}_N(q_3)\gamma^{\mu}u_N(q_1)\biggr]\biggl[\overline{u}_{\chi_1}(q_4)\gamma_{\mu}u_{\chi_1}(q_2)\biggr]
\end{align}
where $B_N^Z$ is the DM-nucleon coupling is related with the quark level coupling $B_q$ is followed by the Eq. \ref{eq:vec-quark-nucleon}. And,
\bea\begin{split}
\overline{|\mathcal{M}_{N\chi_1}^Z|^2}&=\frac{1}{4}\sum_{\rm{all~spin}}|\mathcal{M}^Z_{N\chi_1}|^2\\&=16m_{\chi_1}^2m_N^2|B_N^Z|^2
\end{split}\eea

%Following Eq. \ref{eq:nr-sz}
Now we get the spin-independent $\chi_1 $(WIMP) DM-nucleon scattering cross section in presence of another DM $\phi$ (pFIMP) as \cite{Jungman:1995df,Hisano:2015bma},

\bea\begin{split}
\sigma^{\rm{SI}}_{{\chi_1 N}_{\rm{eff}}}=\frac{\Omega_{\chi_1}}{\Omega_{\chi_1}+\Omega_{\phi}}\frac{\mu_{\chi_1N}^2}{\pi}\left|\beta_N^h+B_N^Z\right|^2
\end{split}
\eea
where $\mu_{\chi_1 N}=m_{\chi_1}m_N/(m_{\chi_1}+m_N)$ is the wimp-nucleon reduced mass, $m_N\sim0.939~\rm GeV$ denotes the nucleon mass.

%The spin-independent Fermion DM (WIMP) nucleus cross-section at zero transfer momentum is \cite{Agrawal:2010fh},
%\begin{align}
% \sigma_{\rm T}^{\rm SI} &=\frac{\mu_{\rm T}^2}{\pi }\left|Z(\beta_p^h+B_p^Z)+(A-Z)(\beta_n^h+B_n^Z)\right|^2
% \end{align}
%where $\rm \mu_T=\frac{m_Tm_{DM}}{m_T+m_{DM}}$ is DM-nucleus reduced mass with $\rm m_T,~m_{DM}$ is the mass of the target nucleus and dark matter respectively.

%%%%%%%%%%%%%%%%%%%%%%%%%%%%%%%%%%%%%%%%%%%%%%%%%%%%%%%%%
\section{Higgs and Z Invisible decay width}
%%%%%%%%%%%%%%%%%%%%%%%%%%%%%%%%%%%%%%%%%%%%%%%%%%%%%%%%%
\label{appendixe}
The observed (expected) upper limit on the invisible branching fraction of the Higgs boson is found to be at $95\%$ confidence level \cite{ATLAS:2022yvh,CMS:2022qva} with total decay width of 125.1 GeV Higgs is $3.2^{+2.8}_{-2.2}\rm~ MeV$ \cite{ParticleDataGroup:2022pth},
\bea\begin{split}
\rm \mathcal{B}_{h\to\rm{invisible}}< \begin{cases}0.145(0.103)~~~\rm( ATLAS)\\0.18(0.10)\rm~~~( CMS)\end{cases}
\end{split}\label{higgs_invisible_decay}\eea
%%%%%%%%%%%%%%%%%%%%%%%%%%%%%%%%%%%%%%%%%%%%%%%%%%%%%%%%%
\begin{align}
&\Gamma_{h\to\phi\phi}=\frac{(\lambda_{\phi H}v+\mathbb{L}_h)^2}{32\pi m_h}\left(1-4\frac{m_{\phi}^2}{m_h^2}\right)^{1/2}\Theta[m_h-2m_{\phi}]\\
&\Gamma_{h\to\chi_1\overline{\chi}_1}=\frac{\sin^4 2\theta}{32\pi v^2}m_h(m_{\chi_1}-m_{\chi_2})^2\left(1-4\frac{m_{\chi_1}^2}{m_h^2}\right)^{3/2}\Theta[m_h-2m_{\chi_1}]
\end{align}
%%%%%%%%%%%%%%%%%%%%%%%%%%%%%%%%%%%%%%%%%%%%%%%%%%%%%%%%%
The recent Z-boson Invisible decay width bound has come from various experiments like \cite{CMS:2021qbc,CMS:2022ett},
\bea\begin{split}
\rm\Gamma_{Z\to invisible}
<  \begin{cases}\rm523\pm 16~MeV~~~( CMS)\\\rm 503\pm 16~MeV~~~(LEP Comb.)\\\rm498\pm 17 MeV~~~(L3)\end{cases}
\end{split}\label{Z_invisible_decay}
\eea

\begin{align}
&\Gamma_{Z\to\phi\phi}=\frac{\mathbb{L}_Z^2m_{Z}}{16\pi}\left(1-4\frac{m_{\phi}^2}{m_Z^2}\right)^{3/2}\Theta[m_Z-2m_{\phi}]\\
&\Gamma_{Z\to\chi_1\overline{\chi}_1}=\frac{m_Z^3\sin^4\theta\sin^4\theta_w}{12\pi v^2}\left(1+2\frac{m_{\chi_1}^2}{m_Z^2}\right)\left(1-4\frac{m_{\chi_1}^2}{m_Z^2}\right)^{1/2}\Theta[m_Z-2m_{\chi_1}]
\end{align}
If the dark matter masses are below resonance then parameter space is also constrained by the invisible Higgs and $Z$ decay constraints.

\end{appendix}

\newpage
\bibliographystyle{JHEP}
\bibliography{wimp_fimp}

\providecommand{\href}[2]{#2}\begingroup\raggedright\begin{thebibliography}{10}

\bibitem{Planck:2018vyg}
{\scshape Planck} collaboration, \emph{{Planck 2018 results. VI. Cosmological
  parameters}},
  \href{https://doi.org/10.1051/0004-6361/201833910}{\emph{Astron. Astrophys.}
  {\bfseries 641} (2020) A6}
  [\href{https://arxiv.org/abs/1807.06209}{{\ttfamily 1807.06209}}].

\bibitem{Gondolo:1990dk}
P.~Gondolo and G.~Gelmini, \emph{{Cosmic abundances of stable particles:
  Improved analysis}},
  \href{https://doi.org/10.1016/0550-3213(91)90438-4}{\emph{Nucl. Phys. B}
  {\bfseries 360} (1991) 145}.

\bibitem{Jungman:1995df}
G.~Jungman, M.~Kamionkowski and K.~Griest, \emph{{Supersymmetric dark matter}},
  \href{https://doi.org/10.1016/0370-1573(95)00058-5}{\emph{Phys. Rept.}
  {\bfseries 267} (1996) 195}
  [\href{https://arxiv.org/abs/hep-ph/9506380}{{\ttfamily hep-ph/9506380}}].

\bibitem{Hochberg:2014dra}
Y.~Hochberg, E.~Kuflik, T.~Volansky and J.G.~Wacker, \emph{{Mechanism for
  Thermal Relic Dark Matter of Strongly Interacting Massive Particles}},
  \href{https://doi.org/10.1103/PhysRevLett.113.171301}{\emph{Phys. Rev. Lett.}
  {\bfseries 113} (2014) 171301}
  [\href{https://arxiv.org/abs/1402.5143}{{\ttfamily 1402.5143}}].

\bibitem{Hall:2009bx}
L.J.~Hall, K.~Jedamzik, J.~March-Russell and S.M.~West, \emph{{Freeze-In
  Production of FIMP Dark Matter}},
  \href{https://doi.org/10.1007/JHEP03(2010)080}{\emph{JHEP} {\bfseries 03}
  (2010) 080} [\href{https://arxiv.org/abs/0911.1120}{{\ttfamily 0911.1120}}].

\bibitem{McDonald:2001vt}
J.~McDonald, \emph{{Thermally generated gauge singlet scalars as
  selfinteracting dark matter}},
  \href{https://doi.org/10.1103/PhysRevLett.88.091304}{\emph{Phys. Rev. Lett.}
  {\bfseries 88} (2002) 091304}
  [\href{https://arxiv.org/abs/hep-ph/0106249}{{\ttfamily hep-ph/0106249}}].

\bibitem{Kaplinghat:2013yxa}
M.~Kaplinghat, S.~Tulin and H.-B.~Yu, \emph{{Direct Detection Portals for
  Self-interacting Dark Matter}},
  \href{https://doi.org/10.1103/PhysRevD.89.035009}{\emph{Phys. Rev. D}
  {\bfseries 89} (2014) 035009}
  [\href{https://arxiv.org/abs/1310.7945}{{\ttfamily 1310.7945}}].

\bibitem{Pappadopulo:2016pkp}
D.~Pappadopulo, J.T.~Ruderman and G.~Trevisan, \emph{{Dark matter freeze-out in
  a nonrelativistic sector}},
  \href{https://doi.org/10.1103/PhysRevD.94.035005}{\emph{Phys. Rev. D}
  {\bfseries 94} (2016) 035005}
  [\href{https://arxiv.org/abs/1602.04219}{{\ttfamily 1602.04219}}].

\bibitem{XENON:2018voc}
{\scshape XENON} collaboration, \emph{{Dark Matter Search Results from a One
  Ton-Year Exposure of XENON1T}},
  \href{https://doi.org/10.1103/PhysRevLett.121.111302}{\emph{Phys. Rev. Lett.}
  {\bfseries 121} (2018) 111302}
  [\href{https://arxiv.org/abs/1805.12562}{{\ttfamily 1805.12562}}].

\bibitem{XENON:2020kmp}
{\scshape XENON} collaboration, \emph{{Projected WIMP sensitivity of the
  XENONnT dark matter experiment}},
  \href{https://doi.org/10.1088/1475-7516/2020/11/031}{\emph{JCAP} {\bfseries
  11} (2020) 031} [\href{https://arxiv.org/abs/2007.08796}{{\ttfamily
  2007.08796}}].

\bibitem{XENONCollaboration:2022kmb}
{\scshape (XENON Collaboration)\textdagger{}\textdagger{}, XENON}
  collaboration, \emph{{Search for New Physics in Electronic Recoil Data from
  XENONnT}}, \href{https://doi.org/10.1103/PhysRevLett.129.161805}{\emph{Phys.
  Rev. Lett.} {\bfseries 129} (2022) 161805}
  [\href{https://arxiv.org/abs/2207.11330}{{\ttfamily 2207.11330}}].

\bibitem{PandaX-II:2021nsg}
{\scshape PandaX-II} collaboration, \emph{{Search for Light Dark
  Matter-Electron Scatterings in the PandaX-II Experiment}},
  \href{https://doi.org/10.1103/PhysRevLett.126.211803}{\emph{Phys. Rev. Lett.}
  {\bfseries 126} (2021) 211803}
  [\href{https://arxiv.org/abs/2101.07479}{{\ttfamily 2101.07479}}].

\bibitem{PandaX-4T:2021bab}
{\scshape PandaX-4T} collaboration, \emph{{Dark Matter Search Results from the
  PandaX-4T Commissioning Run}},
  \href{https://doi.org/10.1103/PhysRevLett.127.261802}{\emph{Phys. Rev. Lett.}
  {\bfseries 127} (2021) 261802}
  [\href{https://arxiv.org/abs/2107.13438}{{\ttfamily 2107.13438}}].

\bibitem{LZ:2022ufs}
{\scshape LZ} collaboration, \emph{{First Dark Matter Search Results from the
  LUX-ZEPLIN (LZ) Experiment}},
  \href{https://arxiv.org/abs/2207.03764}{{\ttfamily 2207.03764}}.

\bibitem{Goodman:2010yf}
J.~Goodman, M.~Ibe, A.~Rajaraman, W.~Shepherd, T.M.P.~Tait and H.-B.~Yu,
  \emph{{Constraints on Light Majorana dark Matter from Colliders}},
  \href{https://doi.org/10.1016/j.physletb.2010.11.009}{\emph{Phys. Lett. B}
  {\bfseries 695} (2011) 185}
  [\href{https://arxiv.org/abs/1005.1286}{{\ttfamily 1005.1286}}].

\bibitem{Goodman:2010ku}
J.~Goodman, M.~Ibe, A.~Rajaraman, W.~Shepherd, T.M.P.~Tait and H.-B.~Yu,
  \emph{{Constraints on Dark Matter from Colliders}},
  \href{https://doi.org/10.1103/PhysRevD.82.116010}{\emph{Phys. Rev. D}
  {\bfseries 82} (2010) 116010}
  [\href{https://arxiv.org/abs/1008.1783}{{\ttfamily 1008.1783}}].

\bibitem{Rajaraman:2011wf}
A.~Rajaraman, W.~Shepherd, T.M.P.~Tait and A.M.~Wijangco, \emph{{LHC Bounds on
  Interactions of Dark Matter}},
  \href{https://doi.org/10.1103/PhysRevD.84.095013}{\emph{Phys. Rev. D}
  {\bfseries 84} (2011) 095013}
  [\href{https://arxiv.org/abs/1108.1196}{{\ttfamily 1108.1196}}].

\bibitem{Fox:2011pm}
P.J.~Fox, R.~Harnik, J.~Kopp and Y.~Tsai, \emph{{Missing Energy Signatures of
  Dark Matter at the LHC}},
  \href{https://doi.org/10.1103/PhysRevD.85.056011}{\emph{Phys. Rev. D}
  {\bfseries 85} (2012) 056011}
  [\href{https://arxiv.org/abs/1109.4398}{{\ttfamily 1109.4398}}].

\bibitem{Buchmueller:2013dya}
O.~Buchmueller, M.J.~Dolan and C.~McCabe, \emph{{Beyond Effective Field Theory
  for Dark Matter Searches at the LHC}},
  \href{https://doi.org/10.1007/JHEP01(2014)025}{\emph{JHEP} {\bfseries 01}
  (2014) 025} [\href{https://arxiv.org/abs/1308.6799}{{\ttfamily 1308.6799}}].

\bibitem{Petrov:2013nia}
A.A.~Petrov and W.~Shepherd, \emph{{Searching for dark matter at LHC with
  Mono-Higgs production}},
  \href{https://doi.org/10.1016/j.physletb.2014.01.051}{\emph{Phys. Lett. B}
  {\bfseries 730} (2014) 178}
  [\href{https://arxiv.org/abs/1311.1511}{{\ttfamily 1311.1511}}].

\bibitem{Altmannshofer:2014cla}
W.~Altmannshofer, P.J.~Fox, R.~Harnik, G.D.~Kribs and N.~Raj, \emph{{Dark
  Matter Signals in Dilepton Production at Hadron Colliders}},
  \href{https://doi.org/10.1103/PhysRevD.91.115006}{\emph{Phys. Rev. D}
  {\bfseries 91} (2015) 115006}
  [\href{https://arxiv.org/abs/1411.6743}{{\ttfamily 1411.6743}}].

\bibitem{Capdevilla:2017doz}
R.M.~Capdevilla, A.~Delgado, A.~Martin and N.~Raj, \emph{{Characterizing dark
  matter at the LHC in Drell-Yan events}},
  \href{https://doi.org/10.1103/PhysRevD.97.035016}{\emph{Phys. Rev. D}
  {\bfseries 97} (2018) 035016}
  [\href{https://arxiv.org/abs/1709.00439}{{\ttfamily 1709.00439}}].

\bibitem{Bell:2015sza}
N.F.~Bell, Y.~Cai, J.B.~Dent, R.K.~Leane and T.J.~Weiler, \emph{{Dark matter at
  the LHC: Effective field theories and gauge invariance}},
  \href{https://doi.org/10.1103/PhysRevD.92.053008}{\emph{Phys. Rev. D}
  {\bfseries 92} (2015) 053008}
  [\href{https://arxiv.org/abs/1503.07874}{{\ttfamily 1503.07874}}].

\bibitem{Yu:2013aca}
Z.-H.~Yu, Q.-S.~Yan and P.-F.~Yin, \emph{{Detecting interactions between dark
  matter and photons at high energy $e^+e^-$ colliders}},
  \href{https://doi.org/10.1103/PhysRevD.88.075015}{\emph{Phys. Rev. D}
  {\bfseries 88} (2013) 075015}
  [\href{https://arxiv.org/abs/1307.5740}{{\ttfamily 1307.5740}}].

\bibitem{Essig:2013vha}
R.~Essig, J.~Mardon, M.~Papucci, T.~Volansky and Y.-M.~Zhong,
  \emph{{Constraining Light Dark Matter with Low-Energy $e^+e^-$ Colliders}},
  \href{https://doi.org/10.1007/JHEP11(2013)167}{\emph{JHEP} {\bfseries 11}
  (2013) 167} [\href{https://arxiv.org/abs/1309.5084}{{\ttfamily 1309.5084}}].

\bibitem{Kadota:2014mea}
K.~Kadota and J.~Silk, \emph{{Constraints on Light Magnetic Dipole Dark Matter
  from the ILC and SN 1987A}},
  \href{https://doi.org/10.1103/PhysRevD.89.103528}{\emph{Phys. Rev. D}
  {\bfseries 89} (2014) 103528}
  [\href{https://arxiv.org/abs/1402.7295}{{\ttfamily 1402.7295}}].

\bibitem{Yu:2014ula}
Z.-H.~Yu, X.-J.~Bi, Q.-S.~Yan and P.-F.~Yin, \emph{{Dark matter searches in the
  mono-$Z$ channel at high energy $e^+e^-$ colliders}},
  \href{https://doi.org/10.1103/PhysRevD.90.055010}{\emph{Phys. Rev. D}
  {\bfseries 90} (2014) 055010}
  [\href{https://arxiv.org/abs/1404.6990}{{\ttfamily 1404.6990}}].

\bibitem{Freitas:2014jla}
A.~Freitas and S.~Westhoff, \emph{{Leptophilic Dark Matter in Lepton
  Interactions at LEP and ILC}},
  \href{https://doi.org/10.1007/JHEP10(2014)116}{\emph{JHEP} {\bfseries 10}
  (2014) 116} [\href{https://arxiv.org/abs/1408.1959}{{\ttfamily 1408.1959}}].

\bibitem{Dutta:2017ljq}
S.~Dutta, D.~Sachdeva and B.~Rawat, \emph{{Signals of Leptophilic Dark Matter
  at the ILC}},
  \href{https://doi.org/10.1140/epjc/s10052-017-5188-8}{\emph{Eur. Phys. J. C}
  {\bfseries 77} (2017) 639}
  [\href{https://arxiv.org/abs/1704.03994}{{\ttfamily 1704.03994}}].

\bibitem{Habermehl:2020njb}
M.~Habermehl, M.~Berggren and J.~List, \emph{{WIMP Dark Matter at the
  International Linear Collider}},
  \href{https://doi.org/10.1103/PhysRevD.101.075053}{\emph{Phys. Rev. D}
  {\bfseries 101} (2020) 075053}
  [\href{https://arxiv.org/abs/2001.03011}{{\ttfamily 2001.03011}}].

\bibitem{Baltz:2006fm}
E.A.~Baltz, M.~Battaglia, M.E.~Peskin and T.~Wizansky, \emph{{Determination of
  dark matter properties at high-energy colliders}},
  \href{https://doi.org/10.1103/PhysRevD.74.103521}{\emph{Phys. Rev. D}
  {\bfseries 74} (2006) 103521}
  [\href{https://arxiv.org/abs/hep-ph/0602187}{{\ttfamily hep-ph/0602187}}].

\bibitem{Belanger:2018sti}
G.~B\'elanger et~al., \emph{{LHC-friendly minimal freeze-in models}},
  \href{https://doi.org/10.1007/JHEP02(2019)186}{\emph{JHEP} {\bfseries 02}
  (2019) 186} [\href{https://arxiv.org/abs/1811.05478}{{\ttfamily
  1811.05478}}].

\bibitem{Boehm:2003bt}
C.~Boehm, D.~Hooper, J.~Silk, M.~Casse and J.~Paul, \emph{{MeV dark matter: Has
  it been detected?}},
  \href{https://doi.org/10.1103/PhysRevLett.92.101301}{\emph{Phys. Rev. Lett.}
  {\bfseries 92} (2004) 101301}
  [\href{https://arxiv.org/abs/astro-ph/0309686}{{\ttfamily
  astro-ph/0309686}}].

\bibitem{Boehm:2002yz}
C.~Boehm, T.A.~Ensslin and J.~Silk, \emph{{Can Annihilating dark matter be
  lighter than a few GeVs?}},
  \href{https://doi.org/10.1088/0954-3899/30/3/004}{\emph{J. Phys. G}
  {\bfseries 30} (2004) 279}
  [\href{https://arxiv.org/abs/astro-ph/0208458}{{\ttfamily
  astro-ph/0208458}}].

\bibitem{Tylka:1989wt}
A.J.~Tylka and F.W.~Stecker, \emph{{$\gamma^-$ ray SIGNATURES OF DARK MATTER
  ANNIHILATION IN THE GALAXY}},  6, 1989.

\bibitem{STEPHENS198955}
S.~Stephens, \emph{Antiprotons in cosmic rays and their implications},
  \href{https://doi.org/https://doi.org/10.1016/0273-1177(89)90308-6}{\emph{Advances
  in Space Research} {\bfseries 9} (1989) 55}.

\bibitem{PhysRev.171.1344}
C.S.~Shen and G.B.~Berkey, \emph{Antiprotons and positrons in cosmic rays},
  \href{https://doi.org/10.1103/PhysRev.171.1344}{\emph{Phys. Rev.} {\bfseries
  171} (1968) 1344}.

\bibitem{osti_6276777}
F.W.~Stecker and A.J.~Tylka, \emph{Cosmic-ray antiproton spectrum from dark
  matter annihilation and its astrophysical implications - a new look},
  \href{https://doi.org/10.1086/185359}{\emph{Astrophys. J.; (United States)}
  }.

\bibitem{Evoli:2011id}
C.~Evoli, I.~Cholis, D.~Grasso, L.~Maccione and P.~Ullio, \emph{{Antiprotons
  from dark matter annihilation in the Galaxy: astrophysical uncertainties}},
  \href{https://doi.org/10.1103/PhysRevD.85.123511}{\emph{Phys. Rev. D}
  {\bfseries 85} (2012) 123511}
  [\href{https://arxiv.org/abs/1108.0664}{{\ttfamily 1108.0664}}].

\bibitem{Delahaye:2007fr}
T.~Delahaye, R.~Lineros, F.~Donato, N.~Fornengo and P.~Salati, \emph{{Positrons
  from dark matter annihilation in the galactic halo: Theoretical
  uncertainties}},
  \href{https://doi.org/10.1103/PhysRevD.77.063527}{\emph{Phys. Rev. D}
  {\bfseries 77} (2008) 063527}
  [\href{https://arxiv.org/abs/0712.2312}{{\ttfamily 0712.2312}}].

\bibitem{Bergstrom:2013jra}
L.~Bergstrom, T.~Bringmann, I.~Cholis, D.~Hooper and C.~Weniger, \emph{{New
  Limits on Dark Matter Annihilation from AMS Cosmic Ray Positron Data}},
  \href{https://doi.org/10.1103/PhysRevLett.111.171101}{\emph{Phys. Rev. Lett.}
  {\bfseries 111} (2013) 171101}
  [\href{https://arxiv.org/abs/1306.3983}{{\ttfamily 1306.3983}}].

\bibitem{Tylka:1989xj}
A.J.~Tylka, \emph{{Cosmic Ray Positrons From Annihilation of Weakly Interacting
  Massive Particles in the Galaxy}},
  \href{https://doi.org/10.1103/PhysRevLett.63.840}{\emph{Phys. Rev. Lett.}
  {\bfseries 63} (1989) 840}.

\bibitem{Cao:2007fy}
Q.-H.~Cao, E.~Ma, J.~Wudka and C.P.~Yuan, \emph{{Multipartite dark matter}},
  \href{https://arxiv.org/abs/0711.3881}{{\ttfamily 0711.3881}}.

\bibitem{Zurek:2008qg}
K.M.~Zurek, \emph{{Multi-Component Dark Matter}},
  \href{https://doi.org/10.1103/PhysRevD.79.115002}{\emph{Phys. Rev. D}
  {\bfseries 79} (2009) 115002}
  [\href{https://arxiv.org/abs/0811.4429}{{\ttfamily 0811.4429}}].

\bibitem{Profumo:2009tb}
S.~Profumo, K.~Sigurdson and L.~Ubaldi, \emph{{Can we discover multi-component
  WIMP dark matter?}},
  \href{https://doi.org/10.1088/1475-7516/2009/12/016}{\emph{JCAP} {\bfseries
  12} (2009) 016} [\href{https://arxiv.org/abs/0907.4374}{{\ttfamily
  0907.4374}}].

\bibitem{Bhattacharya:2013hva}
S.~Bhattacharya, A.~Drozd, B.~Grzadkowski and J.~Wudka, \emph{{Two-Component
  Dark Matter}}, \href{https://doi.org/10.1007/JHEP10(2013)158}{\emph{JHEP}
  {\bfseries 10} (2013) 158} [\href{https://arxiv.org/abs/1309.2986}{{\ttfamily
  1309.2986}}].

\bibitem{DiazSaez:2021pfw}
B.~D\'\i{}az~S\'aez, K.~M\"ohling and D.~St\"ockinger, \emph{{Two real scalar
  WIMP model in the assisted freeze-out scenario}},
  \href{https://doi.org/10.1088/1475-7516/2021/10/027}{\emph{JCAP} {\bfseries
  10} (2021) 027} [\href{https://arxiv.org/abs/2103.17064}{{\ttfamily
  2103.17064}}].

\bibitem{Belanger:2011ww}
G.~Belanger and J.-C.~Park, \emph{{Assisted freeze-out}},
  \href{https://doi.org/10.1088/1475-7516/2012/03/038}{\emph{JCAP} {\bfseries
  03} (2012) 038} [\href{https://arxiv.org/abs/1112.4491}{{\ttfamily
  1112.4491}}].

\bibitem{Maity:2019hre}
T.N.~Maity and T.S.~Ray, \emph{{Exchange driven freeze out of dark matter}},
  \href{https://doi.org/10.1103/PhysRevD.101.103013}{\emph{Phys. Rev. D}
  {\bfseries 101} (2020) 103013}
  [\href{https://arxiv.org/abs/1908.10343}{{\ttfamily 1908.10343}}].

\bibitem{Ghosh:2021wrk}
P.~Ghosh, P.~Konar, A.K.~Saha and S.~Show, \emph{{Self-interacting freeze-in
  dark matter in a singlet doublet scenario}},
  \href{https://doi.org/10.1088/1475-7516/2022/10/017}{\emph{JCAP} {\bfseries
  10} (2022) 017} [\href{https://arxiv.org/abs/2112.09057}{{\ttfamily
  2112.09057}}].

\bibitem{Bhattacharya:2022dco}
S.~Bhattacharya, J.~Lahiri and D.~Pradhan, \emph{{Dynamics of the pseudo-FIMP
  in presence of a thermal Dark Matter}},
  \href{https://arxiv.org/abs/2212.07622}{{\ttfamily 2212.07622}}.

\bibitem{delAguila:1989rq}
F.~del Aguila, L.~Ametller, G.L.~Kane and J.~Vidal, \emph{{Vector Like Fermion
  and Standard Higgs Production at Hadron Colliders}},
  \href{https://doi.org/10.1016/0550-3213(90)90655-W}{\emph{Nucl. Phys. B}
  {\bfseries 334} (1990) 1}.

\bibitem{CMS:2018yfx}
{\scshape CMS} collaboration, \emph{{Search for invisible decays of a Higgs
  boson produced through vector boson fusion in proton-proton collisions at
  $\sqrt{s} =$ 13 TeV}},
  \href{https://doi.org/10.1016/j.physletb.2019.04.025}{\emph{Phys. Lett. B}
  {\bfseries 793} (2019) 520}
  [\href{https://arxiv.org/abs/1809.05937}{{\ttfamily 1809.05937}}].

\bibitem{Griest:1990kh}
K.~Griest and D.~Seckel, \emph{{Three exceptions in the calculation of relic
  abundances}}, \href{https://doi.org/10.1103/PhysRevD.43.3191}{\emph{Phys.
  Rev. D} {\bfseries 43} (1991) 3191}.

\bibitem{PhysRevLett.119.061102}
R.T.~D'Agnolo, D.~Pappadopulo and J.T.~Ruderman, \emph{Fourth exception in the
  calculation of relic abundances},
  \href{https://doi.org/10.1103/PhysRevLett.119.061102}{\emph{Phys. Rev. Lett.}
  {\bfseries 119} (2017) 061102}.

\bibitem{Belanger:2014vza}
G.~B\'elanger, F.~Boudjema, A.~Pukhov and A.~Semenov, \emph{{micrOMEGAs4.1: two
  dark matter candidates}},
  \href{https://doi.org/10.1016/j.cpc.2015.03.003}{\emph{Comput. Phys. Commun.}
  {\bfseries 192} (2015) 322}
  [\href{https://arxiv.org/abs/1407.6129}{{\ttfamily 1407.6129}}].

\bibitem{DAgnolo:2018wcn}
R.T.~D'Agnolo, C.~Mondino, J.T.~Ruderman and P.-J.~Wang, \emph{{Exponentially
  Light Dark Matter from Coannihilation}},
  \href{https://doi.org/10.1007/JHEP08(2018)079}{\emph{JHEP} {\bfseries 08}
  (2018) 079} [\href{https://arxiv.org/abs/1803.02901}{{\ttfamily
  1803.02901}}].

\bibitem{Chua:2013zpa}
C.-K.~Chua and R.-C.~Hsieh, \emph{{Study of Dirac fermionic dark matter}},
  \href{https://doi.org/10.1103/PhysRevD.88.036011}{\emph{Phys. Rev. D}
  {\bfseries 88} (2013) 036011}
  [\href{https://arxiv.org/abs/1305.7008}{{\ttfamily 1305.7008}}].

\bibitem{Edsjo:1997bg}
J.~Edsjo and P.~Gondolo, \emph{{Neutralino relic density including
  coannihilations}},
  \href{https://doi.org/10.1103/PhysRevD.56.1879}{\emph{Phys. Rev. D}
  {\bfseries 56} (1997) 1879}
  [\href{https://arxiv.org/abs/hep-ph/9704361}{{\ttfamily hep-ph/9704361}}].

\bibitem{Duda:2002hf}
G.~Duda, G.~Gelmini, P.~Gondolo, J.~Edsjo and J.~Silk, \emph{{Indirect
  detection of a subdominant density component of cold dark matter}},
  \href{https://doi.org/10.1103/PhysRevD.67.023505}{\emph{Phys. Rev. D}
  {\bfseries 67} (2003) 023505}
  [\href{https://arxiv.org/abs/hep-ph/0209266}{{\ttfamily hep-ph/0209266}}].

\bibitem{Ullio:2002pj}
P.~Ullio, L.~Bergstrom, J.~Edsjo and C.G.~Lacey, \emph{{Cosmological dark
  matter annihilations into gamma-rays - a closer look}},
  \href{https://doi.org/10.1103/PhysRevD.66.123502}{\emph{Phys. Rev. D}
  {\bfseries 66} (2002) 123502}
  [\href{https://arxiv.org/abs/astro-ph/0207125}{{\ttfamily
  astro-ph/0207125}}].

\bibitem{PhysRevD.104.083026}
V.~Gammaldi, J.~P\'erez-Romero, J.~Coronado-Bl\'azquez, M.~Di~Mauro,
  E.V.~Karukes, M.A.~S\'anchez-Conde et~al., \emph{Dark matter search in dwarf
  irregular galaxies with the fermi large area telescope},
  \href{https://doi.org/10.1103/PhysRevD.104.083026}{\emph{Phys. Rev. D}
  {\bfseries 104} (2021) 083026}.

\bibitem{Super-Kamiokande:2020sgt}
{\scshape Super-Kamiokande} collaboration, \emph{{Indirect search for dark
  matter from the Galactic Center and halo with the Super-Kamiokande
  detector}}, \href{https://doi.org/10.1103/PhysRevD.102.072002}{\emph{Phys.
  Rev. D} {\bfseries 102} (2020) 072002}
  [\href{https://arxiv.org/abs/2005.05109}{{\ttfamily 2005.05109}}].

\bibitem{HESS:2022ygk}
{\scshape H.E.S.S.} collaboration, \emph{{Search for Dark Matter Annihilation
  Signals in the H.E.S.S. Inner Galaxy Survey}},
  \href{https://doi.org/10.1103/PhysRevLett.129.111101}{\emph{Phys. Rev. Lett.}
  {\bfseries 129} (2022) 111101}
  [\href{https://arxiv.org/abs/2207.10471}{{\ttfamily 2207.10471}}].

\bibitem{IceCube:2016umi}
{\scshape IceCube} collaboration, \emph{{Observation and Characterization of a
  Cosmic Muon Neutrino Flux from the Northern Hemisphere using six years of
  IceCube data}},
  \href{https://doi.org/10.3847/0004-637X/833/1/3}{\emph{Astrophys. J.}
  {\bfseries 833} (2016) 3} [\href{https://arxiv.org/abs/1607.08006}{{\ttfamily
  1607.08006}}].

\bibitem{IceCube:2022vtr}
{\scshape IceCube} collaboration, \emph{{Searches for Connections between Dark
  Matter and High-Energy Neutrinos with IceCube}},
  \href{https://arxiv.org/abs/2205.12950}{{\ttfamily 2205.12950}}.

\bibitem{IceCube:2021xzo}
{\scshape IceCube} collaboration, \emph{{Search for GeV-scale dark matter
  annihilation in the Sun with IceCube DeepCore}},
  \href{https://doi.org/10.1103/PhysRevD.105.062004}{\emph{Phys. Rev. D}
  {\bfseries 105} (2022) 062004}
  [\href{https://arxiv.org/abs/2111.09970}{{\ttfamily 2111.09970}}].

\bibitem{DiazSaez:2021pmg}
B.~D\'\i{}az~S\'aez, P.~Escalona, S.~Norero and A.R.~Zerwekh, \emph{{Fermion
  singlet dark matter in a pseudoscalar dark matter portal}},
  \href{https://doi.org/10.1007/JHEP10(2021)233}{\emph{JHEP} {\bfseries 10}
  (2021) 233} [\href{https://arxiv.org/abs/2105.04255}{{\ttfamily
  2105.04255}}].

\bibitem{Reinert:2017aga}
A.~Reinert and M.W.~Winkler, \emph{{A Precision Search for WIMPs with Charged
  Cosmic Rays}},
  \href{https://doi.org/10.1088/1475-7516/2018/01/055}{\emph{JCAP} {\bfseries
  01} (2018) 055} [\href{https://arxiv.org/abs/1712.00002}{{\ttfamily
  1712.00002}}].

\bibitem{Fermi-LAT:2015att}
{\scshape Fermi-LAT} collaboration, \emph{{Searching for Dark Matter
  Annihilation from Milky Way Dwarf Spheroidal Galaxies with Six Years of Fermi
  Large Area Telescope Data}},
  \href{https://doi.org/10.1103/PhysRevLett.115.231301}{\emph{Phys. Rev. Lett.}
  {\bfseries 115} (2015) 231301}
  [\href{https://arxiv.org/abs/1503.02641}{{\ttfamily 1503.02641}}].

\bibitem{Schwartz:2014sze}
M.D.~Schwartz, \emph{{Quantum Field Theory and the Standard Model}}, Cambridge
  University Press (3, 2014).

\bibitem{Peskin:1995ev}
M.E.~Peskin and D.V.~Schroeder, \emph{{An Introduction to quantum field
  theory}}, Addison-Wesley, Reading, USA (1995).

\bibitem{Patel:2016fam}
H.H.~Patel, \emph{{Package-X 2.0: A Mathematica package for the analytic
  calculation of one-loop integrals}},
  \href{https://doi.org/10.1016/j.cpc.2017.04.015}{\emph{Comput. Phys. Commun.}
  {\bfseries 218} (2017) 66}
  [\href{https://arxiv.org/abs/1612.00009}{{\ttfamily 1612.00009}}].

\bibitem{Shtabovenko:2020gxv}
V.~Shtabovenko, R.~Mertig and F.~Orellana, \emph{{FeynCalc 9.3: New features
  and improvements}},
  \href{https://doi.org/10.1016/j.cpc.2020.107478}{\emph{Comput. Phys. Commun.}
  {\bfseries 256} (2020) 107478}
  [\href{https://arxiv.org/abs/2001.04407}{{\ttfamily 2001.04407}}].

\bibitem{Bertone:2004pz}
G.~Bertone, D.~Hooper and J.~Silk, \emph{{Particle dark matter: Evidence,
  candidates and constraints}},
  \href{https://doi.org/10.1016/j.physrep.2004.08.031}{\emph{Phys. Rept.}
  {\bfseries 405} (2005) 279}
  [\href{https://arxiv.org/abs/hep-ph/0404175}{{\ttfamily hep-ph/0404175}}].

\bibitem{Ellis:2018dmb}
J.~Ellis, N.~Nagata and K.A.~Olive, \emph{{Uncertainties in WIMP Dark Matter
  Scattering Revisited}},
  \href{https://doi.org/10.1140/epjc/s10052-018-6047-y}{\emph{Eur. Phys. J. C}
  {\bfseries 78} (2018) 569}
  [\href{https://arxiv.org/abs/1805.09795}{{\ttfamily 1805.09795}}].

\bibitem{Agrawal:2010fh}
P.~Agrawal, Z.~Chacko, C.~Kilic and R.K.~Mishra, \emph{{A Classification of
  Dark Matter Candidates with Primarily Spin-Dependent Interactions with
  Matter}},  \href{https://arxiv.org/abs/1003.1912}{{\ttfamily 1003.1912}}.

\bibitem{Lin:2019uvt}
T.~Lin, \emph{{Dark matter models and direct detection}},
  \href{https://doi.org/10.22323/1.333.0009}{\emph{PoS} {\bfseries 333} (2019)
  009} [\href{https://arxiv.org/abs/1904.07915}{{\ttfamily 1904.07915}}].

\bibitem{Hisano:2015bma}
J.~Hisano, R.~Nagai and N.~Nagata, \emph{{Effective Theories for Dark Matter
  Nucleon Scattering}},
  \href{https://doi.org/10.1007/JHEP05(2015)037}{\emph{JHEP} {\bfseries 05}
  (2015) 037} [\href{https://arxiv.org/abs/1502.02244}{{\ttfamily
  1502.02244}}].

\bibitem{ATLAS:2022yvh}
{\scshape ATLAS} collaboration, \emph{{Search for invisible Higgs-boson decays
  in events with vector-boson fusion signatures using 139 fb$^{-1}$ of
  proton-proton data recorded by the ATLAS experiment}},
  \href{https://doi.org/10.1007/JHEP08(2022)104}{\emph{JHEP} {\bfseries 08}
  (2022) 104} [\href{https://arxiv.org/abs/2202.07953}{{\ttfamily
  2202.07953}}].

\bibitem{CMS:2022qva}
{\scshape CMS} collaboration, \emph{{Search for invisible decays of the Higgs
  boson produced via vector boson fusion in proton-proton collisions at
  s=13\,\,TeV}}, \href{https://doi.org/10.1103/PhysRevD.105.092007}{\emph{Phys.
  Rev. D} {\bfseries 105} (2022) 092007}
  [\href{https://arxiv.org/abs/2201.11585}{{\ttfamily 2201.11585}}].

\bibitem{ParticleDataGroup:2022pth}
{\scshape Particle Data Group} collaboration, \emph{{Review of Particle
  Physics}}, \href{https://doi.org/10.1093/ptep/ptac097}{\emph{PTEP} {\bfseries
  2022} (2022) 083C01}.

\bibitem{CMS:2021qbc}
{\scshape CMS} collaboration, \emph{{ Precision measurement of the Z invisible
  width with the CMS experiment in pp collisions at $\sqrt{s}=13~\mathrm{TeV}$
  }}, .

\bibitem{CMS:2022ett}
{\scshape CMS} collaboration, \emph{{Precision measurement of the Z boson
  invisible width in pp collisions at $\sqrt{s}$ = 13 TeV}},
  \href{https://arxiv.org/abs/2206.07110}{{\ttfamily 2206.07110}}.

\end{thebibliography}\endgroup
\end{document}